\documentclass[journal=langmuir,manuscript=article]{achemso}

\usepackage[version=3]{mhchem} 

\usepackage{color}
\usepackage{rotating}
\usepackage{amssymb}
\usepackage{xr} 
\author{Daniela T\"auber}
\email{daniela.taeuber@physik.tu-chemnitz.de}
\author{Ines Trenkmann}
\author{Christian von Borczyskowski}
\affiliation[Chemnitz University of Technology]
{Institute of Physics, Chemnitz University of Technology, Chemnitz,
Germany}

\title[\texttt{achemso} demonstration]
{Influence of van der Waals interactions on morphology and dynamics in ultrathin liquid films at silicon oxide interfaces}

\begin{document}
\begin{abstract}
Single molecule tracer diffusion studies of evaporating (thinning) ultrathin tetrakis-2-ethyl-hexoxysilane (TEHOS) films on silicon with 100~nm thermal oxide reveal a considerable slowdown of the molecular mobility within less than 4~nm  above the substrate (corresponding to a few molecular TEHOS layers). This is related to restricted mobility and structure formation of the liquid in this region, in agreement with information obtained from a long-time ellipsometric study of thinning TEHOS films on silicon substrates with 100~nm thermal or 2~nm native oxide. Both show evidence for the formation of up to four layers. Additionally, on thermal oxide, a lateral flow of the liquid is observed, while the film on the native oxide forms an almost flat surface and shows negligible flow. Thus, on the 2~nm native oxide the liquid mobility is even more restricted in close vicinity to the substrate as compared to the 100~nm thermal oxide. In addition, we found a significantly smaller initial film thickness in case of the native oxide under similar dipcoating conditions. We ascribe these differences to van der Waals interactions with the underlying silicon in case of the native oxide, whereas the thermal oxide suffices to shield those interactions.
\end{abstract}

\section{Introduction}
Thin liquid films are widely present in nature, in particular as surface water layers. Also various technical applications make use of thin liquid films for lubrication, coatings or intermediate layers. Wettability considerations play an important role in the design of porous materials for catalysis and liquid storage\cite{qiao_pressurized_2009, *hummer_water_2001, *holt_fast_2006}. Slippage, diffusion and flow characteristics of confined liquids have to be taken into account in nanoscopic Lab-on-a-Chip fabrication\cite{eijkel_liquid_2007, *mijatovic_technologies_2005, *wang_rapid_2012}. Enhanced flexibility as well as economic considerations have led to an increasing interest in printing almost any kind of electronic devices~\cite{comiskey_electrophoretic_1998, *sheats_manufacturing_2004}. This also includes inkjet printing~\cite{calvert_inkjet_2001}, for which droplet formation and evaporation is an essential feature. From numerous investigations, the surface roughness and chemistry is known to influence droplet shapes, wetting characteristics~\cite{marmur_wetting_2003, *feng_effect_2011, *nosonovsky_materials_2011} and thus also droplet evaporation~\cite{picknett_evaporation_1977, *bourges-monnier_influence_1995, *tadmor_approaches_2011}. 

Deviations from the macroscopic behavior are observed on a molecular scale. For example, liquids show adsorption layers with a thickness of one or two molecular diameters\cite{fisher_experimental_1981}. Partially wetting liquids form a precursor film of similar thickness~\cite{leger_liquid_1992, *xu_molecular_2004}. 
Seemann et al.\ found an influence of long-range capillary forces for nanometric polymer melt droplets in case of small contact angles ($\theta\approx 5^\circ$). For larger contact angles ($\theta>45^\circ$), the shape of the droplets could be  described by surface properties and deviations invoking intermolecular interactions were only seen in the vicinity of the three-phase contact line~\cite{seemann_polystyrene_2001}. Investigations of thin liquid films by ellipsometry and X-ray reflectometry revealed molecular layering of up to four liquid layers at interfaces with solids~\cite{heslot_molecular_1989, *villette_wetting_1996, forcada_molecular_1993, yu_observation_1999, *yu_molecular_2000, *mo_observation_2006, yu_x-ray_2000}, which was further confirmed by simulations~\cite{snook_solvation_1980, *bitsanis_molecular_1990, *stroud_capillary_2001, kaplan_structural_2006}. 

The question, whether the observed liquid layering has an influence on the molecular dynamics of the liquid, motivated optically detected single molecule tracer experiments in confined liquids. Thereby, a slowdown of the tracer diffusion in wetting precursor layers of liquid droplets, in ultrathin liquid films and in a liquid film at a step edge on mica was observed~\cite{schuster_diffusion_2000, *schuster_anisotropic_2003, tauber_single_2009, *trenkmann_investigations_2009, tauber_characterization_2011, patil_combined_2007, *schob_single_2010, grabowski_comparing_2007}.  Furthermore, single molecule tracking (SMT) experiments, revealed a broad distribution of diffusion coefficients~\cite{schuster_diffusion_2000, *schuster_anisotropic_2003, tauber_single_2009, *trenkmann_investigations_2009, tauber_characterization_2011}, which indicates heterogeneous diffusion. The observation of heterogeneous diffusion was attributed to liquid layering~\cite{schuster_diffusion_2000, *schuster_anisotropic_2003, patil_combined_2007, *schob_single_2010, grabowski_comparing_2007}. However, it has to be mentioned, that direct evidence of layering was obtained from investigation (by X-ray and ellipsometry) only for films on silicon substrates with a thin (native) oxide layer. 

On the other hand, for sensitivity reasons, tracer diffusion studies were carried out only on silicon substrates with 100~nm (thermal) oxide, quartz glass or mica. Nevertheless, recent long term single molecule experiments on thinning liquid films on 100~nm oxide further addressed the mutable influence of molecular layering on liquid dynamics~\cite{tauber_single_2009, tauber_characterization_2011}. They showed two discernible diffusion coefficients questioning the extend of the molecular layering to four layers as reported form ellipsometry and X-ray studies~\cite{forcada_molecular_1993, yu_observation_1999}. 

These results initiated considerations about the influence of subsurface substrate properties on liquid dynamics and layering. Until now, no studies have been reported regarding the influence of the interface composition both on liquid layering and on diffusion dynamics close to the solid-liquid interface. However, for thin polymer films the wetting conditions are known to be influenced by subsurface material properties~\cite{seemann_dewetting_2001, *becker_complex_2003, *seemann_dynamics_2005}. Recent studies also demonstrate that the adhesion of proteins, bacteria and setal arrays of geckos on silicon wafers is influenced by the thickness of the silicon oxide interface~\cite{hahl_subsurface_2012, loskill_influence_2012,*loskill_is_2012}.  Additionally, in a previous study, we reported the effect of the subsurface material on the structure and dynamics of thin smectic liquid crystal films~\cite{schulz_influence_2011}.

The above mentioned tracer diffusion experiments on thinning TEHOS (tetrakis-2-ethyl-hexoxy-silane) films on silicon with 100 nm thermal oxide~\cite{tauber_single_2009, tauber_characterization_2011}, also point to an influence of the subsurface substrate composition. However, for this type of multi-component interface no layering studies (neither by ellipsometry nor by X-ray) are known, which prevents from a direct comparison of diffusion and layering experiments. On the other hand, single molecule fluorescence experiments cannot be used to study diffusion in ultrahin liquid films on thin native oxide, since the small distance of the fluorescent tracer molecules to the underlying silicon leads to quenching of the fluorescence by electromagnetic interactions with silicon~\cite{baumgartel_fluorescence_2010}. 

To circumvent these experimental limitations, we will report optical ellipsometry experiments on thinning TEHOS films prepared both on 100~nm thick thermal oxide and on native oxide. Comparing the ellipsometry study with tracer diffusion experiments on thinning TEHOS films on thermal oxide~\cite{tauber_single_2009, tauber_characterization_2011}, we describe the results in the frame work of long-range van der Waals interactions. We find an agreement between reduced tracer mobility with decreasing thinning rates for decreasing film thickness and conclude that the multi-component interface properties also influence diffusion dynamics on a molecular scale.

\section{Experimental}

\subsection{Substrate characterization and sample preparation}
(100)-polished n-doped (phosphor, conductivity 1{\ldots}20 $\Omega$cm) silicon wafers with native oxide (Center for Microtechnologies, Chemnitz) and with 100 nm thermal oxide (CrysTec, Berlin) were rinsed alternately with acetone and ethanol (both technical grade) and sonicated for 10~min in acetone and afterwards in ethanol at 50~$^{\circ}$C (both Merck, spectroscopic grade). In between and at the end the substrates were rinsed with de-ionized water and dried by nitrogen. AFM measurements showed smooth surfaces, with slightly larger average roughness for the thermal oxide $2.2\pm0.2$~\AA\ compared to the native oxide $1.6\pm0.1$~\AA. Macroscopic contact angles of water, diiodmethane and tetrakis-2-ethylhexoxysilane (TEHOS) (see~\ref{tbl:contactangles}) were similar for both types of oxide. Thus, the contact angles depend only on superficial properties of the substrates.
\begin{table}
  \caption{Macroscopic contact angles $\theta$~[$^\circ$] on Si wafers with different oxide thickness}
  \label{tbl:contactangles}
  \begin{tabular}{lcc}
    \hline
    liquid & $\theta$ on native oxide & $\theta$ on 100 nm oxide\\
    \hline
    water & $63\pm2$ & $61.4\pm1.3$ \\
    diiodmethane & $54.7\pm1.7$ & $52\pm3$ \\
    TEHOS & $5.6\pm0.9$ & $7.2\pm0.9$ \\
    \hline
  \end{tabular}
\end{table}

The substrates were stored for 2~h under dry nitrogen atomsphere before preparation of the TEHOS films. Then, the substrates were dipcoated (KSV instruments) in solutions of 0.3\% TEHOS (ABCR GmbH, Germany, for chemical structure see \ref{diffusion} c) in hexane (Merck, Germany, spectroscopic grade) using fast immersion, 20~s waiting time and 5~mm/min withdrawal speed. TEHOS is a branched flexible spherical molecule with a diameter of about 10~\AA. The chemical structure is given in~\ref{diffusion} (c). 

For diffusion measurements, Rhodamine~B (RhB), Rhodamine~6G (R6G) and Oregon Green (OG), were alternatively used as fluorescent tracers, see~\ref{fcs} (b-d) for the chemical structures. The tracer molecules were added to the solution before dipcoating, resulting in nanomolar concentrated solutions in the ultrathin films.

\subsection{Diffusion measurements}
Single molecule tracking (SMT) experiments were carried out with a home-built wide-field microscope described perviously, employing the 514~nm line of an \ce{Ar}-ion laser~\cite{tauber_single_2009, schulz_optical_2010, schulz_influence_2011, tauber_characterization_2011}, and an acquisition frame rate of 50~Hz, yielding a temporal resolution of 20~ms.

For fluorescence correlations spectroscopy (FCS) a previously described home-built scanning confocal microscope was used with a pulsed excitation (40~MHz) at 465~nm, together with a hardware correlator (ALV-5000)~\cite{schulz_optical_2010, tauber_characterization_2011}. This allows for detection of correlation times in the range of 0.1~$\mu$s to $10^{4}$~s. FCS curves were obtained from fluorescence intensity time traces lasting 300~s each.

The film thickness for samples used with diffusion measurements was determined by varying angle ellipsometry with a VASE\texttrademark\ ellipsometer (J.A. Woollam Co., Inc., USA). The diameter of the ellipsometer spot was 3~mm.

\subsection{Mapping ellipsometry}
Mapping varying angle ellipsometry was conducted with a M-2000 ellipsometer (J.A. Woollam Co., Inc., USA), and spot diameter 1~mm. The samples were scanned on a rectangular grid with 1~mm spacing, yielding 81 measuring points on a 10~mm x 10~mm sample. In a first step, the pure substrates were scanned. The obtained ellipsometry data sets were then used during further fitting to retrieve the film thickness of the prepared films. This step is necessary to correct for slightly varying silicon oxide thickness over the sample. The TEHOS film was modeled with a Cauchy model (Cauchy.mat, J.A. Woollam Co., Inc., USA) with the optical parameters $A=1.43$ and $B=0.002$, which had been obtained before from a thick TEHOS film with the same instrument (using the first two terms of the Cauchy power series for the refraction index $n(\lambda)=A+B\lambda^{-2}+\mathcal{O}(\lambda^{-4})$)~\cite{tauber_characterization_2011}. Apart from ellipsometric measurements, the samples were stored in a chamber under a continuous flow of dry nitrogen gas at room temperature $21\pm1^\circ$C.

To obtain film thinning rates, succeeding ellipsometry scans were run on each sample in intervals of about 24~hours. Between succeeding scans an individual thinning rate $r_i=(h_1-h_2)/(\Delta t)$ was calculated for each measured point. Thereby, $h_1$ and $h_2$ are the film thicknesses measured in succeeding scans, and $\Delta t$ is the time lag between these scans. The thus obtained thinning rates were binned according to a method described by Forcada and Mate~\cite{forcada_molecular_1993}: Each individual rate $r_i$ was added to all bins of width 0.5~\AA\ covered by the thickness interval from $h_2$ to $h_1$. In the end, the sum in each bin was divided by the number of its entries, thus yielding an average thinning rate $r(h)$. Given a sufficient amount of data, this method enables the visualization of small changes in the thickness dependent thinning rate $r(h)$, which otherwise would be hidden due to the long intervals between succeeding measurements.

\section{Results and Discussion}
The structure and dynamics of liquids at interfaces with solids can be studied in case of ultrathin liquid films via various experimental methods. Thereby, X-ray reflectometry will reveal the structure of the liquid~\cite{yu_observation_1999}, whereas long-time studies on thinning films using ellipsometry are suited for investigation of the film morphology and slow macroscopic dynamics~\cite{forcada_molecular_1993}. On a microscopic scale, the dynamics within such films can be investigated via tracer diffusion experiments. Thereby, highly diluted fluorescent tracers are used to probe their local environments. Single molecule tracking (SMT) experiments are suitable for detection of diffusion coefficients up to the order of 1~$\rm\mu m^2/s$~\cite{tauber_characterization_2011, schulz_optical_2010, woll_polymers_2009}. From trajectory analysis of SMT, detailed information about average diffusion coefficients $D_{\rm traj}$ along trajectories and  changes in the nature of diffusion (for example, due to confinement to compartments)~\cite{saxton_single-particle_1997} is obtained. With respect to fast diffusion, the temporal resolution of SMT is more efficiently used (resolving diffusion coefficients up to about 10~$\rm\mu m^2/s$), when analyzing probability distributions of diffusivities (scaled square displacements of detected tracers between succeeding frames)~\cite{tauber_single_2009, tauber_characterization_2011, schulz_optical_2010}. Considerable higher temporal resolution can be achieved by fluorescence correlation spectroscopy (FCS)~\cite{tauber_characterization_2011, schulz_optical_2010, woll_polymers_2009}. However, FCS is less sensitive for diffusion slower than 1~$\rm\mu m^2/s$~\cite{ woll_polymers_2009}. We expect tracer diffusion dynamics to occur with diffusion coefficients between 55~$\rm\mu m^2/s$ (bulk value in TEHOS under ambient conditions) and <1~$\rm\mu m^2/s$ as recently reported~\cite{schuster_diffusion_2000, *schuster_anisotropic_2003, schuster_diffusion_2004, grabowski_comparing_2007}.

We start our report with a long-time study of fluorescent tracer diffusion as a function of thinning in ultrathin liquid films on silicon with 100~nm thermal oxide. Thereby, we employ SMT with trajectory as well as diffusivity analysis, and we use FCS on 100~\AA\ thick films to amend our results from the SMT experiment on thinning TEHOS films on thermal oxide. A comparative study of tracer diffusion on thin native oxide is not feasible, due to the in that case small distance of the fluorescent tracer molecules to the underlying silicon (in the order of a few nanometers) which will lead to quenching of the tracer fluorescence~\cite{baumgartel_fluorescence_2010}. We therefore use ellipsometry for comparing the morphology and macroscopic dynamics of thinning films on both kinds of substrates and relate results about film structure and dynamics from tracer diffusion to those from ellipsometry.

\subsection{Tracer diffusion in ultrathin TEHOS films}
Previous work on tracer diffusion in ultrathin TEHOS films on silica substrates with 100~nm oxide revealed a considerable slowdown in comparison to bulk measurements by more than one order in magnitude~\cite{schuster_diffusion_2000, *schuster_anisotropic_2003, patil_combined_2007, grabowski_comparing_2007}. Further information on the diffusion in dependence on the distance to the solid-liquid interface could be obtained from a long-time study on thinning TEHOS films on silicon with 100~nm thermal oxide~\cite{tauber_single_2009, tauber_characterization_2011}. Here we report an extension of this study, especially by FCS experiments on 100~\AA\ thick films, and a discussion in parallel to the results from the ellipsometry study, thus amending the previously published interpretations.

FCS and SMT not only complement each other on the expected range of tracer dynamics in ultrathin TEHOS films. They also differ in sensitivity to adsorption events. If tracers will adsorb to the substrates, these events may dominate $D_{\rm traj}$. To select mobile tracers, a lower limit can be set to the lateral extension (covered area) of trajectories. We name such trajectories "mobile." For long trajectories the average $D_{\rm traj}$ nevertheless may be small, if the dynamics are dominated by slow diffusion and adsorption. Fast diffusing tracers are likely to leave the detection area of about 250~$\rm\mu m^2$, which renders short trajectories. However, determination of $D_{\rm traj}$ from very short trajectories contains large errors~\cite{qian_single_1991}. Within diffusivity analysis of SMT, immobile tracers may be excluded by setting a lower limit to individual diffusion steps~\cite{tauber_single_2009, tauber_characterization_2011}. FCS is not sensitive to completely immobile molecules, since it is based on fluorescence fluctuations. However, ad-/desorption events during acquisition time will contribute and will result in apparent slow diffusion. By carefully comparing and combining results from these different methods, we obtain information about tracer mobility within ultrathin TEHOS films, which will be explained in the following and discussed further in combination with results from ellipsometry.
\begin{figure}[th]
  \begin{minipage}[b]{2.5in}
  \includegraphics[width=2.48in]{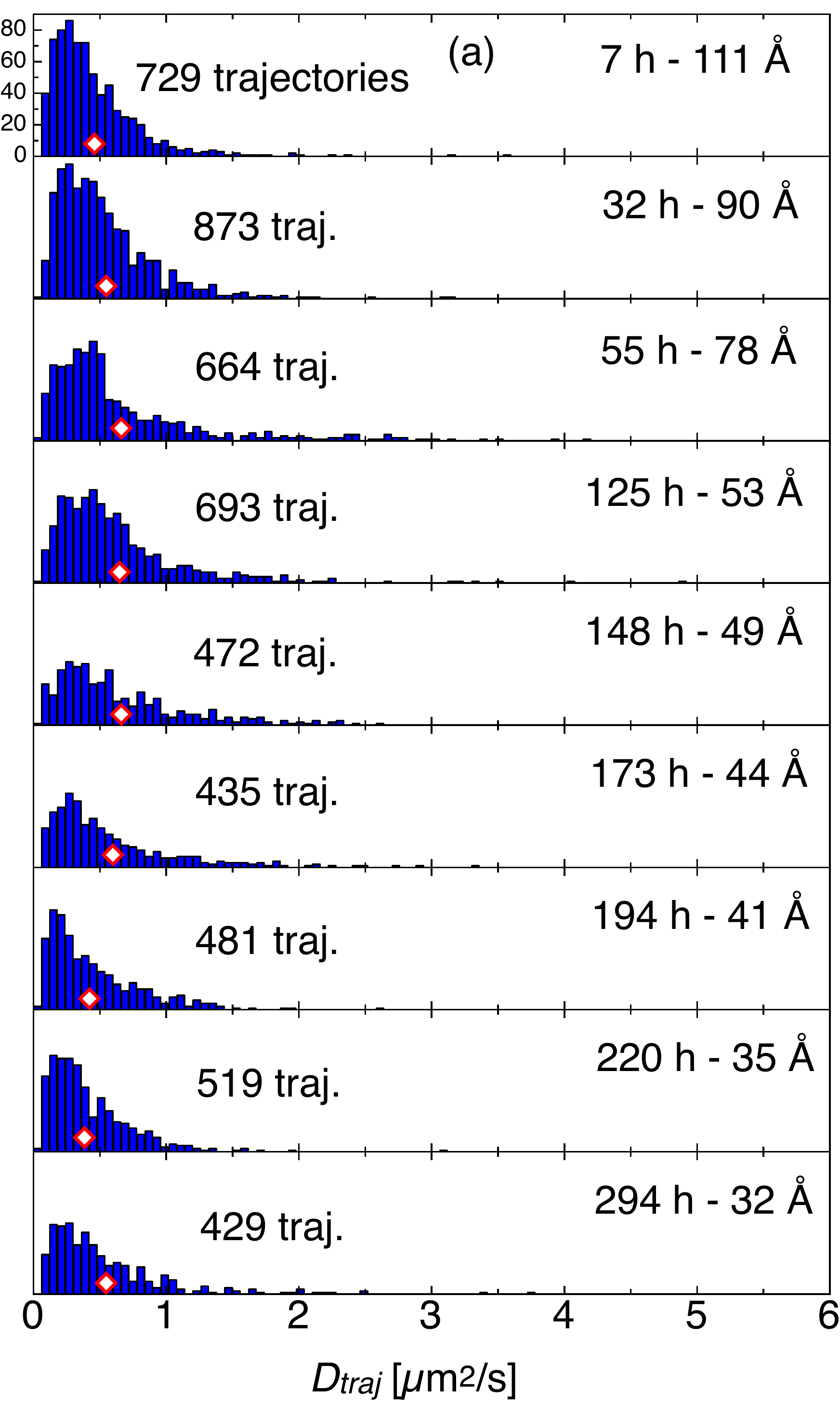}
  \end{minipage}
    \begin{minipage}[b]{2.5in}
  \includegraphics[width=2.48in]{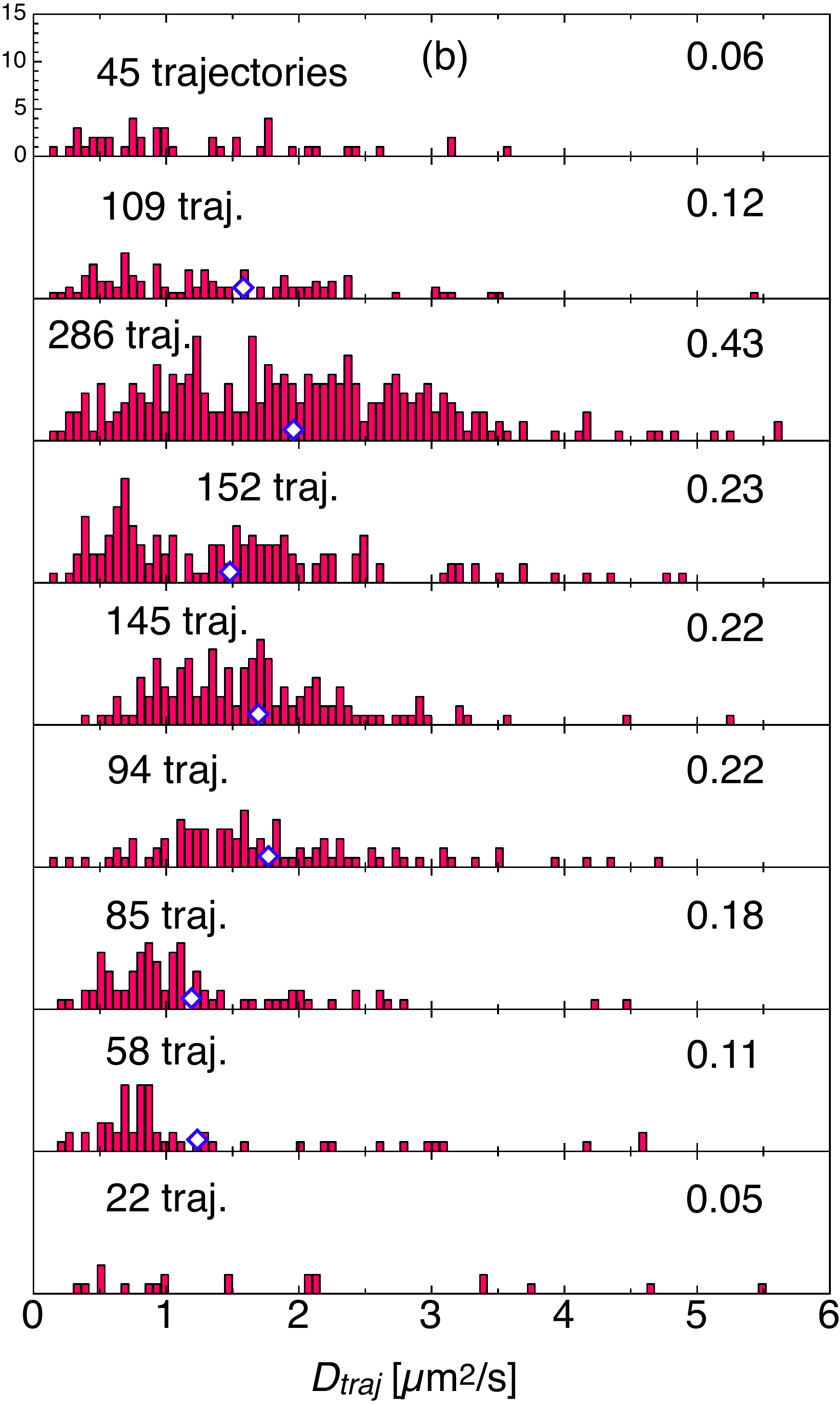}
  \end{minipage}
  \begin{minipage}[b]{1.4in}
  \centering
  \includegraphics[width=1.4in]{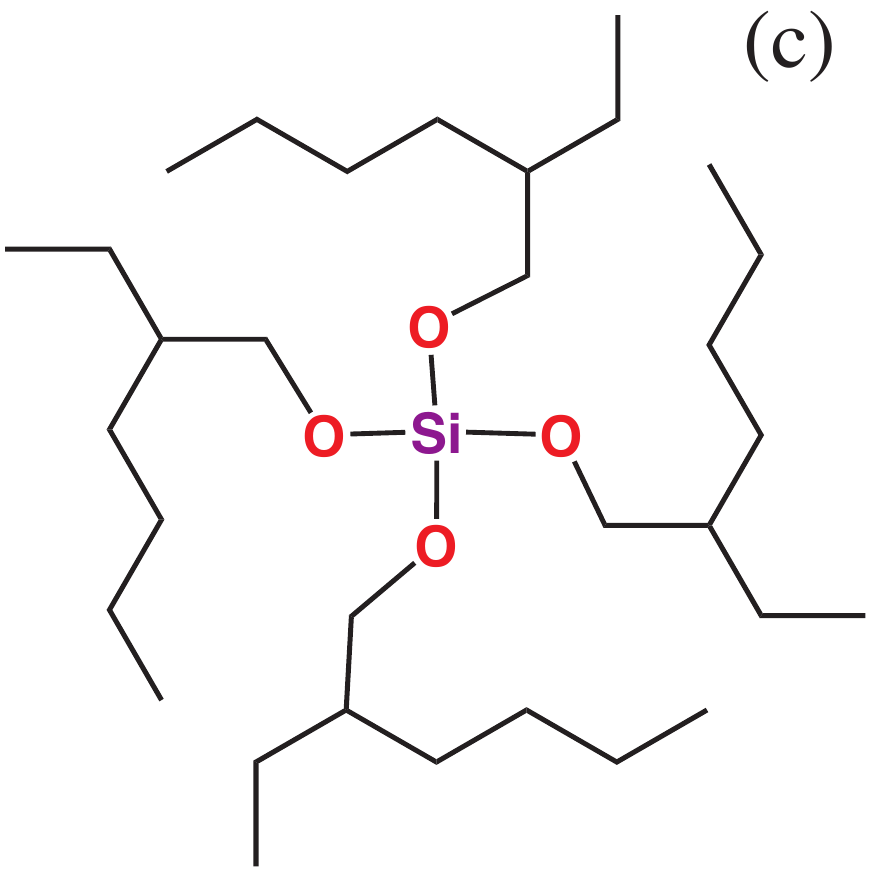}
  \end{minipage}
  \caption{(a,b) Distribution of diffusion coefficients $D_{traj}$ calculated from trajectories of RhB in a thinning TEHOS film on 100~nm oxide. Indicated are the number of trajectories, mean values of $D_{traj}$ ($\Diamond$), the time after preparation in h, and the film thickness in \AA: (a) $D_{traj}$ from all detected trajectories (including immobile ones), (b) $D_{traj}$ for trajectories exceeding a lateral area of 1.3~$\rm\mu m^2$ ("mobile" ones only) and the ratio of these to all trajectories, (a,b) modified from~\cite{tauber_single_2009, tauber_characterization_2011}. (c) Chemical structure of TEHOS.}
  \label{diffusion}
\end{figure}

\begin{figure}[t]
  \begin{minipage}[b]{2.4in}
  \includegraphics[width=2.35in]{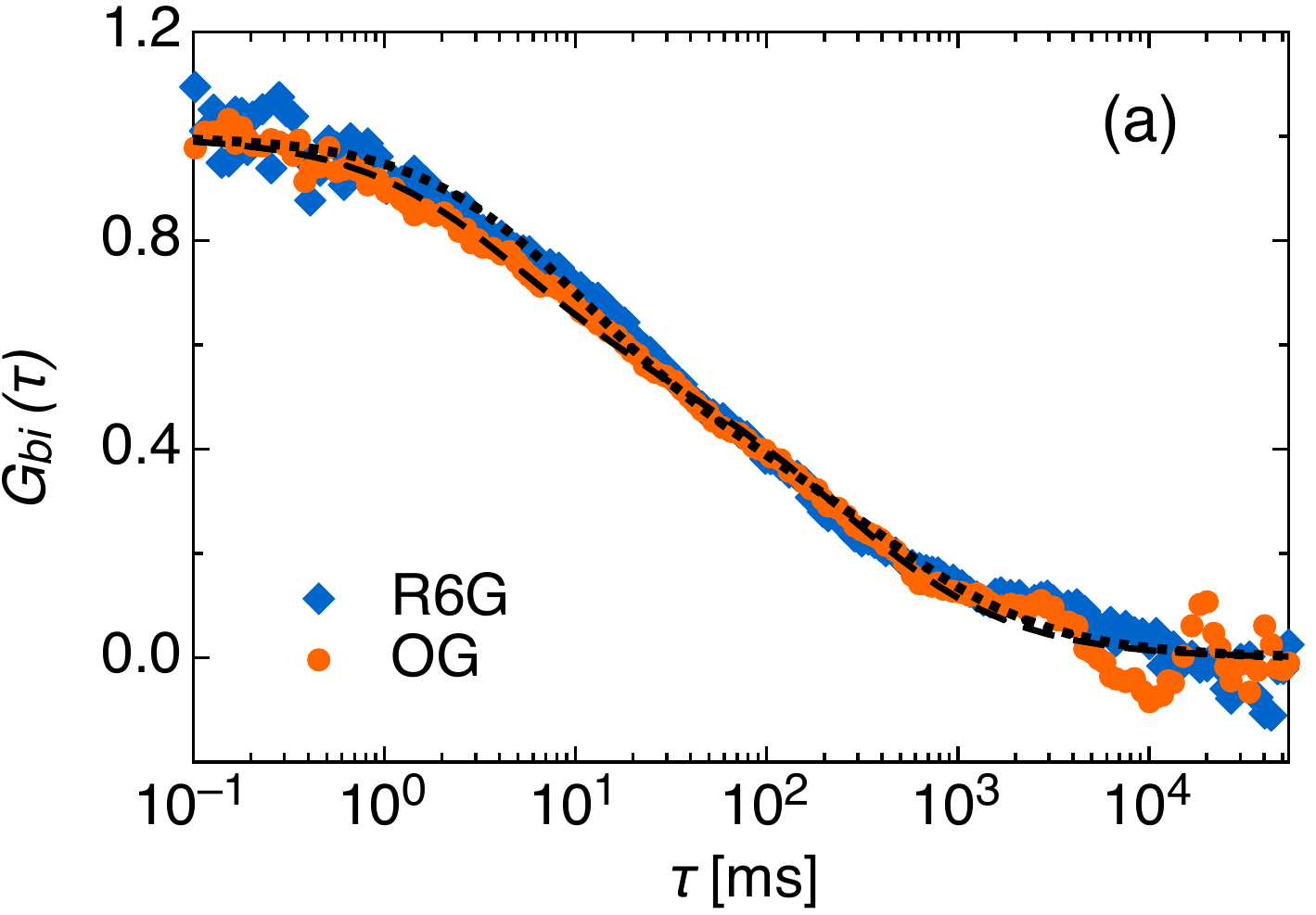}
  \end{minipage}
  \raisebox{0.8cm}{\begin{minipage}[b]{1.21in}
  \makebox(80,33){{\footnotesize (b)}\hspace{0.5cm}R6G}
  \includegraphics[width=1.2in]{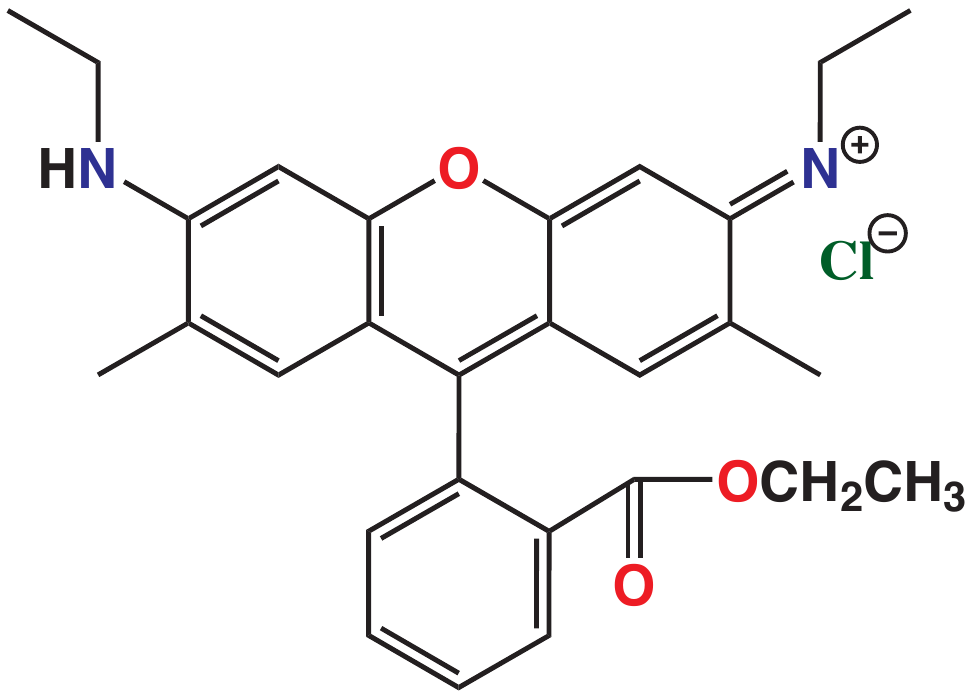}
  \end{minipage}}
  \raisebox{0.8cm}{\begin{minipage}[b]{1.36in}
  \makebox(100,30){{\footnotesize (c)}\hspace{0.5cm}OG}
  \includegraphics[width=1.35in]{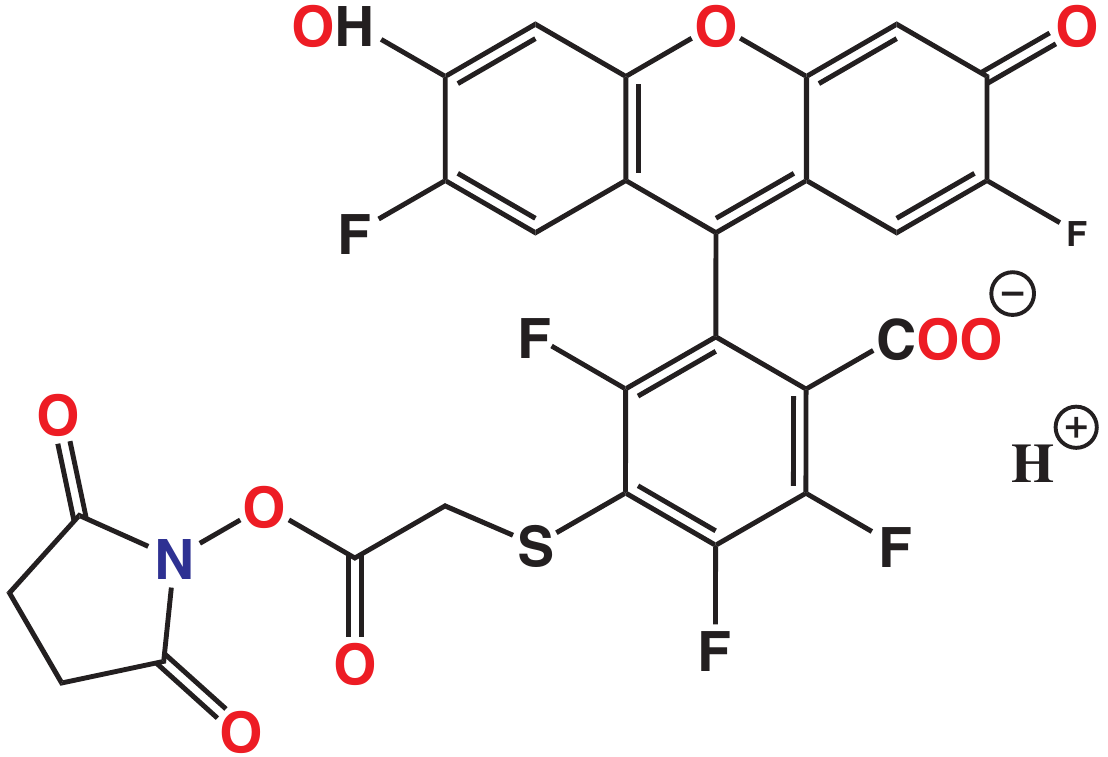}
  \end{minipage}}
  \raisebox{0.8cm}{\begin{minipage}[b]{1.36in}
  \makebox(80,42){{\footnotesize (d)}\hspace{0.5cm}RhB}
  \includegraphics[width=1.35in]{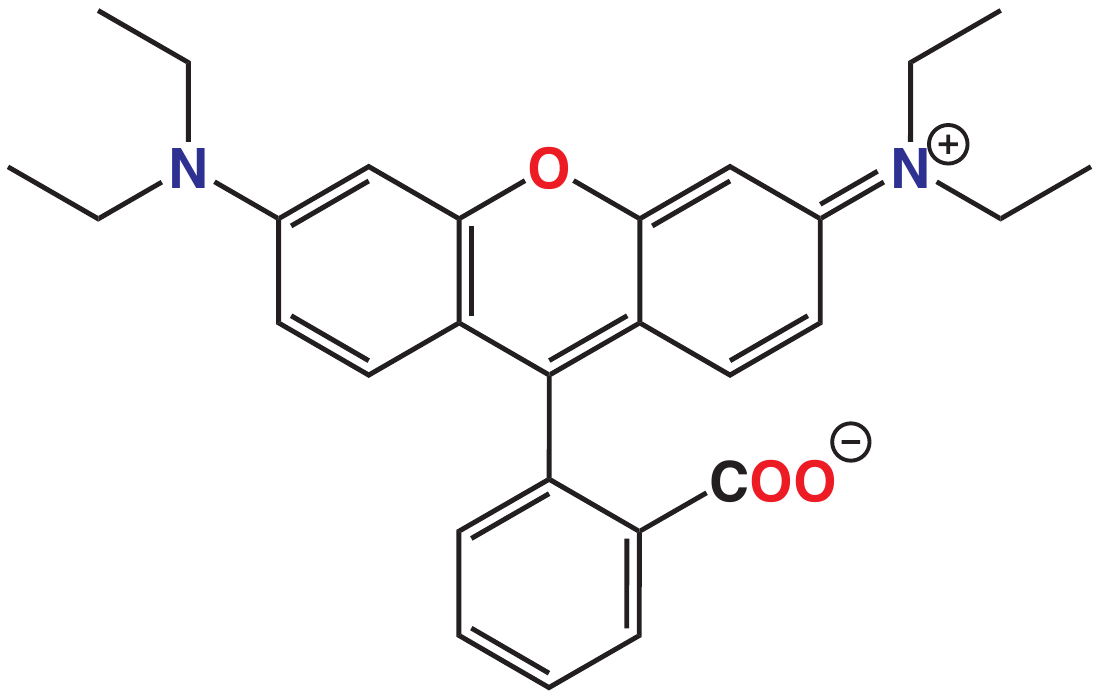}
  \end{minipage}}
  \caption{(a) Normalized FCS autocorrelation curves and two-component fits (- - OG, $\cdot\cdot\cdot$ R6G) in a 10~nm thick TEHOS film on thermal oxide. (b-d) Chemical structure of fluorescent tracers (b) Rhodamine 6G (R6G), (c) Oregon Green (OG), and (d) Rhodamine B (RhB).}
  \label{fcs}
\end{figure}

Single molecule tracking experiments (SMT) using the tracer dye Rhodamine B (RhB, for the chemical structure see~\ref{fcs} d) follow changes in diffusion during film thinning over 294~h from sample preparation, starting with a 111~\AA\ thick film and ending with a 32~\AA\ one. \ref{diffusion}~(a,b) shows the obtained broad distributions of diffusion coefficients $D_{\rm traj}$ calculated via mean square displacements along identified trajectories~\cite{tauber_single_2009, tauber_characterization_2011}. While \ref{diffusion}~(a) collects all trajectories, \ref{diffusion}~(b) puts the emphasis on mobility by selecting those trajectories, which exceed a lateral area of 1.3~$\rm\mu m^2$. The broad distributions of $D_{\rm traj}$ seen in \ref{diffusion}~(a and b) point to heterogeneous tracer diffusion, in agreement with recent reports~\cite{schuster_diffusion_2000, *schuster_anisotropic_2003, patil_combined_2007, grabowski_comparing_2007}. The qualitative overall feature is that diffusion becomes slower with decreasing $h$.

In principle the number of detected trajectories should be constant over the time of the film thinning experiment. However, the number of detected (mobile and immobile) trajectories in~\ref{diffusion}~(a) stays almost constant for $h<50$~\AA, but is considerable higher for larger $h$. As was stated above, SMT, and in particular trajectory analysis of SMT, is restricted with respect to fast diffusion. The loss of fast tracers will artificially increase the number of detected trajectories, because missed out excursions into fast diffusion will split trajectories into several apparently separate trajectories (see supplement for further explanation). 

Therefore, we conducted FCS experiments on 100~\AA\ thick TEHOS films within 24~h after preparation using the Rhodamine derivates Rhodamine 6G (R6G) and Oregon Green (OG). The molecular structure of both dyes is depicted in~\ref{fcs} (b,c). Typical autocorrelation curves for both tracer molecules are shown in~\ref{fcs} (a). The autocorrelation function $G_2(\tau)$ for two-dimensional diffusion~\cite{tauber_characterization_2011, schwille_fluorescence_2002} did not fit to the experimental FCS curves. A better approximation is realized with a two-component function $G_{\rm bi}(\tau)$~\cite{tauber_characterization_2011}, see \ref{2Dcorr_bi} in the supplement. The fits are also indicated in~\ref{fcs}~(a), fitting parameters are given in \ref{fcs_param}, diffusion coefficients in \ref{tbl:fcs}. Both dyes showed similar results within error, whereby the two components $\bar{D}_{1,\rm FCS}=0.1\pm0.05$~$\rm\mu m^2/s$ and $\bar{D}_{2,\rm FCS}=6\pm5$~$\rm\mu m^2/s$ (averaged values from both dyes) were obtained with equal amplitude (see \ref{tbl:fcs}). A comparison with the distributions of $D_{traj}$ for similar film thickness ($h=111$~\AA\ and $h=90$~\AA, top two in \ref{diffusion} a,b) shows only some rare trajectories with $D_{traj}$ in the range of $D_{2,\rm FCS}$. Thus, it can be concluded that a considerable amount of fast diffusion was missed out by the trajectory analysis of the SMT experiment. 

Now let us turn again to the distributions of $D_{\rm traj}$ shown in \ref{diffusion} (b). These trajectories can be considered as mobile, since they cover an area exceeding 1.3~$\rm\mu m^2$. We have already discussed the variation of the total number of trajectories (\ref{diffusion} a) in relation to $h$. This explanation is also valid for the "mobile" trajectories. The number of detected mobile trajectories is largest for $h\approx80$~\AA, and decreases with increasing as well as with decreasing film thickness. For $h\geq90$~\AA, the low number of detected trajectories can be explained by the above discussed loss of fast tracers. With decreasing $h$, the contribution from fast diffusion decreases considerably, increasingly preventing tracers to cover the area threshold set for mobile trajectories, while the total number of trajectories remains constant (\ref{diffusion} a). Keeping these explanations in mind, the ratio of mobile to all detected trajectories can be used to explain tracer mobility. This ratio is indicated in \ref{diffusion}~(b) and will be used later on for comparison with results from ellipsometry.

Below the film thickness of 32~\AA\ (about 3 times the diameter of TEHOS molecules), no "mobile" trajectories could be observed, which corresponds to the absence of mobile tracers up to film thicknesses $h\leq20$~\AA. It has to be mentioned that residual absorption layers and precursor films of wetting droplets are of similar thickness. In the precursor film of a TEHOS droplet on silica, tracer diffusion coefficients of R6G were reported to be in the range of 0.1 to 1~$\rm\mu m^2/s$~\cite{schuster_diffusion_2004}, and thus 2-3 orders of magnitude smaller than the bulk value $D_{\rm bulk}\approx50$~$\rm\mu m^2/s$~\cite{tauber_characterization_2011}. 

\begin{table}[b]
  \caption{Diffusion coefficients $D_{1,2}$~[$\rm\mu m^2/s$] obtained from SMT and FCS on 100 \AA\ thick TEHOS films on thermal oxide.}
  \label{tbl:fcs}
  \begin{tabular}{lccccc}
    \hline
   method & tracer & $a_1$ & $D_1$ & $a_2$ & $D_2$ \\
    \hline
    SMT (diffusivities)& RhB & $0.7$ &$0.3\pm0.1$ & $0.3$ & $4.5\pm0.4$ \\
    FCS & R6G&0.5&$0.06\pm0.05$ &0.5& $4\pm3$\\
    FCS & OG&0.5&$0.14\pm0.05$ &0.5& $8\pm4$ \\
    FCS~\cite{grabowski_comparing_2007}&Alexa~488 && $0.2$ && $13$\\
    \hline
    \multicolumn{6}{l}{\footnotesize error: standard deviation $\sigma$ (SMT), $\sigma$+ uncertainty of focal width (FCS).}\\
  \end{tabular}
\end{table}

Improvement of temporal resolution with respect to trajectory analysis of SMT is gained by analyzing probability distributions of diffusivities ($d_{\rm diff}=r^2/(4\tau)$, i.e.\ scaled square displacements of detected tracer molecules at a fixed time lag $\tau$) for the smallest possible $\tau=\tau_{\rm frame}$ (the inverse acquisition frame rate)~\cite{tauber_single_2009, tauber_characterization_2011, bauer_investigations_2009, *bauer_how_2011, schulz_optical_2010, schulz_influence_2011}. This analysis shows that the tracer diffusion can be approximated by two diffusion coefficients $D_{1,\rm diff}=0.3\pm0.1$~$\rm\mu m^2/s$ and $D_{2,\rm diff}=4.5\pm0.4$~$\rm\mu m^2/s$, which do not notably change in magnitude during film thinning~\cite{tauber_single_2009, tauber_characterization_2011}, see supplementary information for details. The $D_{i,{\rm diff}}$ are shown in \ref{tbl:fcs}. $D_{1,\rm diff}$ is within experimental error comparable to $D_{1,\rm FCS}$ reported for Alexa 488 by Grabowski et al.~\cite{grabowski_comparing_2007}. $D_{1,\rm FCS}$ for R6G and OG is subject to adsorption events which leads to smaller values. The fast components $D_{2,\rm diff}$ and $D_{2,\rm FCS}$ are comparable within experimental error. A quantitative comparison shows that the amplitude $a_{\rm 2,diff}$ of the fast component is only about 0.3 (see \ref{tbl:fcs}), less than the amplitude of $D_{2,\rm FCS}$ ($a_{\rm 2,FCS}=0.5$). This supports our previous argument that some fast diffusing tracers are missed by SMT, even when using diffusivity analysis.
The similar $D_{2,\rm diff}$ and $D_{2,\rm FCS}$, show the influence of the strong vertical confinement on the observed diffusion coefficients, which prevents even FCS from resolving bulk-like tracer diffusion ($D_{\rm bulk}\approx55$~$\rm\mu m^2/s$). $D_2$ rather is an effective diffusion coefficient depending on physical diffusion coefficients and transition probabilities between the heterogeneous regions. For more details about the influence of heterogeneity and confinement on diffusion coefficients~\cite{tauber_characterization_2011, bauer_investigations_2009, *bauer_how_2011} see the supplementary information.

By using the cationic Rhodamine derivate R6G and the anionic OG for FCS experiments, we addressed the question, to which extend the diffusion of tracers is influenced by electrostatic (Coulomb) interactions with the substrate. As stated above, the diffusion coefficients are similar for both tracer molecules~\cite{tauber_characterization_2011}, see~\ref{tbl:fcs}. \ref{tbl:fcs} also gives values for Alexa~488 in ultrathin TEHOS films on quartz substrates~\cite{grabowski_comparing_2007}. In contrast to RhB, R6G and OG, Alexa~488, does not adsorb to the substrate~\cite{grabowski_comparing_2007, tauber_characterization_2011}, thus leading to somewhat larger values $D_{1,\rm FCS}=0.2$~$\rm\mu m^2/s$ and $D_{2,\rm FCS}=13$~$\rm\mu m^2/s$~\cite{grabowski_comparing_2007}. The similar diffusion coefficients obtained for the differently charged tracer molecules, indicate that Coulomb interactions are in the present case negligible for tracer diffusion in ultrathin liquid films. Similarly, H\"ahl et al.\ found that Coulomb interactions play only a minor role in protein adsorption~\cite{hahl_subsurface_2012}. 

To summarize, the presence of at least two different diffusion coefficients indicates an inhomogeneous film morphology. According to diffusivity analysis of the film thinning experiment, this heterogeneity lasts down to a film thickness of about 30~\AA. The slow diffusion coefficients determined via FCS and SMT can be assigned to diffusion processes in a near-surface region. Trajectory analysis of SMT indicates two significant changes in tracer dynamics and thus in liquid mobility. Above $h\approx80$~\AA\ diffusion becomes very fast, indicated by the loss of fast trajectories. Below $h\approx30$~\AA\ diffusion becomes considerable slow seen by the lack of "mobile" trajectories.
Since the fast component from diffusivity analysis is still present for $h=32$~\AA\ (corresponding to not much more than 3 molecular diameters of TEHOS), the near-surface region of reduced mobility is limited to 1-2 diameters of TEHOS (about 10-20 \AA)~\cite{tauber_single_2009, tauber_characterization_2011}. This suggests an only weak layering structure of TEHOS on thermal oxide of probably only two molecular layers. In contrast, up to four liquid layers have been reported for TEHOS on silicon with native oxide~\cite{forcada_molecular_1993}.  To examine, whether these findings are related to the film morphology, we compare the diffusion experiments with the ellipsometry study of TEHOS films on substrates with different oxide thickness reported in the following. A combined discussion of changes in mobility seen from tracer diffusion with features seen from film thinning rates will be given in the section on film thinning rates along \ref{Evaporation} (b). 

\subsection{Ellipsometry study of film thinning and morphology}
The recent development of simultaneous multi-wavelength ellipsometry together with an improved lateral resolution, facilitates long-time studies of thinning TEHOS films using mapping ellipsometry. In this section, maps of film thickness for both types of substrate are shown and discussed. Additionally, thickness dependent film thinning rates are calculated from the mapping ellipsometry data.
\begin{figure}[hb]
  \begin{minipage}[b]{0.2in}
    \makebox(0,200){\begin{sideways}native\hspace{0.9in} thermal\end{sideways}}
  \end{minipage}
  \begin{minipage}[b]{1.4in}
    \includegraphics[width=1.4in]{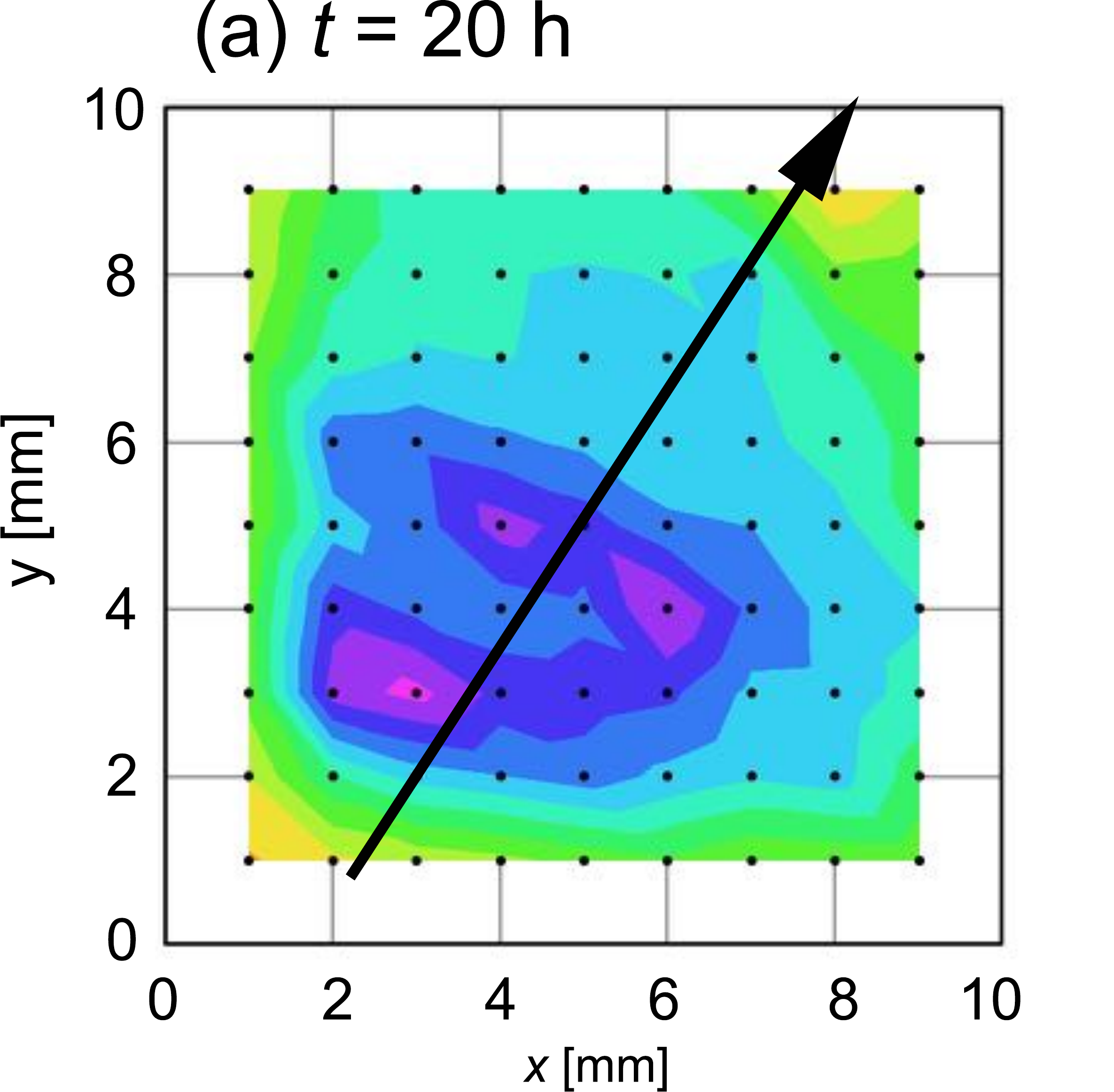}
    \includegraphics[width=1.4in]{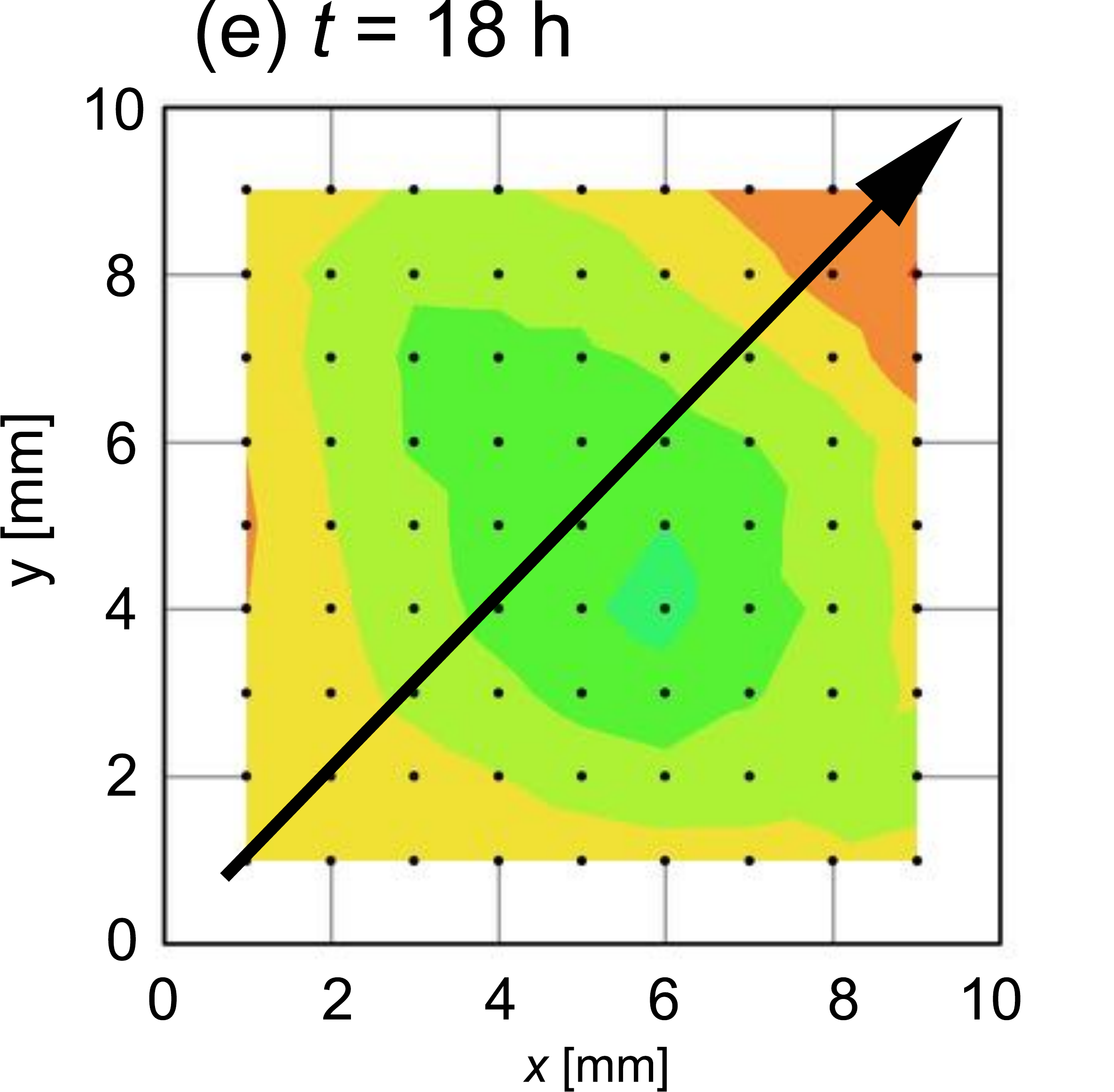}
  \end{minipage}
  \begin{minipage}[b]{1.4in}
    \includegraphics[width=1.4in]{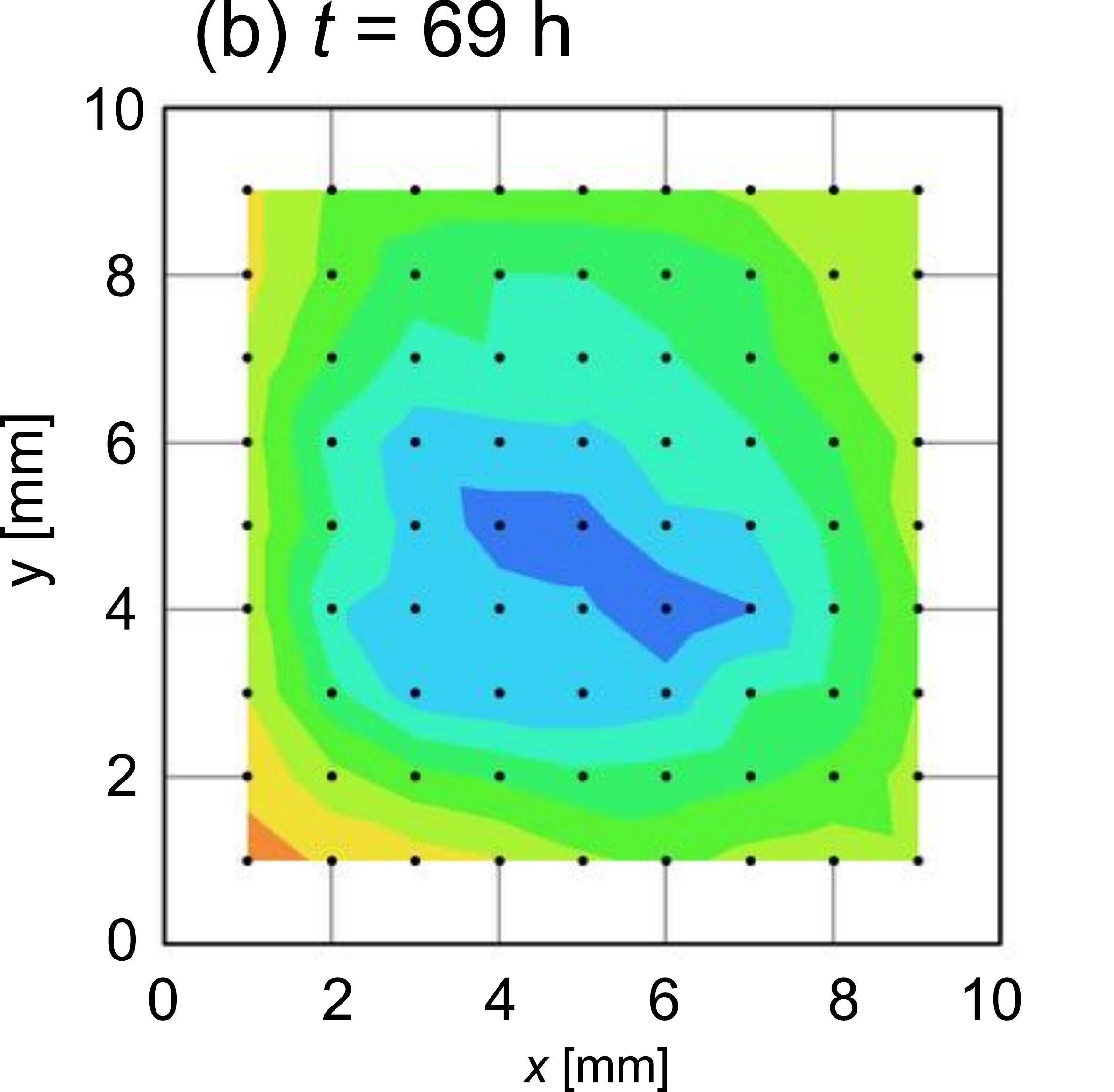}
    \includegraphics[width=1.4in]{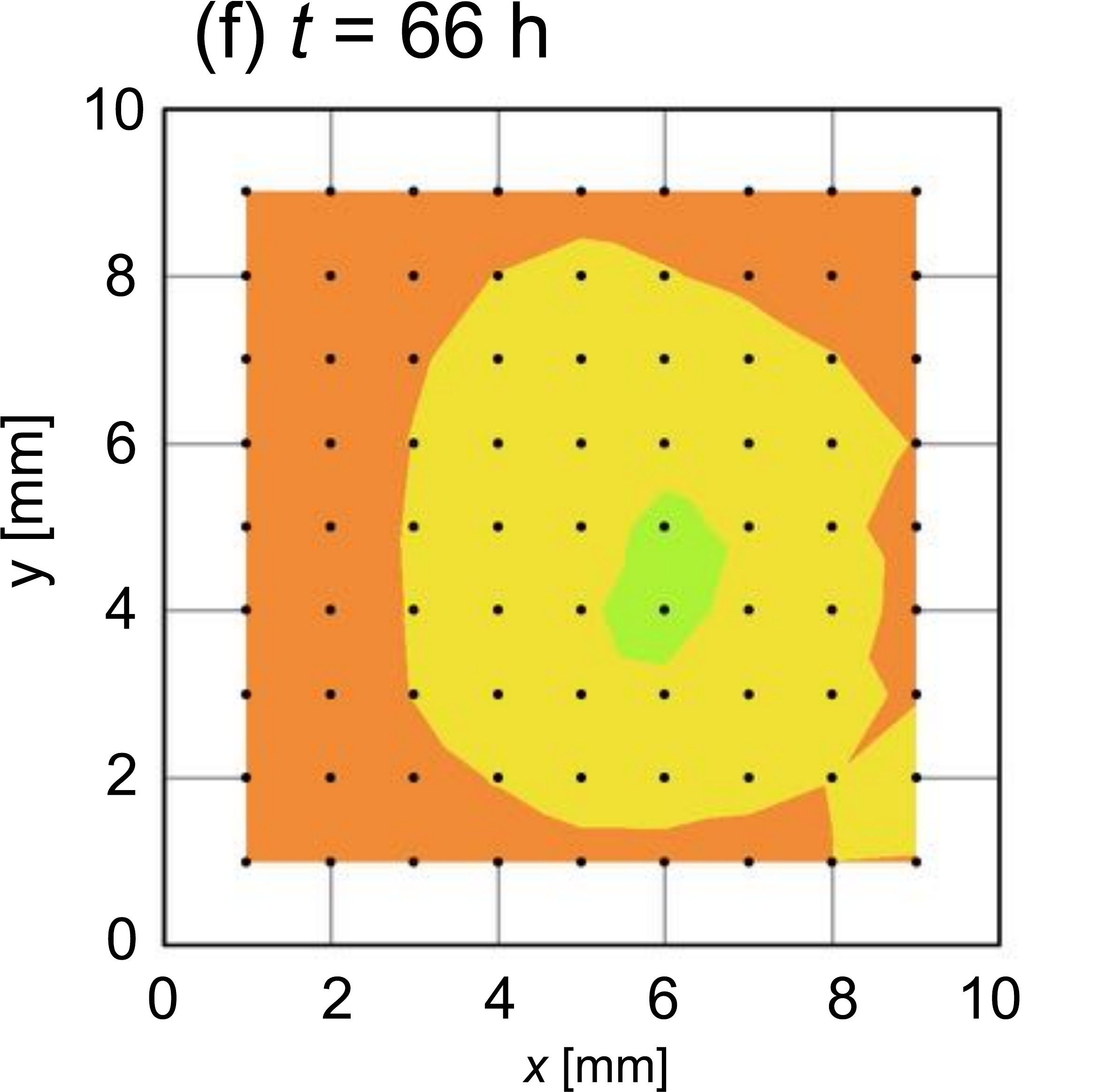}
  \end{minipage}
  \begin{minipage}[b]{1.4in}
    \includegraphics[width=1.4in]{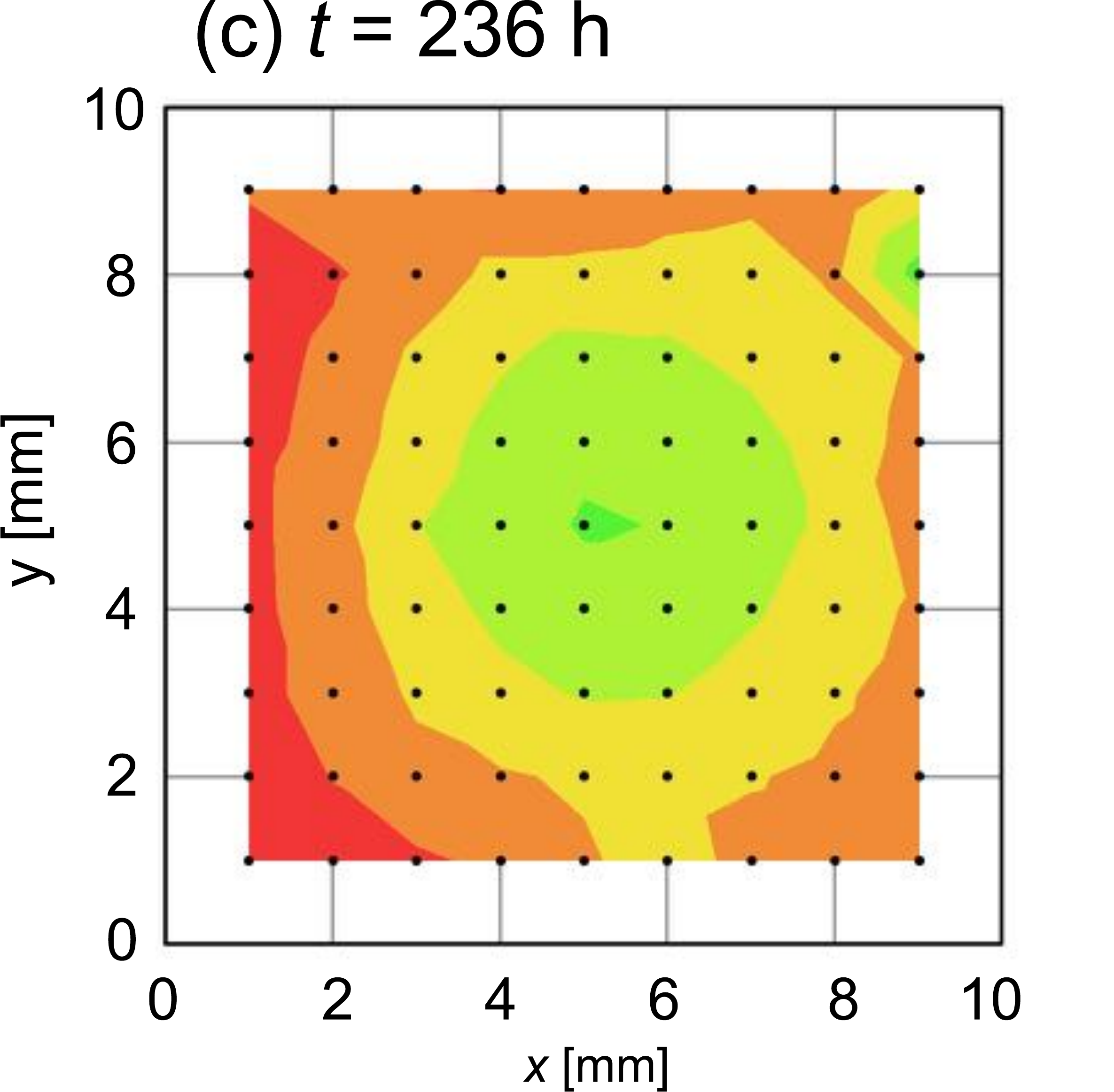}
    \includegraphics[width=1.4in]{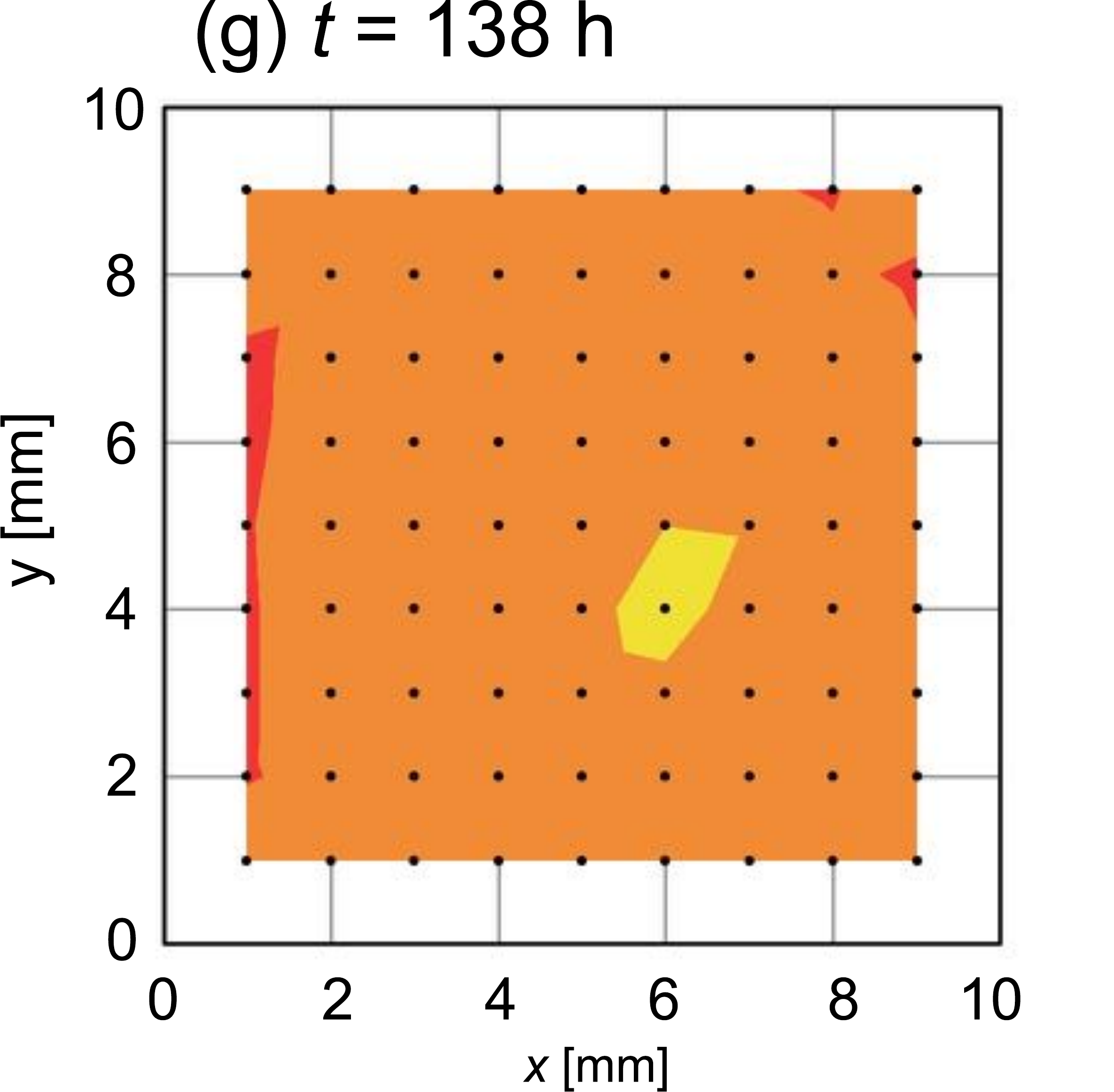}
  \end{minipage}
  \begin{minipage}[b]{1.4in}
    \includegraphics[width=1.4in]{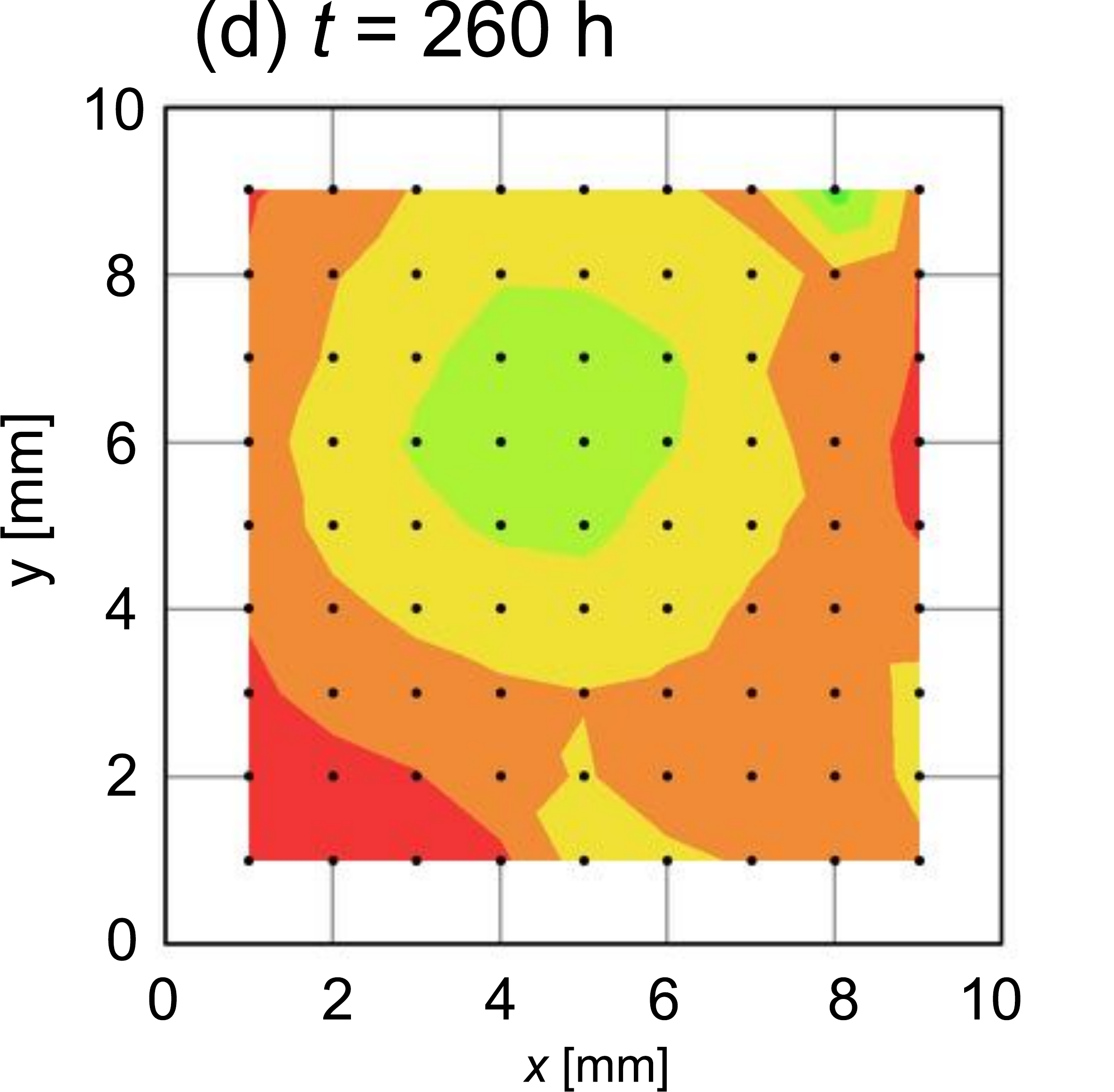}
    \includegraphics[width=1.4in]{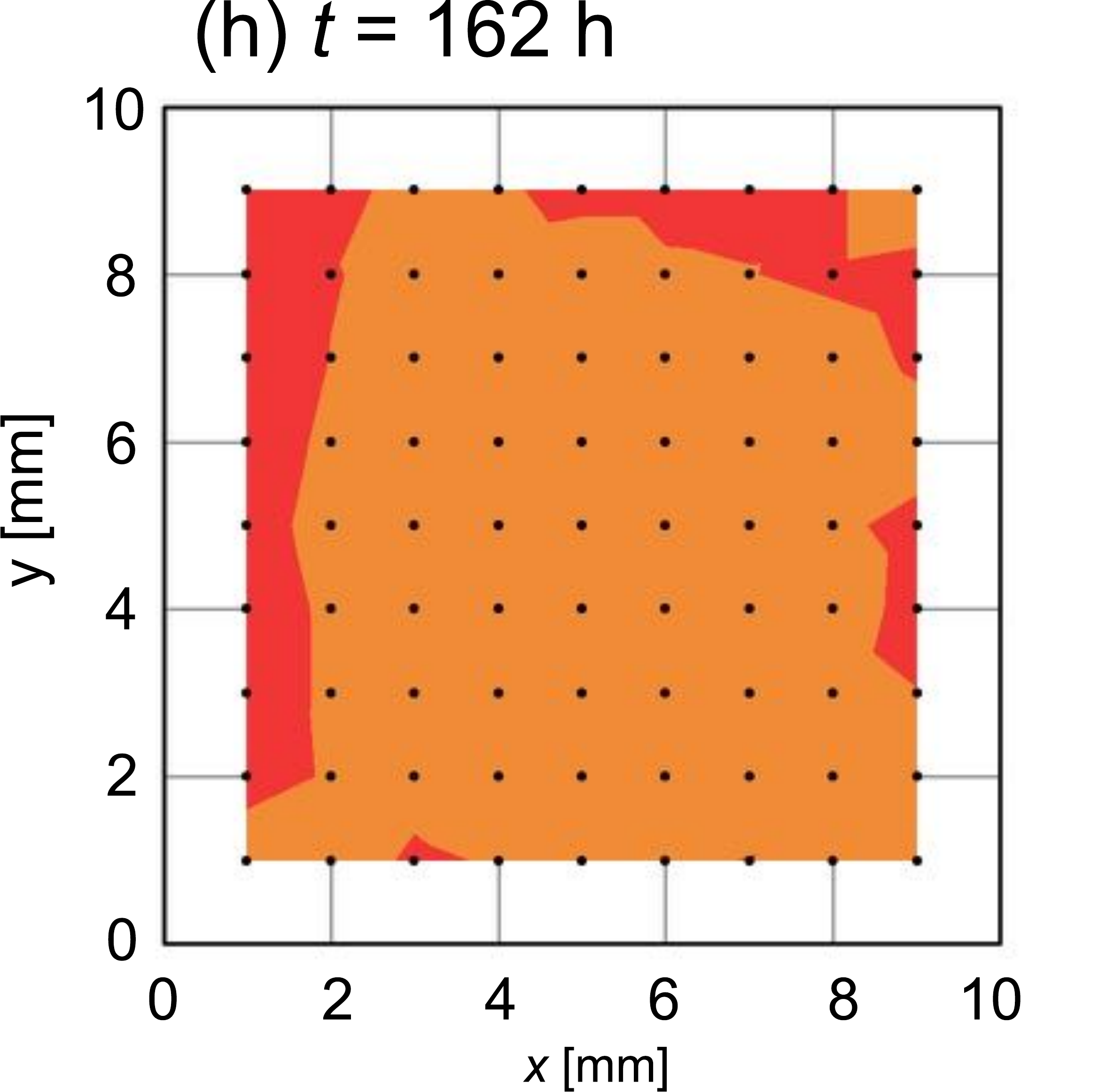}
  \end{minipage}
  \begin{minipage}[b]{0.4in}
    \includegraphics[width=0.4in]{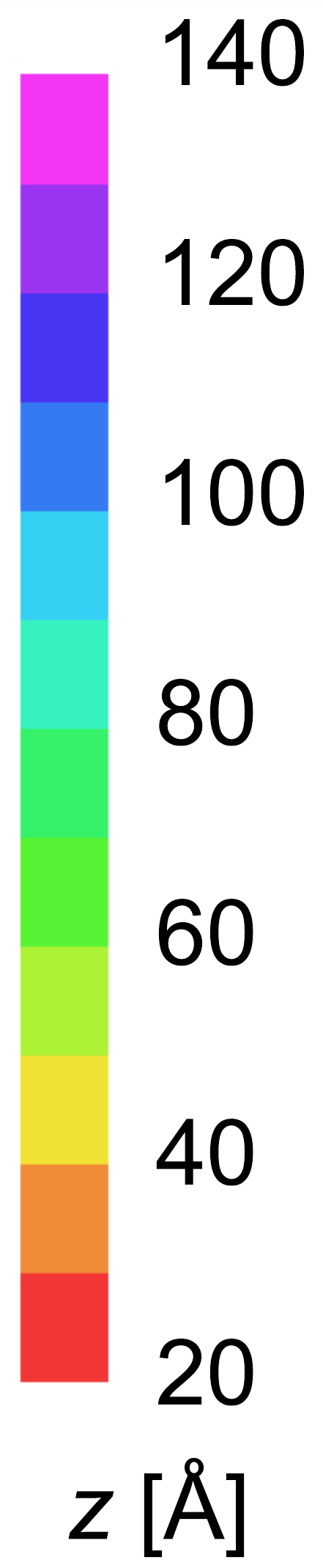}
  \end{minipage}  
  \caption{Maps (10 x 10 mm$^2$) of TEHOS film height $z$ as a function of time (in h) after preparation on (top) 100~nm thermal and (bottom) native oxide. Arrows denote the direction of withdrawal during dipcoating. Maps are shown for thinning films from left to right. A false color code for the height $z$ is given on the right.}
  \label{Maps}
\end{figure}

\ref{Maps} shows maps of TEHOS film thicknesses on 100 nm thermal (top row) and native (bottom row) oxide for two samples prepared under similar conditions. It is noticeable that the initially obtained TEHOS film on native oxide is thinner than the one on thermal oxide despite identical preparation. This is a general observation within our experiments. \ref{Maps}~(a and e) show thickness maps from the first ellipsometry runs on these samples 18~h and 20~h, respectively, after preparation. On 100 nm thermal oxide, the largest local film thickness was about 140~\AA, whereas on the native oxide the largest value (about 80~\AA) was about a factor of 1.8 smaller. We will explain these findings in a further section in terms of long-range electromagnetic forces.

Both TEHOS films exhibited a convex surface across the sample. The partly observed elliptical shape of both films is related to the preparation, since the dipcoating direction (indicated by arrows in~\ref{Maps} a,e) coincides with the small axes of the ellipses. The film on thermal oxide initially shows an inhomogeneous surface, which smoothens within the first 70~h. After this smoothening, the radius of curvature of the film surface on thermal oxide remains smaller than that of the film on native oxide. This can be seen from the extension of the color steps in the map for the thermal oxide 260~h after preparation in~\ref{Maps} (b) in comparison to that in the map for the native oxide 66~h after preparation (f), which show similar similar film thickness. Further evidence for the different curvature is given by linear hight profiles, as shown in \ref{profiles}. Seemann et al.\ observed an influence of the interface potential on the shape of polystyrene nanodroplets in case of small contact angles\cite{seemann_polystyrene_2001}. Our observation points to an influence of the interface potential on the shape of ultrathin film surfaces as well. Considerations and calculations of interface potentials will be postponed to a further section.

Moreover, the film on native oxide shows a height maximum, which remains in the same position ($x\approx6$~mm, $y\approx4$~mm) during film thinning (see \ref{Maps} bottom row, and \ref{tbl:thickness} + discussion in the supplement). This is not the case for the film on thermal oxide. Here a shift of the maximum is observed. All recorded maximum values and positions are given in\ref{tbl:thickness} in the supplement. This behavior corresponds to a flow of the liquid. The absence of TEHOS flow on native oxide is in accordance with the observation reported by Forcada and Mate, who found negligible flow at the edge of a TEHOS film generated by etching part of the film~\cite{forcada_molecular_1993}. Further information on the structure of the films can be obtained by calculating thickness dependent film thinning rates~\cite{forcada_molecular_1993}, which will be discussed in the following.

\subsection{Film thinning rates}
\begin{figure}[t]
  \begin{minipage}[b]{2.6in}
  \includegraphics[width=2.6in]{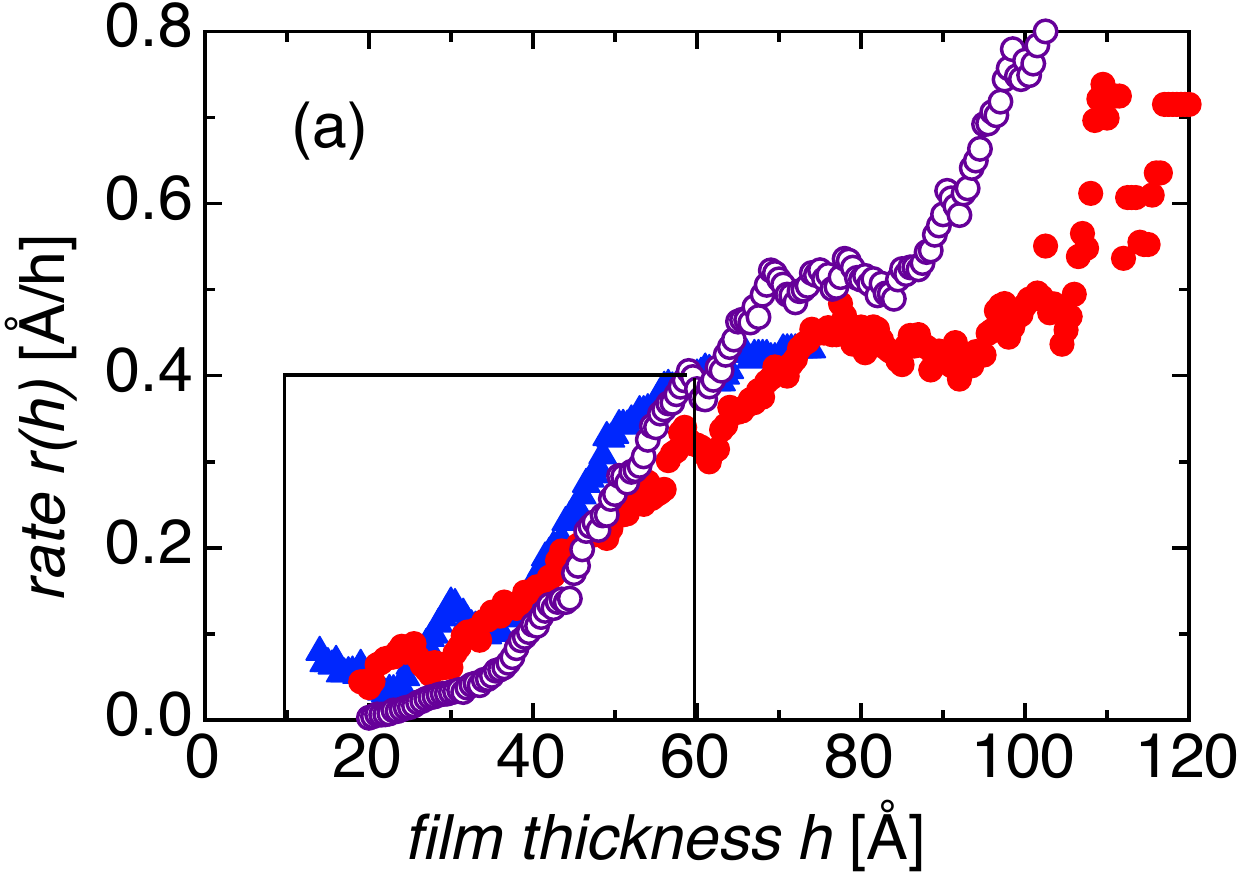}
  \includegraphics[width=2.6in]{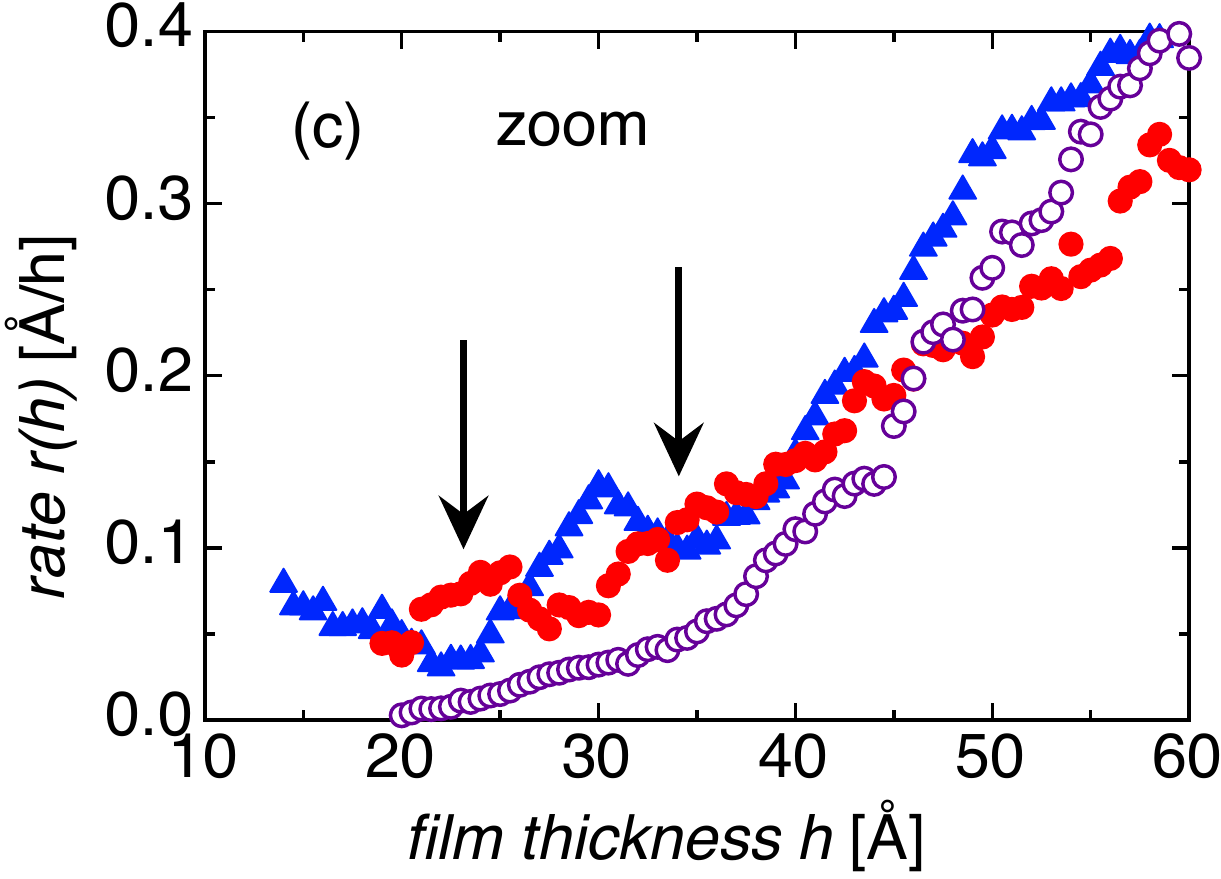}
    \end{minipage}
  \begin{minipage}[b]{2.8in}
  \includegraphics[width=2.9in]{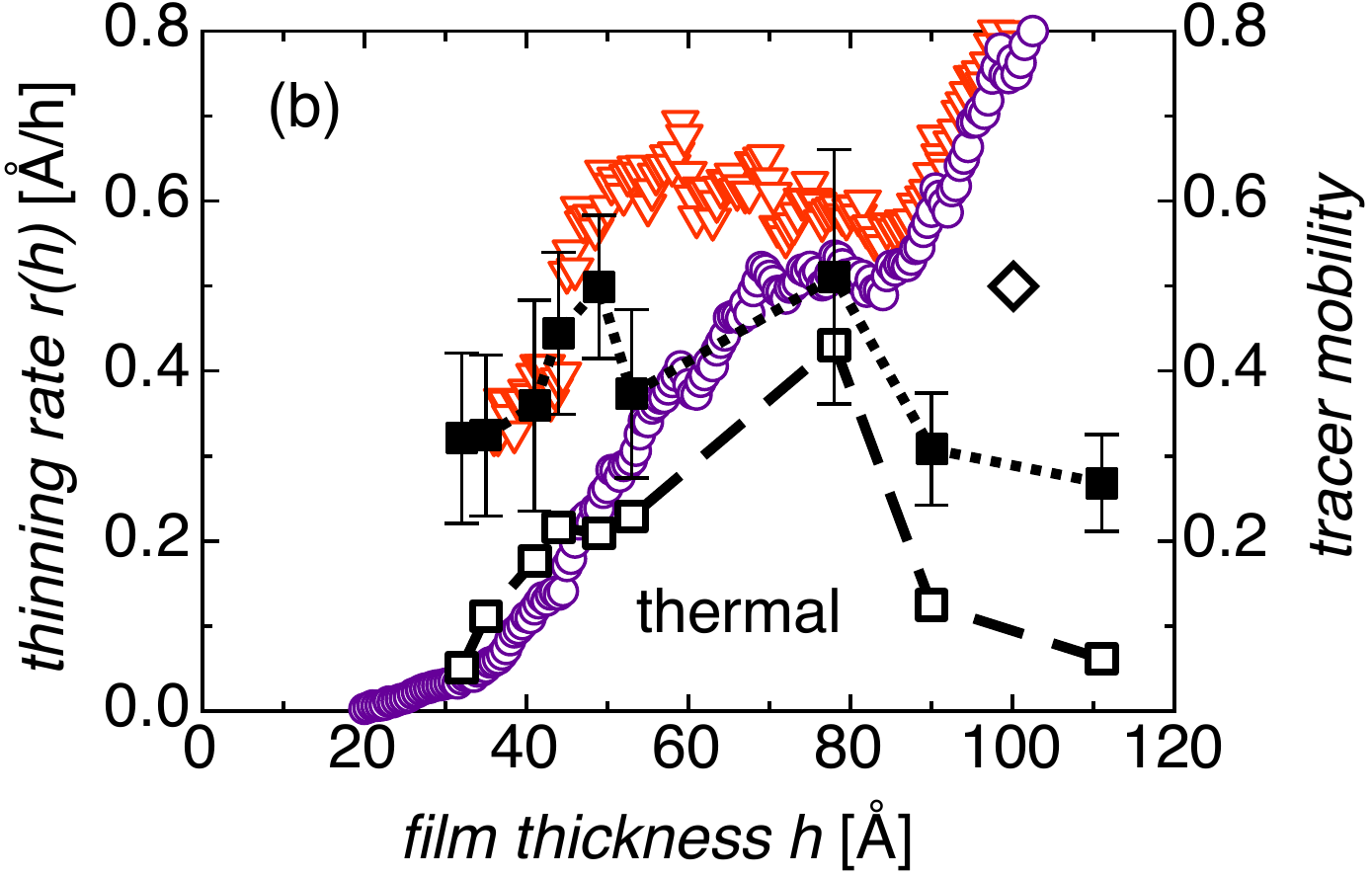}
  \includegraphics[width=2.6in]{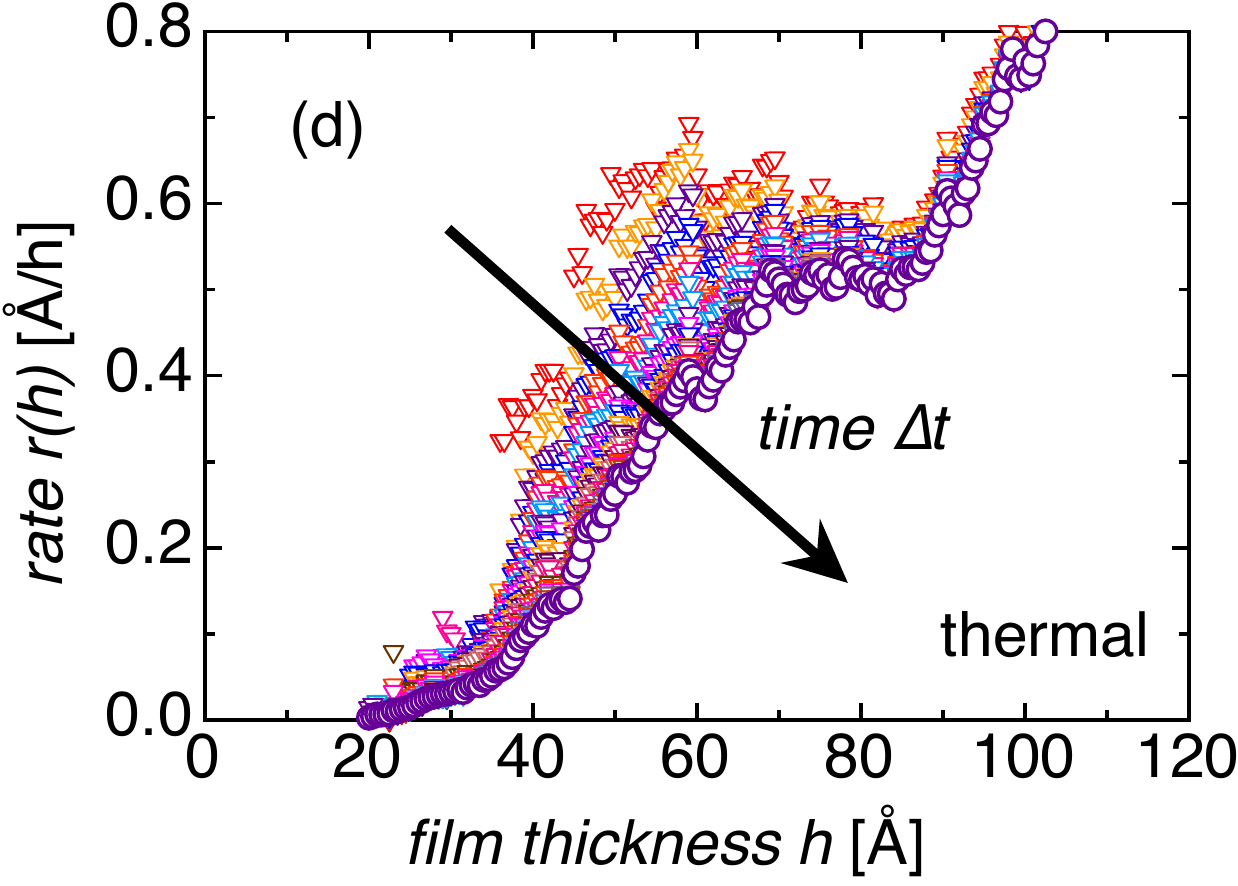}
   \end{minipage}
  \caption{Film thinning rates for a TEHOS film on native oxide ($\blacktriangle$) and two different films on 100 nm thermal oxide ($\bullet$, $\circ$). (a) Thinning rates for the full data sets obtained over time spans of some weeks. (b) Thinning rates for one of the films on thermal oxide obtained after ($\triangledown$) one week and ($\circ$) more than four weeks, together with tracer mobility: ($\square$) ratio of numbers of mobile to all trajectories, ($\blacksquare$) amplitudes $a_2$ of fast components from diffusivity analysis - error bars denote standard deviations, and ($\Diamond$) amplitude of fast component from FCS. (c) Data shown in (a) on a magnified thickness range, arrows indicate minima of rate on native oxide ($\blacktriangle$). (d) Thinning rate for one of the samples on thermal oxide shown in (a, $\circ$) together with thinning rates calculated from several subsets ($\triangledown$, differing in colors) of the full data set ($\circ$). The subsets contain increasing numbers of succeeding measurements after sample preparation, and thus cover increasing time spans $\Delta t$ of the total time of investigation.}
\label{Evaporation}
\end{figure}

TEHOS films on native oxide show similar characteristics as those reported by Forcada and Mate~\cite{forcada_molecular_1993}. In contrast, flow of the liquid is observed for TEHOS only on thermal oxide. We used two different samples of TEHOS films on thermal oxide for comparison with our sample of a TEHOS film on native oxide, to examine how this finding depends on preparation conditions.  \ref{Evaporation} shows thickness related thinning rates for the two films on thermal oxide ($\bullet$, $\circ$) and the film on native oxide ($\blacktriangle$). After a period of about two weeks, the ellipsometric measurements on two of the samples ($\bullet$, $\blacktriangle$, corresponding maps are shown in~\ref{Maps} c,b) had to be stopped due to experimental constraints. Thus the smallest studied film thickness $h$ is 13~\AA\ on native and 19~\AA\ on thermal oxide. The other thermal oxide sample ($\circ$) has been studied over a period of more than four weeks. 

In general, the thickness dependent thinning rates $r$ (given in \AA/h) are in a similar range for films on native and on thermal oxide. In particular, a constant $r\approx 0.4-0.5$~\AA/h is seen in \ref{Evaporation}(a) for a thickness interval between 90~\AA\ and 70~\AA, followed by an almost linear decrease to $r\approx 0.1-0.05$~\AA/h for $h\le 40$~\AA. Below about 40~\AA, the thinning rates decrease more slowly. Thus, two critical film thicknesses at $h\approx 80$~\AA\ and $h\approx 40$~\AA\ emerge, very similar to those previously discussed in diffusion experiments.
The similar general range of $r$ for films on native and on thermal oxide also agrees with the observation of similar macroscopic contact angles of TEHOS on both kinds of oxide (see~\ref{tbl:contactangles}), which are related to the liquid-vapor surface tension by Young's equation. The latter is related to the boiling point of the liquid~\cite{israelachvili_intermolecular_2011} and thus to evaporation. For the film on native oxide ($\blacktriangle$) the constant thinning rate $r$ starts at the largest observed local film thickness. Whereas, for the two films on thermal oxide, larger film thicknesses with much higher thinning rates are observed.

The different regions of mobility on thermal oxide represented by the changes in thinning rate~$r$ are also seen from tracer mobility, as can be seen in \ref{Evaporation}~(b). For $h\approx100$~\AA\ the ratio of mobile to all trajectories ($\square$) as well as the amplitude $a_2$ of the fast component from diffusivitiy analysis of SMT ($\blacksquare$) are low. The value obtained by FCS ($\Diamond$) is also included, indicating the loss of fast diffusion by SMT. This observation from tracer diffusion coincides with faster thinning rates ($\bullet$, $\circ$) at $h\geq90$~\AA. The fast diffusion in this region probably is bulk-like, however, due to the strong vertical confinement the diffusion coefficient cannot be resolved by tracer diffusion, see supplementary information.
For $h\approx80$~\AA, both, the ratio of mobile to all trajectories ($\square$) as well as $a_2$ ($\blacksquare$) are high. The ratio of mobile to all trajectories ($\square$) decreases for $h<80$~\AA, while $a_2$ ($\blacksquare$) stays constant within error until $h\approx50$~\AA. As was explained previously, the selection for mobile trajectories by an area threshold is rather arbitrary and in particular will miss out some mobile tracers at small $h$. The broad distributions of $D_{\rm traj}$ from all detected trajectories even for $h=32$~\AA\ (see \ref{diffusion} a), points to considerable heterogeneous tracer diffusion down to this film thickness. Within the range of 80 to 40~\AA, the thinning rates obtained after one week ($\triangledown$) and after more than four weeks ($\circ$) differ considerably. As we will see later when discussing \ref{Evaporation}~(d), this can be attributed to the influence of liquid flow on the thinning rates on thermal oxide, which hampers the evaluation of thickness dependent evaporation rates. Nevertheless, the thinning rates show still considerable liquid mobility for $h$ down to 40~\AA. This agrees with the mobility from diffusivity analysis of SMT ($\blacksquare$). The change in mobility seen from thinning rates and from diffusivity analysis between 50 and 40~\AA, might point to the onset of layering, since the extend of layering was previously reported to be about 4 liquid molecule diameters~\cite{heslot_molecular_1989, *villette_wetting_1996, forcada_molecular_1993, yu_observation_1999, *yu_molecular_2000, *mo_observation_2006, yu_x-ray_2000, snook_solvation_1980, *bitsanis_molecular_1990, *stroud_capillary_2001, kaplan_structural_2006}.

For better visualization of potentially layer dependent thinning rates $r$, the data of~\ref{Evaporation} (a) are shown on a magnified thickness range in~\ref{Evaporation} (c). The thinning rate of the TEHOS film on native oxide ($\blacktriangle$) shows two minima at 23 and 33~\AA\ (indicated by arrows) with a separation of about 10~\AA\ (see~\ref{Evaporation} c). A further minimum is expected for $h\approx 13$~\AA, which could not be reached due to experimental constraints. Nevertheless, the observation of local minima on native oxide with a separation matching the liquid molecule diameter agrees with the observation of Forcada et al., who attributed it to liquid layering~\cite{forcada_molecular_1993}.

In contrast, for the films on thermal oxide (\ref{Evaporation} (c) $\bullet$, $\circ$) no clear local maxima and minima can be seen. The thinning rate for one of the films ($\bullet$), shows a slight undulation for $h\leq30$~\AA. But, for this sample, the end of measurement renders low statistics in the range below 30~\AA. The other sample on thermal oxide ($\circ$), shows no undulations of the thinning rate for $h<40$~\AA. For this sample, the film thickness does not decrease below 20~\AA, although the sample was studied over a period of more than four weeks. For $h\leq45$~\AA, the thinning rate $r(h)$ of this sample is much lower than those of the other two samples. Yet, for TEHOS on thermal oxide, lateral flow is observed, which is expected to influence the thickness dependent thinning rates. 

Let us assume flow of liquid from position A to B between succeeding measurements. Then the local thinning rate for A appears to be higher than the average value for the corresponding thickness in absence of flow, whereas the rate at B is smaller than the average rate corresponding to the thickness at B. A flow from thicker regions to lower regions thus will explain the low thinning rates observed for this sample. If liquid flow is responsible for the here observed low thinning rates, a change in thinning rates should be seen, when average thinning rates are calculated from different time intervals $\Delta t$ of the more than four weeks total period of measurement. 

\ref{Evaporation} (d) shows such thinning rates calculated from several subsets ($\triangledown$, subsets differ by color) for the film on thermal oxide shown in~\ref{Evaporation} (a-c, $\circ$). Thereby, all subsets start from sample preparation and contain an increasing number of succeeding ellipsometry scans, thus, spanning increasing time intervals $\Delta t$ of the total period of measurement. The arrow marks the increasing time span $\Delta t$ covered by the subsets. The full data set is denoted by ($\circ$). Obviously, for the thickness interval between 80 and 20~\AA\, the apparent film thinning rates decrease with increasing $\Delta t$. This observations can be explained by a flow of the liquid from thicker regions as well as laterally outside regions into the region of measurement, which is reasonable, since also the edges and the back side of the sample are covered by TEHOS. 1150~h after sample preparation, still a residual TEHOS film of 20~\AA\ thickness is observed. 

From our film thinning experiments under ambient conditions in air, we generally observed residual layers of thicknesses between 10 to 20~\AA. The actual residual thickness depends on changing ambient conditions. For the film on native oxide, no changes over time were seen in the calculated thinning rates (see \ref{Evaporation2} (a) in the supplement). For the other thermal oxide (\ref{Evaporation} a,c, $\bullet$), the change over time of the apparent thinning was less pronounced (\ref{Evaporation2} (b) in the supplement). For this sample, liquid flow is observed also from evaluation of height profiles (see \ref{Maps} c-d, and supplementary information), which contributes to the apparent thinning rates. According to our observation, calculated thinning rates on thermal oxide contain contributions from evaporation and from liquid flow at any particular interval of the total measurement time. However, on this microscopic scale, the extend and direction of flow is induced by thermal fluctuations and defects. Thus the ratio of both components contributing to the thinning rate may change over time. On thermal oxide, the contribution from liquid flow to the apparent thinning rates smears out any dependence on layering, whereas on native oxide the absence of liquid flow~\cite{forcada_molecular_1993} enables the observation of discrete steps related to layer dependent thinning rates. This is further supported by the larger radius of curvature (see previous section) for the film on native oxide, where the almost flat film surface facilitates the observation of changes in thinning rate due to layering.

For the film on native oxide, molecular layering can be seen for thicknesses below 40~\AA, in accordance with the report by Forcada and Mate~\cite{forcada_molecular_1993}. However, from our experiments, it could not be determined, to which extend molecular layering is present in TEHOS films on thermal oxide. Nevertheless, our ellipsometry studies reveal a thinner initial film thickness on  native oxide and a higher mobility of TEHOS on  thermal oxide. The larger radius of curvature for the film on native oxide, is in agreement with observations by Seemann et al., who found an influence of long-range capillary forces on the shape of nanometric polymer melt droplets in the case of small contact angles ($\theta\approx 5^\circ$)~\cite{seemann_polystyrene_2001}. Since, both kinds of oxides displayed similar macroscopic contact angles (see~\ref{tbl:contactangles}) and surface roughnesses, the observed differences for the liquid films cannot be explained by superficial properties of the substrates. However, long-range van der Waals (vdW) interactions may account for the observation, because they are different for both kinds of substrates. Thus, we will now turn to evaluate the contribution from vdW forces to the interface potentials for the here studied ultrathin TEHOS films.

\subsection{Interface potentials}
\begin{figure}[ht]
  \includegraphics[width=2.6in]{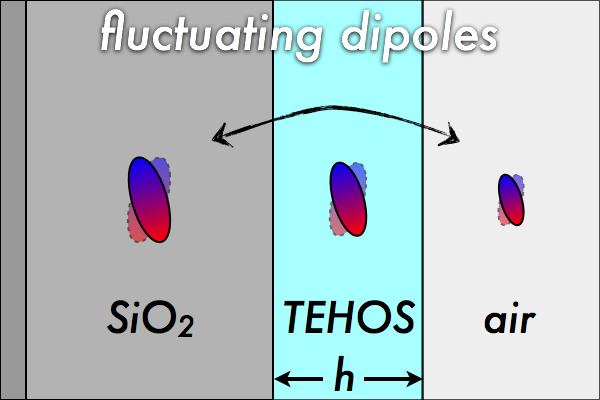}
  \hspace{0.2in}
  \includegraphics[width=2.6in]{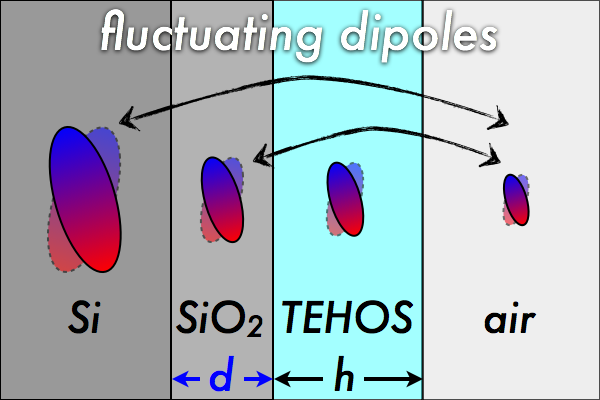}
  \caption{Scheme of van der Waals interactions for (left) a TEHOS film with thickness $h$ on thick thermal oxide (3-layer system, since the underlying silicon is shielded by the thick oxide), and (right) a TEHOS film with thickness $h$ on silicon with a thin native oxide of thickness $d$ (4-layer system).}
  \label{Layers}
\end{figure}

Van der Waals forces arise from electromagnetic fluctuations coupling between different (or like) materials across a medium. The direction and strength of the forces result from the existence or absence of resonances of these fluctuations in the relevant materials. VdW interactions combine zero frequency fluctuations of permanent dipoles with pure dispersive fluctuations (of induced dipoles only), see schematic presentation in~\ref{Layers}. These forces are non-additive, apart from diluted gases. In 1956, Lifshitz applied a continuum approach in quantum field theory to describe vdW forces~\cite{israelachvili_intermolecular_2011}. His theory was simplified using a summation of oscillator frequencies~\cite{parsegian_van_2006}. However, its application acquires extended knowledge of optical material properties, which are hardly available for solids, and which are even harder to achieve for soft matter. Nevertheless, an approximation may be obtained using the dielectric properties of the relevant materials in the visible frequency range~\cite{israelachvili_intermolecular_2011}. 

A thin liquid film of thickness $h$ on a homogeneous substrate in air corresponds to a 3-layer system, (see~\ref{Layers}, left), for which the effective vdW potential is given by~\cite{israelachvili_intermolecular_2011,parsegian_van_2006,jacobs_stability_2008}
\begin{equation}
\Phi_{vdW}(h)=-\frac{A_{12-23}}{12\pi h^2}\, ,
\label{3Layer}
\end{equation}
where $A_{12-23}$ is the material dependent effective Hamaker coefficient (details in the supplement). In case of an additional coating material on the substrate, \ref{3Layer} has to be extended including the interaction of the additional layer with the liquid film~\cite{israelachvili_intermolecular_2011, parsegian_van_2006,jacobs_stability_2008}. For the system of a thin liquid film of thickness $h$ prepared on a silicon wafer with an oxide layer of thickness $d$ (see~\ref{Layers}, right), this results in 
\begin{equation}
\Phi_{vdW}(h)=-\frac{A_{SiO2}}{12\pi h^2}-\frac{A_{Si}-A_{SiO2}}{12\pi (h+d)^2}\, ,
\label{4Layer}
\end{equation}
where $A_{SiO2}$ and $A_{Si}$ are the effective Hamaker coefficients for the system \ce{SiO2}/film/air and  Si/film/air, respectively~\cite{seemann_dewetting_2001, jacobs_stability_2008}. This approximation yields an one oder of magnitude stronger effective interface potential for a TEHOS film on native oxide in comparison to a TEHOS film on thick thermal oxide. The difference is caused by the influence of the strong polarizability of the underlying silicon, which is shielded in case of  thick thermal oxide, but contributes in case of only 2~nm native oxide (\ref{Layers}). A more detailed description is given in the supplementary information. \ref{Potentials} (a) shows the thus obtained effective interface potentials $\Phi(h)$ for TEHOS films on both kinds of substrate. 
\begin{figure}
  \centering
  \begin{minipage}[b]{2.6in}
  \includegraphics[width=2.3in]{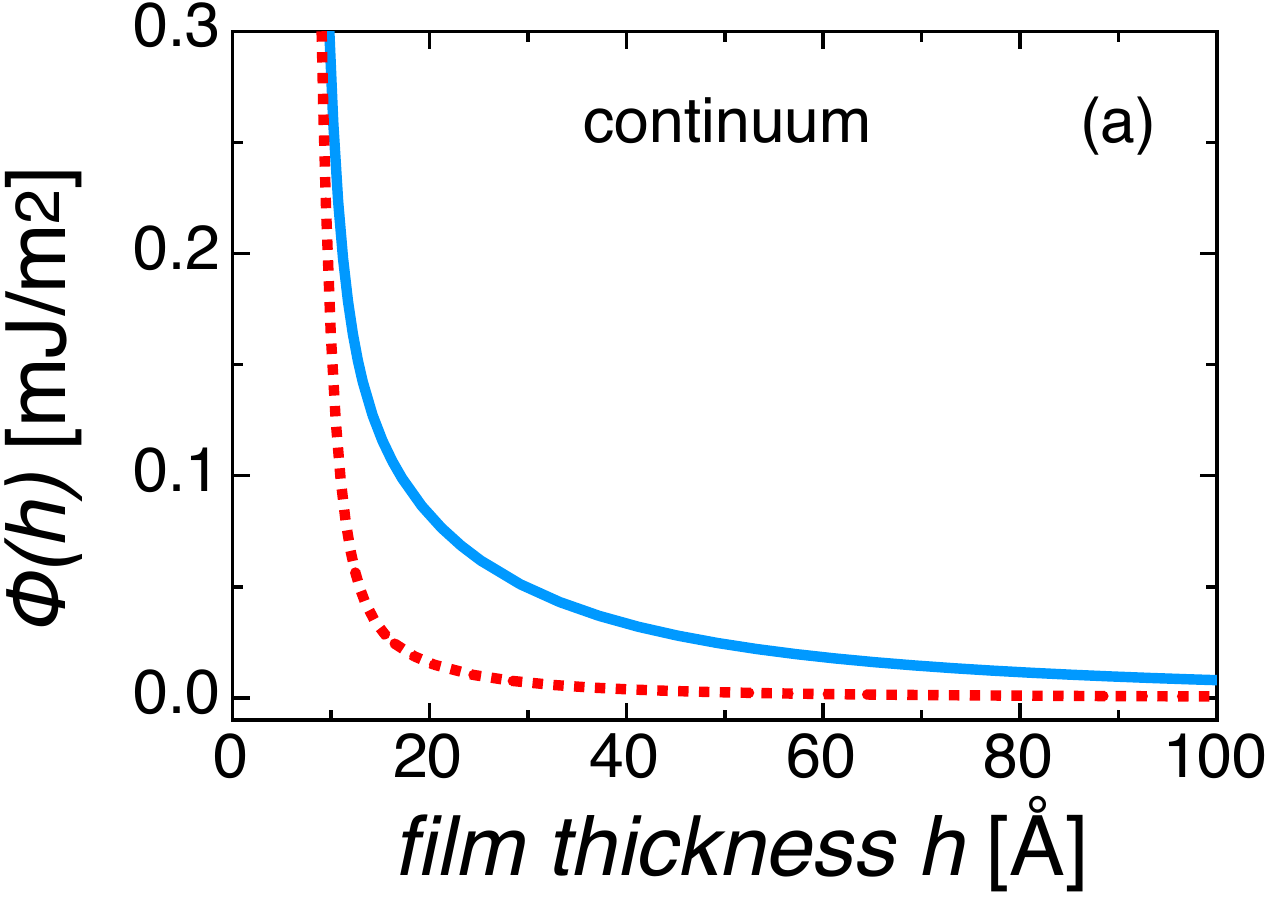}
 \includegraphics[width=2.58in]{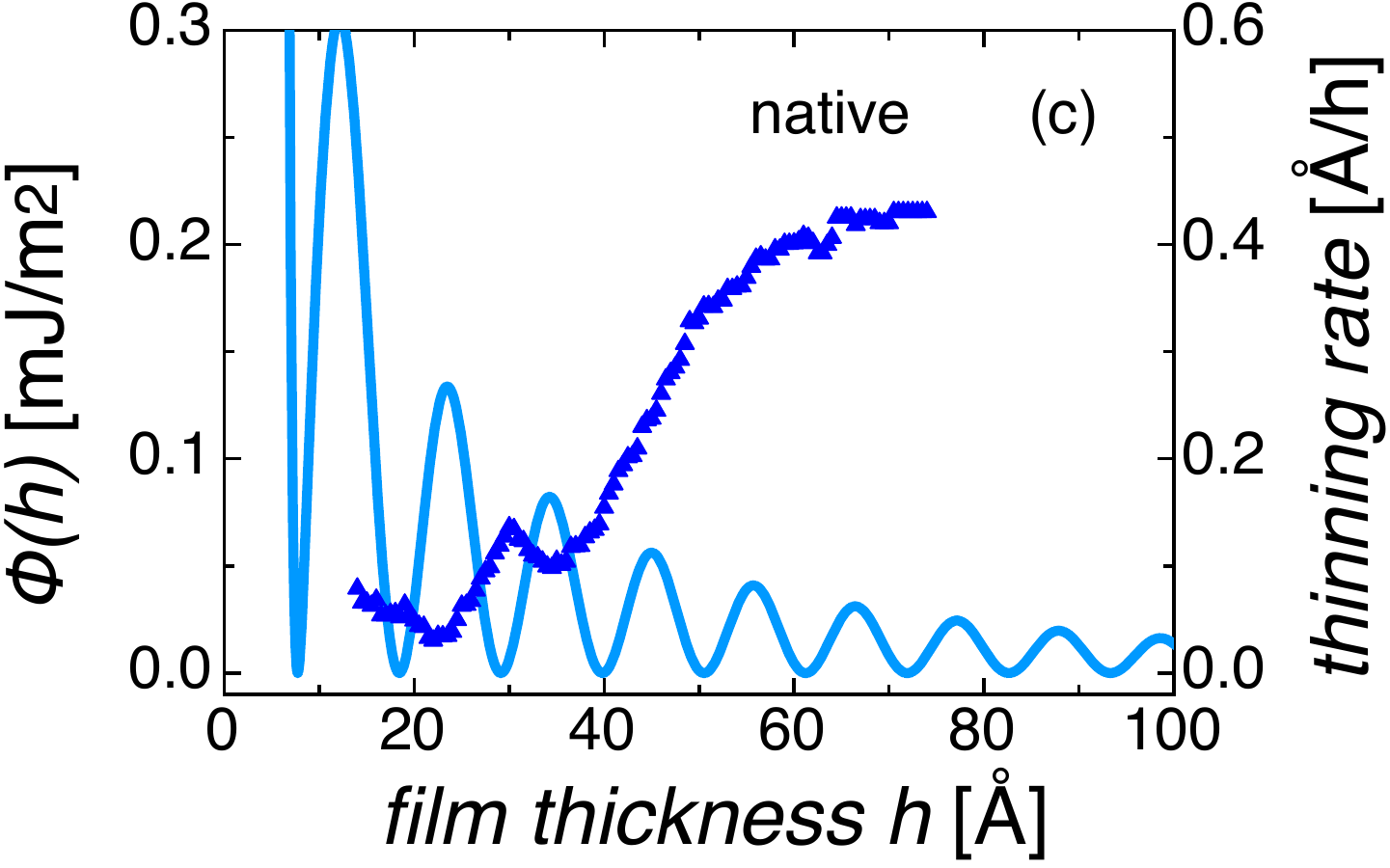}
  \end{minipage}
  \begin{minipage}[b]{2.6in}
  \includegraphics[width=2.3in]{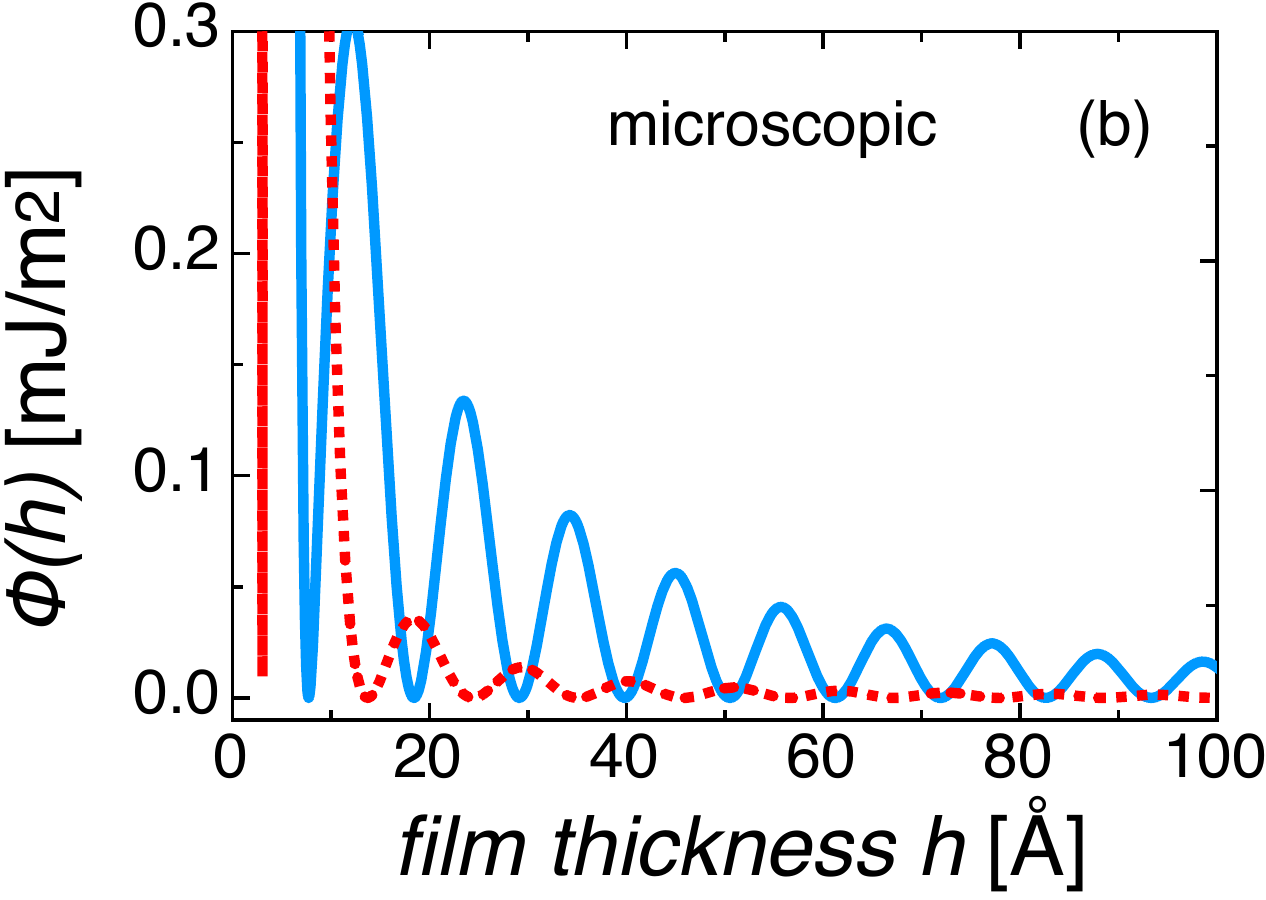}
  \includegraphics[width=2.58in]{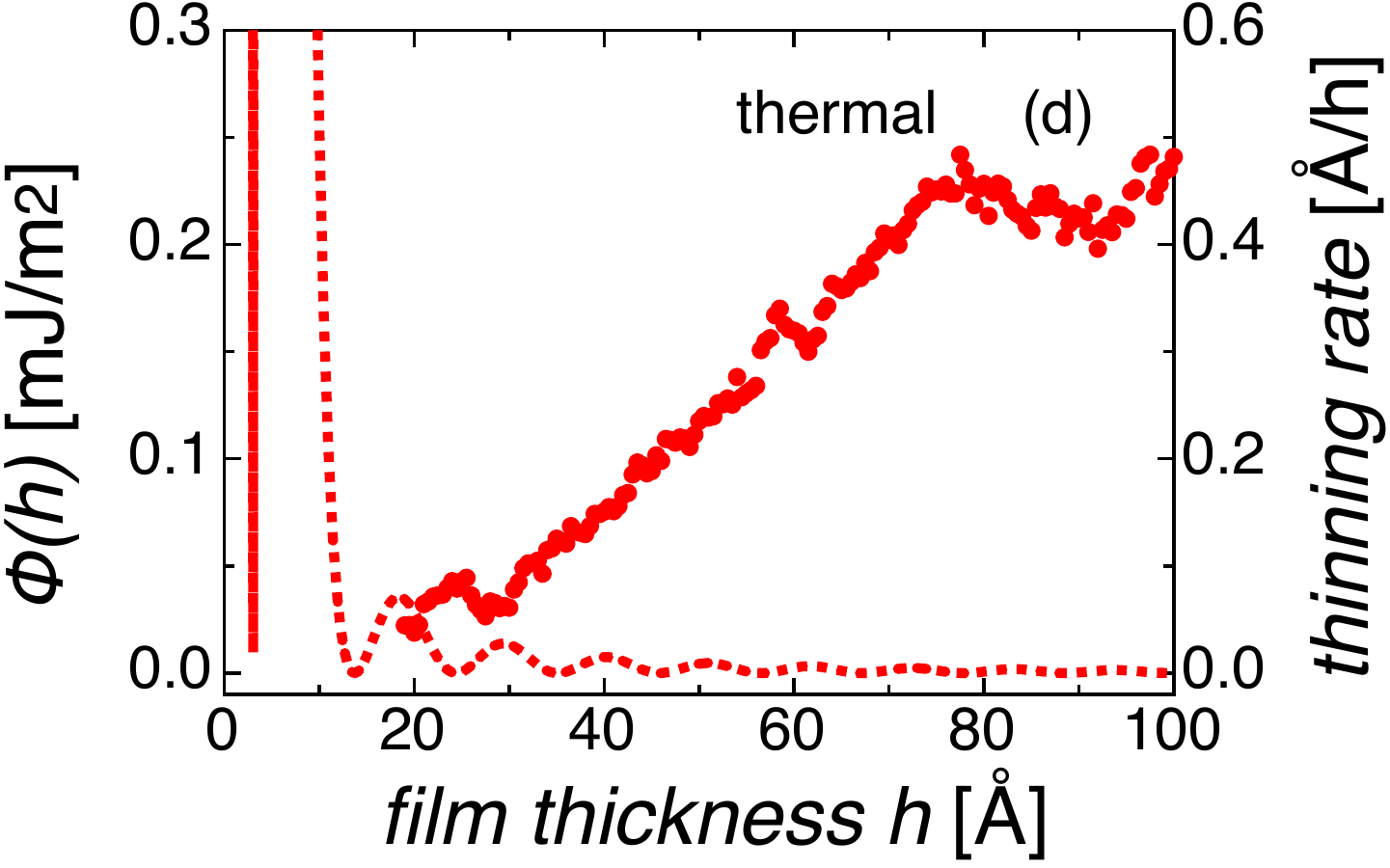}
  \end{minipage}
  \caption{Effective interface potentials $\Phi(h)$ for TEHOS films on silicon wafers with (---) native and ($\cdot\cdot\cdot$) 100~nm thermal oxide (a) calculated from continuum theory~\ref{interface_potential}, (b) with oscillations from layering, (c,d) compared to film thinning rates on (c) native and (d) thermal oxide (shown in~\ref{Evaporation} a).}
  \label{Potentials}
\end{figure}

VdW potentials calculated from this simplified Lifshitz theory and its approximations are derived from (homogeneous) continuum material properties and thus do not account for modifications of these properties on a molecular scale (for example due to molecular layering). In particular, the refractive index is related to material density. Therefore, density oscillations derived from X-ray reflectometry in thin liquid films close to solid-liquid interfaces can be used to improve the determination of the effective interface potential~\cite{israelachvili_intermolecular_2011}. Yu et al.\ studied molecular layering of TEHOS films on silicon with native oxide. They found electron density oscillations with a period of 11~\AA~\cite{yu_x-ray_2000}, which corresponds approximately to the molecular diameter of TEHOS molecules~\cite{heslot_molecular_1989, yu_x-ray_2000}. The amplitude of these oscillations decayed within several tens of nanometers. Thus, their findings resemble a liquid layering up to four molecular diameters above the substrate, whereby, the localization of the molecules within the layers decreases with distance from the substrate~\cite{yu_x-ray_2000}. Yu et al.\ showed that the experimental data could be modeled by a damped sine function. For this, they
used a model independent fit as well as a damped sine function to fit the data. For a distance larger than 10~\AA, both gave very similar results. Only very close to the substrate, the electron density cannot be reproduced by the damped sine function~\cite{yu_x-ray_2000}. Thus, we use a sine function of the same period (11~\AA) to modulate the interface potentials obtained from continuum material properties. The resulting potentials are shown in  \ref{Potentials} (b). Although X-ray measurements are reported only for TEHOS on native oxide, a similar modulation is also superposed on the effective interface potential for thermal oxide. However, it is not straightforward to decide about the exact position of the absolute phase due to possible conformational changes of the branched and thus soft TEHOS molecules in the layer next to the substrate~\cite{forcada_molecular_1993, heslot_molecular_1989}. Due to this lack of information about the thickness of the first molecular layer, the phase shifts of the sine functions are chosen in such a way that the 
minima of the thinning rates at $h\approx30$~\AA\ (thermal) and $h\approx35$~\AA\ (native) are met by maxima of the potentials, as can be seen in the comparisons in~\ref{Potentials} (c,d). As was shown by Yu et al., the damped since function cannot represent the interface potential below 10~\AA~\cite{yu_x-ray_2000}. Therefore, the modulated potentials are only considered for the discussion for $h\ge10$~\AA.

The differing strength of the interface potentials for ultrathin TEHOS films on native and on thick thermal oxide could already be seen from continuum theory (\ref{Potentials} a). Yet, from the comparison of the layer-modulated potentials (\ref{Potentials} b-d), similarities and differences can be seen even more clearly. Both potentials are large for $h=10$~\AA, and thus support a residual (frozen) adsorption layer. The potential maxima at $h\approx20$~\AA\ and $h\approx30$~\AA\ seen for thermal oxide are much weaker as the maxima for native oxide in the range below $h=40$~\AA. This may explain the higher curvature on thermal oxide and the flattened surface of the ultrathin TEHOS film, as well as the restricted macroscopic fluidity on native oxide. 

Despite the differences in potential strength, the thinning rates for $h=40$~\AA\ are similar on both substrates. We found that the thinning rate for thermal oxide is just an apparent thinning rate influenced by evaporation as well as by liquid flow, and thus the evaporation rate may even be higher, as is seen in \ref{Evaporation} (b) for thinning rates obtained from shorter spans of measurement time $\Delta t$. Nevertheless, in the range between 40 and 70~\AA, the thinning rates increase to a constant value $r\approx0.4-0.5$~\AA/h. This points to geometrical constraints influencing the film thinning rates, in particular, molecular layering may also extend to a four molecules distance on thermal oxide. This suggestion is further supported by the drop in the fast component of tracer diffusion observed at this film thickness (see~\ref{Evaporation} b, $\blacksquare$). Our calculations did not take the finite velocity of light into account, which leads to retardation of the interaction for a film thickness $h\ge 50$~\AA, implying an even faster decay of the interface potentials with $h$~\cite{parsegian_van_2006, israelachvili_intermolecular_2011}. The increase of the thinning rate $r(h)$ starting at $h\approx40$~\AA, even for native oxide, points to a general crossover from an evaporation rate mainly influenced by $\Phi(h)$, to an evaporation rate influenced only by the surface tension of TEHOS in air.

The stronger potential $\Phi(h)$ in general, does not necessarily explain the observed smaller overall film thickness from preparation on native oxide. Here, a closer look at the contributions to the potential may provide further information.~\ref{Potentials} shows effective interface potentials containing contributions from van der Waals interactions related to both, permanent dipoles (zero frequency fluctuations) and dispersion, i.e.\ fluctuations from induced dipoles. The contribution of the fluctuations from permanent dipoles to the effective Hamaker coefficients is negative in case of the system Si/TEHOS/air (enhancing film stability), but it is positive for the system \ce{SiO2}/TEHOS/air (see~\ref{tbl:effHam}in the supplementary information). A similar trend is observed for systems containing hexane instead of TEHOS (see supplement).

Silicon wafers are cut from highly ordered crystals. Therefore, its permanent dipoles are also highly ordered. In technical crystal grow, the so called "van der Waals epitaxy" uses ordering properties of van der Waals forces to grow crystals on substrate crystals in a highly ordered structure in spite of a mismatch in lattice properties~\cite{koma_fabrication_1984,*koma_van_1999}. Atomistic simulations of liquids at interfaces with crystalline solids showed substantial in-plane ordering within liquid layers in case of compatible structures of the liquid phase and the solid. For incompatible liquid phases the liquid formed layers with a liquid-like in plane structure~\cite{kaplan_structural_2006}. It is reasonable that not only solids, but also liquids are influenced by the crystal structure of the underlying substrate crystal in the range of several tens of nanometers, which is likely to lead to restricted rotational and translational motion of liquid molecules in this interface region. By this, the restricted mobility and highly ordered structure of ultrathin TEHOS films on \ce{Si} wafers with native oxide can be explained. Furthermore, the strong ordering and reduced mobility may allow attachment of TEHOS molecules only at preferable sites within the growing film during dipcoating, and therefore may account for the thinner TEHOS films on native oxide compared to those on 100 nm thick thermal oxide. This suggestion is further supported by the recent observation of the coordinated mobility of aggregated fullerenes on weakly interacting substrates~\cite{svec_van_2012}.

The observed flow of TEHOS on the thick thermal oxide may be related to thermally induced thickness fluctuations and thus to capillary waves at the free surface with air. In case of thin polymer films, surface modes with growing amplitudes are responsible for spinodal dewetting. These growing modes will occur, if the interface potential contains a local minimum and a negative second derivative around a critical film thickness $h_c$~\cite{vrij_possible_1966,seemann_dewetting_2001, jacobs_stability_2008}. The interface potentials $\Phi(h)$ for the here studied TEHOS films, however, do not contain such minima (since $sin``(x)=-sin(x)$). Yet, the surface of the TEHOS films on thick thermal oxide was considerably convex. Thus, the derivation given by Vrij for the increase of free energy $\Delta F$ associated with thickness fluctuations of a film of otherwise homogenous thickness~\cite{vrij_possible_1966} does not fully apply. In particular, the contributions from the local thickness modulations to $\Delta F$ will override any contribution from the considerably weak $\Phi(h)$ for film thicknesses $h>20$~\AA.

In hydrodynamics, liquid flow at interfaces of solids is discussed in terms of slip or no-slip boundary conditions. Thereby, the occurrence of slip is usually related to superficial properties like interface roughness and chemistry. The above reported results emphasize that also in this case long-range interactions with subsurface materials have to be considered. According to theoretical considerations by Bocquet and Barrat, a reduction of $\Phi(h)$ will enhance the slip length and thus the flow of the liquid\cite{bocquet_flow_2007}, which is in accordance with our above described observations. The observed impact of the interface potential on liquid dynamics, may also contribute to the mobility of polymer chains at interfaces even besides dewetting situations. Thus recently observed deviations of polymer properties in thin films from those in the corresponding bulk material~\cite{hall_small_1998, *lin_reduced_1999, *hou_characterization_2000, *araoz_cage_2012} should be discussed also in the light of these long-range vdW interactions.

\section{Conclusions}
This contribution aims at the understanding of the influence of interface properties on formation of structure and molecular mobilities in liquid films on a nanometer scale. The observations from film thinning of ultrathin TEHOS films on silicon with either native or 100~nm thermal oxide show different behavior of the films on the two kinds of substrate. In accordance with reports from literature~\cite{forcada_molecular_1993}, TEHOS films on native oxide are highly structured and show negligible flow. In contrast, on thick thermal oxide flow was observed for TEHOS films down to a residual film thickness $h\approx20$~\AA. This flow may conceal layering from detection by our ellipsometry study. Nevertheless, film thinning rates showed an overall similar thickness dependence on both kinds of substrates. In particular, the thinning rates $r$ decreased from $r\approx0.4$~\AA/h at $h\approx70$~\AA\ to $r\approx0.1$~\AA/h at $h\approx40$~\AA. This points to a similar structural change on both kinds of substrates in the region of $h\approx40$~\AA. Despite the higher mobility and surface curvature for films on thermal oxide, TEHOS may be organized into molecular layers also in this case. This suggestion is further supported by our findings from tracer diffusion in thinning TEHOS films on thermal oxide. These experiments show changes at $h\approx80$~\AA\ (for thicker films, faster diffusion occurs), and $h\approx50$~\AA\ (number of detected trajectories decreases to a constant value and amplitude $a_2$ of fast component starts decreasing with $h$). The latter corresponds to the expected range of molecular layering. Further information may be obtained from a comparative X-ray study.

The here reported observation of subsurface composition influencing both the structure and dynamics at solid-liquid interfaces, points to the necessity of taking also long-range vdW interactions into account. Pairwise additivity of Lennard Jones interactions may overestimate the extend of layering reported from simple Monte Carlo simulations~\cite{snook_solvation_1980, *bitsanis_molecular_1990, *stroud_capillary_2001}. In a similar way, for modeling adsorption of organic molecules on metals it was found necessary to include vdW interactions in DFT calculations~\cite{hauschild_normal-incidence_2010,*ruiz_density-functional_2012}. From our findings, it can be concluded that the structure and mobility of liquids in vicinity of solid-liquid interfaces is influenced not only by superficial properties of the substrate, but also by its subsurface composition. In particular, liquid structure formation and diffusion dynamics observed for one specific kind of substrate, may not be generalized to all substrates containing similar surface chemistry and roughness, while neglecting mutable influence of different subsurface composition. On the other hand, this very influence of the subsurface material may be used for tailoring structure and dynamics within technical applications without the need of changing the superficial properties of the substrates.

\acknowledgement
We thank D.\ R.\ T.\ Zahn, TU Chemnitz, for the possibility to use the M-2000 ellipsometer and Ovidiu Gordan and Michael Fronk, TU Chemnitz, for instructions and help with the ellipsometry equipment. Support from the DFG within the research group FOR 877 "From local constraints to macroscopic transport" is gratefully acknowledged.

\suppinfo

\bibliography{VdW}

\providecommand*\mcitethebibliography{\thebibliography}
\csname @ifundefined\endcsname{endmcitethebibliography}
  {\let\endmcitethebibliography\endthebibliography}{}
\begin{mcitethebibliography}{69}
\providecommand*\natexlab[1]{#1}
\providecommand*\mciteSetBstSublistMode[1]{}
\providecommand*\mciteSetBstMaxWidthForm[2]{}
\providecommand*\mciteBstWouldAddEndPuncttrue
  {\def\EndOfBibitem{\unskip.}}
\providecommand*\mciteBstWouldAddEndPunctfalse
  {\let\EndOfBibitem\relax}
\providecommand*\mciteSetBstMidEndSepPunct[3]{}
\providecommand*\mciteSetBstSublistLabelBeginEnd[3]{}
\providecommand*\EndOfBibitem{}
\mciteSetBstSublistMode{f}
\mciteSetBstMaxWidthForm{subitem}{(\alph{mcitesubitemcount})}
\mciteSetBstSublistLabelBeginEnd
  {\mcitemaxwidthsubitemform\space}
  {\relax}
  {\relax}

\bibitem[Qiao et~al.(2009)Qiao, Liu, and Chen]{qiao_pressurized_2009}
Qiao,~Y.; Liu,~L.; Chen,~X. Pressurized Liquid in Nanopores: A Modified
  {Laplace-Young} Equation. \emph{Nano Letters} \textbf{2009}, \emph{9},
  984--988\relax
\mciteBstWouldAddEndPuncttrue
\mciteSetBstMidEndSepPunct{\mcitedefaultmidpunct}
{\mcitedefaultendpunct}{\mcitedefaultseppunct}\relax
\EndOfBibitem
\bibitem[Hummer et~al.(2001)Hummer, Rasaiah, and Noworyta]{hummer_water_2001}
Hummer,~G.; Rasaiah,~J.~C.; Noworyta,~J.~P. Water conduction through the
  hydrophobic channel of a carbon nanotube. \emph{Nature} \textbf{2001},
  \emph{414}, 188--190\relax
\mciteBstWouldAddEndPuncttrue
\mciteSetBstMidEndSepPunct{\mcitedefaultmidpunct}
{\mcitedefaultendpunct}{\mcitedefaultseppunct}\relax
\EndOfBibitem
\bibitem[Holt et~al.(2006)Holt, Park, Wang, Stadermann, Artyukhin,
  Grigoropoulos, Noy, and Bakajin]{holt_fast_2006}
Holt,~J.~K.; Park,~H.~G.; Wang,~Y.; Stadermann,~M.; Artyukhin,~A.~B.;
  Grigoropoulos,~C.~P.; Noy,~A.; Bakajin,~O. Fast Mass Transport Through
  {Sub-2-Nanometer} Carbon Nanotubes. \emph{Science} \textbf{2006}, \emph{312},
  1034 --1037\relax
\mciteBstWouldAddEndPuncttrue
\mciteSetBstMidEndSepPunct{\mcitedefaultmidpunct}
{\mcitedefaultendpunct}{\mcitedefaultseppunct}\relax
\EndOfBibitem
\bibitem[Eijkel(2007)]{eijkel_liquid_2007}
Eijkel,~J. C.~T. Liquid slip in micro- and nanofluidics: recent research and
  its possible implications. \emph{Lab Chip} \textbf{2007}, \emph{7},
  299--301\relax
\mciteBstWouldAddEndPuncttrue
\mciteSetBstMidEndSepPunct{\mcitedefaultmidpunct}
{\mcitedefaultendpunct}{\mcitedefaultseppunct}\relax
\EndOfBibitem
\bibitem[Mijatovic et~al.(2005)Mijatovic, Eijkel, and van~den
  Berg]{mijatovic_technologies_2005}
Mijatovic,~D.; Eijkel,~J. C.~T.; van~den Berg,~A. Technologies for nanofluidic
  systems: top-down vs. bottom-up-a review. \emph{Lab Chip} \textbf{2005},
  \emph{5}, 492--500\relax
\mciteBstWouldAddEndPuncttrue
\mciteSetBstMidEndSepPunct{\mcitedefaultmidpunct}
{\mcitedefaultendpunct}{\mcitedefaultseppunct}\relax
\EndOfBibitem
\bibitem[Wang et~al.(2012)Wang, Ouyang, Ye, Xu, Chen, and Xia]{wang_rapid_2012}
Wang,~C.; Ouyang,~J.; Ye,~D.; Xu,~J.; Chen,~H.; Xia,~X. Rapid protein
  concentration, efficient fluorescence labeling and purification on a
  micro/nanofluidics chip. \emph{Lab Chip} \textbf{2012}, \emph{12},
  2664--2671\relax
\mciteBstWouldAddEndPuncttrue
\mciteSetBstMidEndSepPunct{\mcitedefaultmidpunct}
{\mcitedefaultendpunct}{\mcitedefaultseppunct}\relax
\EndOfBibitem
\bibitem[Comiskey et~al.(1998)Comiskey, Albert, Yoshizawa, and
  Jacobson]{comiskey_electrophoretic_1998}
Comiskey,~B.; Albert,~J.~D.; Yoshizawa,~H.; Jacobson,~J. An electrophoretic ink
  for all-printed reflective electronic displays. \emph{Nature} \textbf{1998},
  \emph{394}, 253--255\relax
\mciteBstWouldAddEndPuncttrue
\mciteSetBstMidEndSepPunct{\mcitedefaultmidpunct}
{\mcitedefaultendpunct}{\mcitedefaultseppunct}\relax
\EndOfBibitem
\bibitem[Sheats(2004)]{sheats_manufacturing_2004}
Sheats,~J.~R. Manufacturing and commercialization issues in organic
  electronics. \emph{Journal of Materials Research} \textbf{2004}, \emph{19},
  1974--1989\relax
\mciteBstWouldAddEndPuncttrue
\mciteSetBstMidEndSepPunct{\mcitedefaultmidpunct}
{\mcitedefaultendpunct}{\mcitedefaultseppunct}\relax
\EndOfBibitem
\bibitem[Calvert(2001)]{calvert_inkjet_2001}
Calvert,~P. Inkjet Printing for Materials and Devices. \emph{Chemistry of
  Materials} \textbf{2001}, \emph{13}, 3299--3305\relax
\mciteBstWouldAddEndPuncttrue
\mciteSetBstMidEndSepPunct{\mcitedefaultmidpunct}
{\mcitedefaultendpunct}{\mcitedefaultseppunct}\relax
\EndOfBibitem
\bibitem[Marmur(2003)]{marmur_wetting_2003}
Marmur,~A. Wetting on Hydrophobic Rough Surfaces: To Be Heterogeneous or Not To
  Be? \emph{Langmuir} \textbf{2003}, \emph{19}, 8343--8348\relax
\mciteBstWouldAddEndPuncttrue
\mciteSetBstMidEndSepPunct{\mcitedefaultmidpunct}
{\mcitedefaultendpunct}{\mcitedefaultseppunct}\relax
\EndOfBibitem
\bibitem[Feng et~al.(2011)Feng, Zhang, Cao, Ye, and Jiang]{feng_effect_2011}
Feng,~L.; Zhang,~Y.; Cao,~Y.; Ye,~X.; Jiang,~L. The effect of surface
  microstructures and surface compositions on the wettabilities of flower
  petals. \emph{Soft Matter} \textbf{2011}, \emph{7}, 2977--2980\relax
\mciteBstWouldAddEndPuncttrue
\mciteSetBstMidEndSepPunct{\mcitedefaultmidpunct}
{\mcitedefaultendpunct}{\mcitedefaultseppunct}\relax
\EndOfBibitem
\bibitem[Nosonovsky(2011)]{nosonovsky_materials_2011}
Nosonovsky,~M. Materials science: Slippery when wetted. \emph{Nature}
  \textbf{2011}, \emph{477}, 412--413\relax
\mciteBstWouldAddEndPuncttrue
\mciteSetBstMidEndSepPunct{\mcitedefaultmidpunct}
{\mcitedefaultendpunct}{\mcitedefaultseppunct}\relax
\EndOfBibitem
\bibitem[Picknett and Bexon(1977)Picknett, and
  Bexon]{picknett_evaporation_1977}
Picknett,~R.; Bexon,~R. The evaporation of sessile or pendant drops in still
  air. \emph{Journal of Colloid and Interface Science} \textbf{1977},
  \emph{61}, 336--350\relax
\mciteBstWouldAddEndPuncttrue
\mciteSetBstMidEndSepPunct{\mcitedefaultmidpunct}
{\mcitedefaultendpunct}{\mcitedefaultseppunct}\relax
\EndOfBibitem
\bibitem[{Bourges-Monnier} and Shanahan(1995){Bourges-Monnier}, and
  Shanahan]{bourges-monnier_influence_1995}
{Bourges-Monnier},~C.; Shanahan,~M. E.~R. Influence of Evaporation on Contact
  Angle. \emph{Langmuir} \textbf{1995}, \emph{11}, 2820--2829\relax
\mciteBstWouldAddEndPuncttrue
\mciteSetBstMidEndSepPunct{\mcitedefaultmidpunct}
{\mcitedefaultendpunct}{\mcitedefaultseppunct}\relax
\EndOfBibitem
\bibitem[Tadmor(2011)]{tadmor_approaches_2011}
Tadmor,~R. Approaches in wetting phenomena. \emph{Soft Matter} \textbf{2011},
  \emph{7}, 1577--1580\relax
\mciteBstWouldAddEndPuncttrue
\mciteSetBstMidEndSepPunct{\mcitedefaultmidpunct}
{\mcitedefaultendpunct}{\mcitedefaultseppunct}\relax
\EndOfBibitem
\bibitem[Fisher and Israelachvili(1981)Fisher, and
  Israelachvili]{fisher_experimental_1981}
Fisher,~L.~R.; Israelachvili,~J.~N. Experimental studies on the applicability
  of the Kelvin equation to highly curved concave menisci. \emph{Journal of
  Colloid and Interface Science} \textbf{1981}, \emph{80}, 528--541\relax
\mciteBstWouldAddEndPuncttrue
\mciteSetBstMidEndSepPunct{\mcitedefaultmidpunct}
{\mcitedefaultendpunct}{\mcitedefaultseppunct}\relax
\EndOfBibitem
\bibitem[Leger and Joanny(1992)Leger, and Joanny]{leger_liquid_1992}
Leger,~L.; Joanny,~J.~F. Liquid spreading. \emph{Reports on Progress in
  Physics} \textbf{1992}, \emph{55}, 431\relax
\mciteBstWouldAddEndPuncttrue
\mciteSetBstMidEndSepPunct{\mcitedefaultmidpunct}
{\mcitedefaultendpunct}{\mcitedefaultseppunct}\relax
\EndOfBibitem
\bibitem[Xu et~al.(2004)Xu, Shirvanyants, Beers, Matyjaszewski, Rubinstein, and
  Sheiko]{xu_molecular_2004}
Xu,~H.; Shirvanyants,~D.; Beers,~K.; Matyjaszewski,~K.; Rubinstein,~M.;
  Sheiko,~S.~S. Molecular Motion in a Spreading Precursor Film. \emph{Physical
  Review Letters} \textbf{2004}, \emph{93}, 206103\relax
\mciteBstWouldAddEndPuncttrue
\mciteSetBstMidEndSepPunct{\mcitedefaultmidpunct}
{\mcitedefaultendpunct}{\mcitedefaultseppunct}\relax
\EndOfBibitem
\bibitem[Seemann et~al.(2001)Seemann, Jacobs, and
  Blossey]{seemann_polystyrene_2001}
Seemann,~R.; Jacobs,~K.; Blossey,~R. Polystyrene nanodroplets. \emph{Journal of
  Physics: Condensed Matter} \textbf{2001}, \emph{13}, 4915\relax
\mciteBstWouldAddEndPuncttrue
\mciteSetBstMidEndSepPunct{\mcitedefaultmidpunct}
{\mcitedefaultendpunct}{\mcitedefaultseppunct}\relax
\EndOfBibitem
\bibitem[Heslot et~al.(1989)Heslot, Fraysse, and
  Cazabat]{heslot_molecular_1989}
Heslot,~F.; Fraysse,~N.; Cazabat,~A.~M. Molecular layering in the spreading of
  wetting liquid drops. \emph{Nature} \textbf{1989}, \emph{338}, 640--642\relax
\mciteBstWouldAddEndPuncttrue
\mciteSetBstMidEndSepPunct{\mcitedefaultmidpunct}
{\mcitedefaultendpunct}{\mcitedefaultseppunct}\relax
\EndOfBibitem
\bibitem[Villette et~al.(1996)Villette, Valignat, Cazabat, Jullien, and
  Tiberg]{villette_wetting_1996}
Villette,~S.; Valignat,~M.~P.; Cazabat,~A.~M.; Jullien,~L.; Tiberg,~F. Wetting
  on the Molecular Scale and the Role of Water. A Case Study of Wetting of
  Hydrophilic Silica Surfaces. \emph{Langmuir} \textbf{1996}, \emph{12},
  825--830\relax
\mciteBstWouldAddEndPuncttrue
\mciteSetBstMidEndSepPunct{\mcitedefaultmidpunct}
{\mcitedefaultendpunct}{\mcitedefaultseppunct}\relax
\EndOfBibitem
\bibitem[Forcada and Mate(1993)Forcada, and Mate]{forcada_molecular_1993}
Forcada,~M.~L.; Mate,~C.~M. Molecular layering during evaporation of ultrathin
  liquid films. \emph{Nature} \textbf{1993}, \emph{363}, 527\relax
\mciteBstWouldAddEndPuncttrue
\mciteSetBstMidEndSepPunct{\mcitedefaultmidpunct}
{\mcitedefaultendpunct}{\mcitedefaultseppunct}\relax
\EndOfBibitem
\bibitem[Yu et~al.(1999)Yu, Richter, Datta, Durbin, and
  Dutta]{yu_observation_1999}
Yu,~C.; Richter,~A.~G.; Datta,~A.; Durbin,~M.~K.; Dutta,~P. Observation of
  Molecular Layering in Thin Liquid Films Using {X-Ray} Reflectivity.
  \emph{Physical Review Letters} \textbf{1999}, \emph{82}, 2326\relax
\mciteBstWouldAddEndPuncttrue
\mciteSetBstMidEndSepPunct{\mcitedefaultmidpunct}
{\mcitedefaultendpunct}{\mcitedefaultseppunct}\relax
\EndOfBibitem
\bibitem[Yu et~al.(2000)Yu, Richter, Datta, Durbin, and
  Dutta]{yu_molecular_2000}
Yu,~C.~J.; Richter,~A.~G.; Datta,~A.; Durbin,~M.~K.; Dutta,~P. Molecular
  layering in a liquid on a solid substrate: an X-ray reflectivity study.
  \emph{Physica B: Condensed Matter} \textbf{2000}, \emph{283}, 27--31\relax
\mciteBstWouldAddEndPuncttrue
\mciteSetBstMidEndSepPunct{\mcitedefaultmidpunct}
{\mcitedefaultendpunct}{\mcitedefaultseppunct}\relax
\EndOfBibitem
\bibitem[Mo et~al.(2006)Mo, Evmenenko, Kewalramani, Kim, Ehrlich, and
  Dutta]{mo_observation_2006}
Mo,~H.; Evmenenko,~G.; Kewalramani,~S.; Kim,~K.; Ehrlich,~S.~N.; Dutta,~P.
  Observation of Surface Layering in a Nonmetallic Liquid. \emph{Physical
  Review Letters} \textbf{2006}, \emph{96}, 096107\relax
\mciteBstWouldAddEndPuncttrue
\mciteSetBstMidEndSepPunct{\mcitedefaultmidpunct}
{\mcitedefaultendpunct}{\mcitedefaultseppunct}\relax
\EndOfBibitem
\bibitem[Yu et~al.(2000)Yu, Richter, Kmetko, Datta, and Dutta]{yu_x-ray_2000}
Yu,~C.~J.; Richter,~A.~G.; Kmetko,~J.; Datta,~A.; Dutta,~P. X-ray diffraction
  evidence of ordering in a normal liquid near the solid-liquid interface.
  \emph{Europhysics Letters} \textbf{2000}, \emph{50}, 487--493\relax
\mciteBstWouldAddEndPuncttrue
\mciteSetBstMidEndSepPunct{\mcitedefaultmidpunct}
{\mcitedefaultendpunct}{\mcitedefaultseppunct}\relax
\EndOfBibitem
\bibitem[Snook and Megen(1980)Snook, and Megen]{snook_solvation_1980}
Snook,~I.~K.; Megen,~W.~v. Solvation forces in simple dense fluids. I.
  \emph{The Journal of Chemical Physics} \textbf{1980}, \emph{72},
  2907--2913\relax
\mciteBstWouldAddEndPuncttrue
\mciteSetBstMidEndSepPunct{\mcitedefaultmidpunct}
{\mcitedefaultendpunct}{\mcitedefaultseppunct}\relax
\EndOfBibitem
\bibitem[Bitsanis and Hadziioannou(1990)Bitsanis, and
  Hadziioannou]{bitsanis_molecular_1990}
Bitsanis,~I.; Hadziioannou,~G. Molecular dynamics simulations of the structure
  and dynamics of confined polymer melts. \emph{The Journal of Chemical
  Physics} \textbf{1990}, \emph{92}, 3827--3847\relax
\mciteBstWouldAddEndPuncttrue
\mciteSetBstMidEndSepPunct{\mcitedefaultmidpunct}
{\mcitedefaultendpunct}{\mcitedefaultseppunct}\relax
\EndOfBibitem
\bibitem[Stroud et~al.(2001)Stroud, Curry, and Cushman]{stroud_capillary_2001}
Stroud,~W.~J.; Curry,~J.~E.; Cushman,~J.~H. Capillary Condensation and Snap-off
  in Nanoscale Contacts. \emph{Langmuir} \textbf{2001}, \emph{17},
  688--698\relax
\mciteBstWouldAddEndPuncttrue
\mciteSetBstMidEndSepPunct{\mcitedefaultmidpunct}
{\mcitedefaultendpunct}{\mcitedefaultseppunct}\relax
\EndOfBibitem
\bibitem[Kaplan and Kauffmann(2006)Kaplan, and
  Kauffmann]{kaplan_structural_2006}
Kaplan,~W.; Kauffmann,~Y. Structural Order in Liquids Induced by Interfaces
  with Crystals. \emph{Annual Review of Materials Science} \textbf{2006},
  \emph{36}, 1--48\relax
\mciteBstWouldAddEndPuncttrue
\mciteSetBstMidEndSepPunct{\mcitedefaultmidpunct}
{\mcitedefaultendpunct}{\mcitedefaultseppunct}\relax
\EndOfBibitem
\bibitem[Schuster et~al.(2000)Schuster, Cichos, Wrachtrup, and von
  Borzcyskowski]{schuster_diffusion_2000}
Schuster,~J.; Cichos,~F.; Wrachtrup,~J.; von Borzcyskowski,~C. Diffusion of
  Single Molecules Close to Interfaces. \emph{Single Molecules} \textbf{2000},
  \emph{1}, 299--305\relax
\mciteBstWouldAddEndPuncttrue
\mciteSetBstMidEndSepPunct{\mcitedefaultmidpunct}
{\mcitedefaultendpunct}{\mcitedefaultseppunct}\relax
\EndOfBibitem
\bibitem[Schuster et~al.(2003)Schuster, Cichos, and von
  Borczyskowski]{schuster_anisotropic_2003}
Schuster,~J.; Cichos,~F.; von Borczyskowski,~C. Anisotropic diffusion of single
  molecules in thin liquid films. \emph{The European Physical Journal E: Soft
  Matter and Biological Physics} \textbf{2003}, \emph{12}, 75--80\relax
\mciteBstWouldAddEndPuncttrue
\mciteSetBstMidEndSepPunct{\mcitedefaultmidpunct}
{\mcitedefaultendpunct}{\mcitedefaultseppunct}\relax
\EndOfBibitem
\bibitem[T\"auber et~al.(2009)T\"auber, Heidern\"atsch, Bauer, Radons,
  Schuster, and von Borczyskowski]{tauber_single_2009}
T\"auber,~D.; Heidern\"atsch,~M.; Bauer,~M.; Radons,~G.; Schuster,~J.; von
  Borczyskowski,~C. Single molecule tracking of the molecular mobility in
  thinning liquid films on thermally grown {SiO2}. \emph{Diffusion Fundamentals
  Journal} \textbf{2009}, \emph{11}, 107(10)\relax
\mciteBstWouldAddEndPuncttrue
\mciteSetBstMidEndSepPunct{\mcitedefaultmidpunct}
{\mcitedefaultendpunct}{\mcitedefaultseppunct}\relax
\EndOfBibitem
\bibitem[Trenkmann et~al.(2009)Trenkmann, T\"auber, Bauer, Schuster, Bok,
  Gangopadhyhay, and von Borczyskowski]{trenkmann_investigations_2009}
Trenkmann,~I.; T\"auber,~D.; Bauer,~M.; Schuster,~J.; Bok,~S.;
  Gangopadhyhay,~S.; von Borczyskowski,~C. Investigations of solid liquid
  interfaces in ultra-thin liquid films via single particle tracking of silica
  nanoparticles. \emph{Diffusion Fundamentals Journal} \textbf{2009},
  \emph{11}, 108(12)\relax
\mciteBstWouldAddEndPuncttrue
\mciteSetBstMidEndSepPunct{\mcitedefaultmidpunct}
{\mcitedefaultendpunct}{\mcitedefaultseppunct}\relax
\EndOfBibitem
\bibitem[T\"auber(2011)]{tauber_characterization_2011}
T\"auber,~D. Characterization of heterogeneous diffusion in confined soft
  matter. PhD Thesis, {TU} Chemnitz, Chemnitz, 2011\relax
\mciteBstWouldAddEndPuncttrue
\mciteSetBstMidEndSepPunct{\mcitedefaultmidpunct}
{\mcitedefaultendpunct}{\mcitedefaultseppunct}\relax
\EndOfBibitem
\bibitem[Patil et~al.(2007)Patil, Matei, Grabowski, Hoffmann, and
  Mukhopadhyay]{patil_combined_2007}
Patil,~S.; Matei,~G.; Grabowski,~C.~A.; Hoffmann,~P.~M.; Mukhopadhyay,~A.
  Combined Atomic Force Microscopy and Fluorescence Correlation Spectroscopy
  Measurements to Study the Dynamical Structure of Interfacial Fluids.
  \emph{Langmuir} \textbf{2007}, \emph{23}, 4988--4992\relax
\mciteBstWouldAddEndPuncttrue
\mciteSetBstMidEndSepPunct{\mcitedefaultmidpunct}
{\mcitedefaultendpunct}{\mcitedefaultseppunct}\relax
\EndOfBibitem
\bibitem[Schob and Cichos(2010)Schob, and Cichos]{schob_single_2010}
Schob,~A.; Cichos,~F. Single molecule diffusion at step edges. \emph{Chemical
  Physics Letters} \textbf{2010}, \emph{484}, 192--196\relax
\mciteBstWouldAddEndPuncttrue
\mciteSetBstMidEndSepPunct{\mcitedefaultmidpunct}
{\mcitedefaultendpunct}{\mcitedefaultseppunct}\relax
\EndOfBibitem
\bibitem[Grabowski and Mukhopadhyay(2007)Grabowski, and
  Mukhopadhyay]{grabowski_comparing_2007}
Grabowski,~C.~A.; Mukhopadhyay,~A. Comparing the activation energy of diffusion
  in bulk and ultrathin fluid films. \emph{The Journal of Chemical Physics}
  \textbf{2007}, \emph{127}, 171101\relax
\mciteBstWouldAddEndPuncttrue
\mciteSetBstMidEndSepPunct{\mcitedefaultmidpunct}
{\mcitedefaultendpunct}{\mcitedefaultseppunct}\relax
\EndOfBibitem
\bibitem[Seemann et~al.(2001)Seemann, Herminghaus, and
  Jacobs]{seemann_dewetting_2001}
Seemann,~R.; Herminghaus,~S.; Jacobs,~K. Dewetting Patterns and Molecular
  Forces: A Reconciliation. \emph{Phys. Rev. Lett.} \textbf{2001}, \emph{86},
  5534--5537\relax
\mciteBstWouldAddEndPuncttrue
\mciteSetBstMidEndSepPunct{\mcitedefaultmidpunct}
{\mcitedefaultendpunct}{\mcitedefaultseppunct}\relax
\EndOfBibitem
\bibitem[Becker et~al.(2003)Becker, Grun, Seemann, Mantz, Jacobs, Mecke, and
  Blossey]{becker_complex_2003}
Becker,~J.; Grun,~G.; Seemann,~R.; Mantz,~H.; Jacobs,~K.; Mecke,~K.~R.;
  Blossey,~R. Complex dewetting scenarios captured by thin-film models.
  \emph{Nat Mater} \textbf{2003}, \emph{2}, 59--63\relax
\mciteBstWouldAddEndPuncttrue
\mciteSetBstMidEndSepPunct{\mcitedefaultmidpunct}
{\mcitedefaultendpunct}{\mcitedefaultseppunct}\relax
\EndOfBibitem
\bibitem[Seemann et~al.(2005)Seemann, Herminghaus, Neto, Schlagowski, Podzimek,
  Konrad, Mantz, and Jacobs]{seemann_dynamics_2005}
Seemann,~R.; Herminghaus,~S.; Neto,~C.; Schlagowski,~S.; Podzimek,~D.;
  Konrad,~R.; Mantz,~H.; Jacobs,~K. Dynamics and structure formation in thin
  polymer melt films. \emph{Journal of Physics: Condensed Matter}
  \textbf{2005}, \emph{17}, S267\relax
\mciteBstWouldAddEndPuncttrue
\mciteSetBstMidEndSepPunct{\mcitedefaultmidpunct}
{\mcitedefaultendpunct}{\mcitedefaultseppunct}\relax
\EndOfBibitem
\bibitem[H\"ahl et~al.(2012)H\"ahl, Evers, Grandthyll, Paulus, Sternemann,
  Loskill, Lessel, H\"usecken, Brenner, Tolan, and
  Jacobs]{hahl_subsurface_2012}
H\"ahl,~H.; Evers,~F.; Grandthyll,~S.; Paulus,~M.; Sternemann,~C.; Loskill,~P.;
  Lessel,~M.; H\"usecken,~A.~K.; Brenner,~T.; Tolan,~M.; Jacobs,~K. Subsurface
  Influence on the Structure of Protein Adsorbates as Revealed by in Situ X-ray
  Reflectivity. \emph{Langmuir} \textbf{2012}, \emph{28}, 7747--7756\relax
\mciteBstWouldAddEndPuncttrue
\mciteSetBstMidEndSepPunct{\mcitedefaultmidpunct}
{\mcitedefaultendpunct}{\mcitedefaultseppunct}\relax
\EndOfBibitem
\bibitem[Loskill et~al.(2012)Loskill, H\"ahl, Thewes, Kreis, Bischoff,
  Herrmann, and Jacobs]{loskill_influence_2012}
Loskill,~P.; H\"ahl,~H.; Thewes,~N.; Kreis,~C.~T.; Bischoff,~M.; Herrmann,~M.;
  Jacobs,~K. Influence of the Subsurface Composition of a Material on the
  Adhesion of Staphylococci. \emph{Langmuir} \textbf{2012}, \emph{28},
  7242--7248\relax
\mciteBstWouldAddEndPuncttrue
\mciteSetBstMidEndSepPunct{\mcitedefaultmidpunct}
{\mcitedefaultendpunct}{\mcitedefaultseppunct}\relax
\EndOfBibitem
\bibitem[Loskill et~al.(2012)Loskill, H\"ahl, Faidt, Grandthyll, M\"uller, and
  Jacobs]{loskill_is_2012}
Loskill,~P.; H\"ahl,~H.; Faidt,~T.; Grandthyll,~S.; M\"uller,~F.; Jacobs,~K. Is
  adhesion superficial? Silicon wafers as a model system to study van der Waals
  interactions. \emph{Advances in Colloid and Interface Science} \textbf{2012},
  a10.1016/j.cis.2012.06.006\relax
\mciteBstWouldAddEndPuncttrue
\mciteSetBstMidEndSepPunct{\mcitedefaultmidpunct}
{\mcitedefaultendpunct}{\mcitedefaultseppunct}\relax
\EndOfBibitem
\bibitem[Schulz et~al.(2011)Schulz, T\"auber, Schuster, Baumg\"artel, and von
  Borczyskowski]{schulz_influence_2011}
Schulz,~B.; T\"auber,~D.; Schuster,~J.; Baumg\"artel,~T.; von Borczyskowski,~C.
  Influence of mesoscopic structures on single molecule dynamics in thin
  smectic liquid crystal films. \emph{Soft Matter} \textbf{2011}, \emph{7},
  7431--7440\relax
\mciteBstWouldAddEndPuncttrue
\mciteSetBstMidEndSepPunct{\mcitedefaultmidpunct}
{\mcitedefaultendpunct}{\mcitedefaultseppunct}\relax
\EndOfBibitem
\bibitem[Baumg\"artel et~al.(2010)Baumg\"artel, {von Borczyskowski}, and
  Graaf]{baumgartel_fluorescence_2010}
Baumg\"artel,~T.; {von Borczyskowski},~C.; Graaf,~H. Fluorescence studies of
  Rhodamine {6G} functionalized silicon oxide nanostructures.
  \emph{Nanotechnology} \textbf{2010}, \emph{21}, 475205\relax
\mciteBstWouldAddEndPuncttrue
\mciteSetBstMidEndSepPunct{\mcitedefaultmidpunct}
{\mcitedefaultendpunct}{\mcitedefaultseppunct}\relax
\EndOfBibitem
\bibitem[Schulz et~al.(2010)Schulz, T\"auber, Friedriszik, Graaf, Schuster, and
  {von Borczyskowski}]{schulz_optical_2010}
Schulz,~B.; T\"auber,~D.; Friedriszik,~F.; Graaf,~H.; Schuster,~J.; {von
  Borczyskowski},~C. Optical detection of heterogeneous single molecule
  diffusion in thin liquid crystal films. \emph{Phys. Chem. Chem. Phys.}
  \textbf{2010}, \emph{12}, 11555--11564\relax
\mciteBstWouldAddEndPuncttrue
\mciteSetBstMidEndSepPunct{\mcitedefaultmidpunct}
{\mcitedefaultendpunct}{\mcitedefaultseppunct}\relax
\EndOfBibitem
\bibitem[W\"oll et~al.(2009)W\"oll, Braeken, Deres, {DeSchryver}, Uji-i, and
  Hofkens]{woll_polymers_2009}
W\"oll,~D.; Braeken,~E.; Deres,~A.; {DeSchryver},~F. C.~D.; Uji-i,~H.;
  Hofkens,~J. Polymers and single molecule fluorescence spectroscopy, what can
  we learn?. \emph{Chemical Society Reviews} \textbf{2009}, \emph{38},
  313--328\relax
\mciteBstWouldAddEndPuncttrue
\mciteSetBstMidEndSepPunct{\mcitedefaultmidpunct}
{\mcitedefaultendpunct}{\mcitedefaultseppunct}\relax
\EndOfBibitem
\bibitem[Saxton and Jacobson(1997)Saxton, and
  Jacobson]{saxton_single-particle_1997}
Saxton,~M.~J.; Jacobson,~K. {SINGLE-PARTICLE} {TRACKING:Applications} to
  Membrane Dynamics. \emph{Annual Review of Biophysics and Biomolecular
  Structure} \textbf{1997}, \emph{26}, 373--399\relax
\mciteBstWouldAddEndPuncttrue
\mciteSetBstMidEndSepPunct{\mcitedefaultmidpunct}
{\mcitedefaultendpunct}{\mcitedefaultseppunct}\relax
\EndOfBibitem
\bibitem[Schuster et~al.(2004)Schuster, Cichos, and {von
  Borzcyskowski}]{schuster_diffusion_2004}
Schuster,~J.; Cichos,~F.; {von Borzcyskowski},~C. Diffusion in ultrathin liquid
  films. \emph{European Polymer Journal} \textbf{2004}, \emph{40},
  993--999\relax
\mciteBstWouldAddEndPuncttrue
\mciteSetBstMidEndSepPunct{\mcitedefaultmidpunct}
{\mcitedefaultendpunct}{\mcitedefaultseppunct}\relax
\EndOfBibitem
\bibitem[Qian et~al.(1991)Qian, Sheetz, and Elson]{qian_single_1991}
Qian,~H.; Sheetz,~M.~P.; Elson,~E.~L. Single particle tracking. Analysis of
  diffusion and flow in two-dimensional systems. \emph{Biophysical Journal}
  \textbf{1991}, \emph{60}, 910--921\relax
\mciteBstWouldAddEndPuncttrue
\mciteSetBstMidEndSepPunct{\mcitedefaultmidpunct}
{\mcitedefaultendpunct}{\mcitedefaultseppunct}\relax
\EndOfBibitem
\bibitem[Schwille and Haustein(2002)Schwille, and
  Haustein]{schwille_fluorescence_2002}
Schwille,~P.; Haustein,~E. \emph{Single Molecule Techniques}; Biophysical
  Society: Bethesda, {MD}, 2002\relax
\mciteBstWouldAddEndPuncttrue
\mciteSetBstMidEndSepPunct{\mcitedefaultmidpunct}
{\mcitedefaultendpunct}{\mcitedefaultseppunct}\relax
\EndOfBibitem
\bibitem[Bauer et~al.(2009)Bauer, Heidern\"atsch, T\"auber, {von
  Borczyskowski}, and Radons]{bauer_investigations_2009}
Bauer,~M.; Heidern\"atsch,~M.; T\"auber,~D.; {von Borczyskowski},~C.;
  Radons,~G. Investigations of heterogeneous diffusion based on the probability
  density of scaled squared displacements observed from single molecules in
  ultra-thin liquid films. \emph{Diffusion Fundamentals Journal} \textbf{2009},
  \emph{11}, 104\relax
\mciteBstWouldAddEndPuncttrue
\mciteSetBstMidEndSepPunct{\mcitedefaultmidpunct}
{\mcitedefaultendpunct}{\mcitedefaultseppunct}\relax
\EndOfBibitem
\bibitem[Bauer et~al.(2011)Bauer, Valiullin, Radons, and
  K\"arger]{bauer_how_2011}
Bauer,~M.; Valiullin,~R.; Radons,~G.; K\"arger,~J. How to compare diffusion
  processes assessed by single-particle tracking and pulsed field gradient
  nuclear magnetic resonance. \emph{The Journal of Chemical Physics}
  \textbf{2011}, \emph{135}, 144118\relax
\mciteBstWouldAddEndPuncttrue
\mciteSetBstMidEndSepPunct{\mcitedefaultmidpunct}
{\mcitedefaultendpunct}{\mcitedefaultseppunct}\relax
\EndOfBibitem
\bibitem[Israelachvili(2011)]{israelachvili_intermolecular_2011}
Israelachvili,~J.~N. \emph{Intermolecular and surface forces}, 3rd ed.;
  Academic Press: Amsterdam ; Heidelberg [etc], 2011\relax
\mciteBstWouldAddEndPuncttrue
\mciteSetBstMidEndSepPunct{\mcitedefaultmidpunct}
{\mcitedefaultendpunct}{\mcitedefaultseppunct}\relax
\EndOfBibitem
\bibitem[Parsegian(2006)]{parsegian_van_2006}
Parsegian,~V.~A. \emph{Van der Waals Forces - A Handbook for Biologists,
  Chemists, Engineers, and Physicists}; Cambridge University Press, 2006\relax
\mciteBstWouldAddEndPuncttrue
\mciteSetBstMidEndSepPunct{\mcitedefaultmidpunct}
{\mcitedefaultendpunct}{\mcitedefaultseppunct}\relax
\EndOfBibitem
\bibitem[Jacobs et~al.(2008)Jacobs, Seemann, and
  Herminghaus]{jacobs_stability_2008}
Jacobs,~K.; Seemann,~R.; Herminghaus,~S. \emph{Polymer thin films}; Series in
  soft condensed matter; World Scientific: New Jersey, 2008; Vol.~1; pp
  243--266\relax
\mciteBstWouldAddEndPuncttrue
\mciteSetBstMidEndSepPunct{\mcitedefaultmidpunct}
{\mcitedefaultendpunct}{\mcitedefaultseppunct}\relax
\EndOfBibitem
\bibitem[Koma et~al.(1984)Koma, Sunouchi, and Miyajima]{koma_fabrication_1984}
Koma,~A.; Sunouchi,~K.; Miyajima,~T. Fabrication and characterization of
  heterostructures with subnanometer thickness. \emph{Microelectronic
  Engineering} \textbf{1984}, \emph{2}, 129--136\relax
\mciteBstWouldAddEndPuncttrue
\mciteSetBstMidEndSepPunct{\mcitedefaultmidpunct}
{\mcitedefaultendpunct}{\mcitedefaultseppunct}\relax
\EndOfBibitem
\bibitem[Koma(1999)]{koma_van_1999}
Koma,~A. Van der Waals epitaxy for highly lattice-mismatched systems.
  \emph{Journal of Crystal Growth} \textbf{1999}, \emph{201--202},
  236--241\relax
\mciteBstWouldAddEndPuncttrue
\mciteSetBstMidEndSepPunct{\mcitedefaultmidpunct}
{\mcitedefaultendpunct}{\mcitedefaultseppunct}\relax
\EndOfBibitem
\bibitem[Svec et~al.(2012)Svec, Merino, Dappe, Gonzalez, Abad, Jelinek, and
  Martin~Gago]{svec_van_2012}
Svec,~M.; Merino,~P.; Dappe,~Y.~J.; Gonzalez,~C.; Abad,~E.; Jelinek,~P.;
  Martin~Gago,~J.~A. van der Waals interactions mediating the cohesion of
  fullerenes on graphene. \emph{Phys. Rev. B} \textbf{2012}, \emph{86},
  121407\relax
\mciteBstWouldAddEndPuncttrue
\mciteSetBstMidEndSepPunct{\mcitedefaultmidpunct}
{\mcitedefaultendpunct}{\mcitedefaultseppunct}\relax
\EndOfBibitem
\bibitem[Vrij(1966)]{vrij_possible_1966}
Vrij,~A. Possible mechanism for the spontaneous rupture of thin, free liquid
  films. \textbf{1966}, \emph{42}, 23--33\relax
\mciteBstWouldAddEndPuncttrue
\mciteSetBstMidEndSepPunct{\mcitedefaultmidpunct}
{\mcitedefaultendpunct}{\mcitedefaultseppunct}\relax
\EndOfBibitem
\bibitem[Bocquet and Barrat(2007)Bocquet, and Barrat]{bocquet_flow_2007}
Bocquet,~L.; Barrat,~J. Flow boundary conditions from nano- to micro-scales.
  \emph{Soft Matter} \textbf{2007}, \emph{3}, 685--693\relax
\mciteBstWouldAddEndPuncttrue
\mciteSetBstMidEndSepPunct{\mcitedefaultmidpunct}
{\mcitedefaultendpunct}{\mcitedefaultseppunct}\relax
\EndOfBibitem
\bibitem[Hall and Torkelson(1998)Hall, and Torkelson]{hall_small_1998}
Hall,~D.~B.; Torkelson,~J.~M. Small Molecule Probe Diffusion in Thin and
  Ultrathin Supported Polymer Films. \emph{Macromolecules} \textbf{1998},
  \emph{31}, 8817--8825\relax
\mciteBstWouldAddEndPuncttrue
\mciteSetBstMidEndSepPunct{\mcitedefaultmidpunct}
{\mcitedefaultendpunct}{\mcitedefaultseppunct}\relax
\EndOfBibitem
\bibitem[Lin et~al.(1999)Lin, Kolb, Satija, and Wu]{lin_reduced_1999}
Lin,~E.~K.; Kolb,~R.; Satija,~S.~K.; Wu,~W.-l. Reduced Polymer Mobility near
  the {Polymer/Solid} Interface as Measured by Neutron Reflectivity.
  \emph{Macromolecules} \textbf{1999}, \emph{32}, 3753--3757\relax
\mciteBstWouldAddEndPuncttrue
\mciteSetBstMidEndSepPunct{\mcitedefaultmidpunct}
{\mcitedefaultendpunct}{\mcitedefaultseppunct}\relax
\EndOfBibitem
\bibitem[Hou et~al.(2000)Hou, Bardo, Martinez, and
  Higgins]{hou_characterization_2000}
Hou,~Y.; Bardo,~A.~M.; Martinez,~C.; Higgins,~D.~A. Characterization of
  Molecular Scale Environments in Polymer Films by Single Molecule
  Spectroscopy. \emph{The Journal of Physical Chemistry B} \textbf{2000},
  \emph{104}, 212--219\relax
\mciteBstWouldAddEndPuncttrue
\mciteSetBstMidEndSepPunct{\mcitedefaultmidpunct}
{\mcitedefaultendpunct}{\mcitedefaultseppunct}\relax
\EndOfBibitem
\bibitem[Araoz et~al.(2012)Araoz, T\"auber, von Borczyskowski, and
  Aramendia]{araoz_cage_2012}
Araoz,~B.; T\"auber,~D.; von Borczyskowski,~C.; Aramendia,~P.~F. Cage Effect in
  Poly(alkyl methacrylate) Thin Films Studied by Nile Red Single Molecule
  Fluorescence Spectroscopy. \emph{The Journal of Physical Chemistry C}
  \textbf{2012}, \emph{116}, 7573--7580\relax
\mciteBstWouldAddEndPuncttrue
\mciteSetBstMidEndSepPunct{\mcitedefaultmidpunct}
{\mcitedefaultendpunct}{\mcitedefaultseppunct}\relax
\EndOfBibitem
\bibitem[Hauschild et~al.(2010)Hauschild, Temirov, Soubatch, Bauer, Sch\"oll,
  Cowie, Lee, Tautz, and Sokolowski]{hauschild_normal-incidence_2010}
Hauschild,~A.; Temirov,~R.; Soubatch,~S.; Bauer,~O.; Sch\"oll,~A.; Cowie,~B.
  C.~C.; Lee,~T.; Tautz,~F.~S.; Sokolowski,~M. Normal-incidence x-ray
  standing-wave determination of the adsorption geometry of {PTCDA} on Ag(111):
  Comparison of the ordered room-temperature and disordered low-temperature
  phases. \emph{Phys. Rev. B} \textbf{2010}, \emph{81}, 125432\relax
\mciteBstWouldAddEndPuncttrue
\mciteSetBstMidEndSepPunct{\mcitedefaultmidpunct}
{\mcitedefaultendpunct}{\mcitedefaultseppunct}\relax
\EndOfBibitem
\bibitem[Ruiz et~al.(2012)Ruiz, Liu, Zojer, Scheffler, and
  Tkatchenko]{ruiz_density-functional_2012}
Ruiz,~V.~G.; Liu,~W.; Zojer,~E.; Scheffler,~M.; Tkatchenko,~A.
  {Density-Functional} Theory with Screened van der Waals Interactions for the
  Modeling of Hybrid {Inorganic-Organic} Systems. \emph{Phys. Rev. Lett.}
  \textbf{2012}, \emph{108}, 146103\relax
\mciteBstWouldAddEndPuncttrue
\mciteSetBstMidEndSepPunct{\mcitedefaultmidpunct}
{\mcitedefaultendpunct}{\mcitedefaultseppunct}\relax
\EndOfBibitem
\end{mcitethebibliography}


\providecommand*\mcitethebibliography{\thebibliography}
\csname @ifundefined\endcsname{endmcitethebibliography}
  {\let\endmcitethebibliography\endthebibliography}{}
\begin{mcitethebibliography}{15}
\providecommand*\natexlab[1]{#1}
\providecommand*\mciteSetBstSublistMode[1]{}
\providecommand*\mciteSetBstMaxWidthForm[2]{}
\providecommand*\mciteBstWouldAddEndPuncttrue
  {\def\EndOfBibitem{\unskip.}}
\providecommand*\mciteBstWouldAddEndPunctfalse
  {\let\EndOfBibitem\relax}
\providecommand*\mciteSetBstMidEndSepPunct[3]{}
\providecommand*\mciteSetBstSublistLabelBeginEnd[3]{}
\providecommand*\EndOfBibitem{}
\mciteSetBstSublistMode{f}
\mciteSetBstMaxWidthForm{subitem}{(\alph{mcitesubitemcount})}
\mciteSetBstSublistLabelBeginEnd
  {\mcitemaxwidthsubitemform\space}
  {\relax}
  {\relax}

\bibitem[Bauer et~al.(2009)Bauer, Heidern\"atsch, T\"auber, {von
  Borczyskowski}, and Radons]{bauer_investigations_2009}
Bauer,~M.; Heidern\"atsch,~M.; T\"auber,~D.; {von Borczyskowski},~C.;
  Radons,~G. Investigations of heterogeneous diffusion based on the probability
  density of scaled squared displacements observed from single molecules in
  ultra-thin liquid films. \emph{Diffusion Fundamentals Journal} \textbf{2009},
  \emph{11}, 104\relax
\mciteBstWouldAddEndPuncttrue
\mciteSetBstMidEndSepPunct{\mcitedefaultmidpunct}
{\mcitedefaultendpunct}{\mcitedefaultseppunct}\relax
\EndOfBibitem
\bibitem[Bauer et~al.(2011)Bauer, Valiullin, Radons, and
  K\"arger]{bauer_how_2011}
Bauer,~M.; Valiullin,~R.; Radons,~G.; K\"arger,~J. How to compare diffusion
  processes assessed by single-particle tracking and pulsed field gradient
  nuclear magnetic resonance. \emph{The Journal of Chemical Physics}
  \textbf{2011}, \emph{135}, 144118\relax
\mciteBstWouldAddEndPuncttrue
\mciteSetBstMidEndSepPunct{\mcitedefaultmidpunct}
{\mcitedefaultendpunct}{\mcitedefaultseppunct}\relax
\EndOfBibitem
\bibitem[T\"auber(2011)]{tauber_characterization_2011}
T\"auber,~D. Characterization of heterogeneous diffusion in confined soft
  matter. PhD Thesis, {TU} Chemnitz, Chemnitz, 2011\relax
\mciteBstWouldAddEndPuncttrue
\mciteSetBstMidEndSepPunct{\mcitedefaultmidpunct}
{\mcitedefaultendpunct}{\mcitedefaultseppunct}\relax
\EndOfBibitem
\bibitem[Montiel et~al.(2006)Montiel, Cang, and
  Yang]{montiel_quantitative_2006}
Montiel,~D.; Cang,~H.; Yang,~H. Quantitative Characterization of Changes in
  Dynamical Behavior for {Single-Particle} Tracking Studies. \emph{The Journal
  of Physical Chemistry B} \textbf{2006}, \emph{110}, 19763--19770\relax
\mciteBstWouldAddEndPuncttrue
\mciteSetBstMidEndSepPunct{\mcitedefaultmidpunct}
{\mcitedefaultendpunct}{\mcitedefaultseppunct}\relax
\EndOfBibitem
\bibitem[Heidern\"atsch(2009)]{heidernaetsch_development_2009}
Heidern\"atsch,~M. Development of a Computer Program for Simulation and
  Analysis of Particle Movement in thin Liquid Films. Master Thesis, {TU}
  Chemnitz, 2009\relax
\mciteBstWouldAddEndPuncttrue
\mciteSetBstMidEndSepPunct{\mcitedefaultmidpunct}
{\mcitedefaultendpunct}{\mcitedefaultseppunct}\relax
\EndOfBibitem
\bibitem[Grabowski and Mukhopadhyay(2007)Grabowski, and
  Mukhopadhyay]{grabowski_comparing_2007}
Grabowski,~C.~A.; Mukhopadhyay,~A. Comparing the activation energy of diffusion
  in bulk and ultrathin fluid films. \emph{The Journal of Chemical Physics}
  \textbf{2007}, \emph{127}, 171101\relax
\mciteBstWouldAddEndPuncttrue
\mciteSetBstMidEndSepPunct{\mcitedefaultmidpunct}
{\mcitedefaultendpunct}{\mcitedefaultseppunct}\relax
\EndOfBibitem
\bibitem[Schwille and Haustein(2002)Schwille, and
  Haustein]{schwille_fluorescence_2002}
Schwille,~P.; Haustein,~E. \emph{Single Molecule Techniques}; Biophysical
  Society: Bethesda, {MD}, 2002\relax
\mciteBstWouldAddEndPuncttrue
\mciteSetBstMidEndSepPunct{\mcitedefaultmidpunct}
{\mcitedefaultendpunct}{\mcitedefaultseppunct}\relax
\EndOfBibitem
\bibitem[Hac et~al.(2005)Hac, Seeger, Fidorra, and
  Heimburg]{hac_diffusion_2005}
Hac,~A.~E.; Seeger,~H.~M.; Fidorra,~M.; Heimburg,~T. Diffusion in
  {Two-Component} Lipid Membranes - A Fluorescence Correlation Spectroscopy and
  Monte Carlo Simulation Study. \emph{Biophysical Journal} \textbf{2005},
  \emph{88}, 317--333\relax
\mciteBstWouldAddEndPuncttrue
\mciteSetBstMidEndSepPunct{\mcitedefaultmidpunct}
{\mcitedefaultendpunct}{\mcitedefaultseppunct}\relax
\EndOfBibitem
\bibitem[Schulz et~al.(2010)Schulz, T\"auber, Friedriszik, Graaf, Schuster, and
  {von Borczyskowski}]{schulz_optical_2010}
Schulz,~B.; T\"auber,~D.; Friedriszik,~F.; Graaf,~H.; Schuster,~J.; {von
  Borczyskowski},~C. Optical detection of heterogeneous single molecule
  diffusion in thin liquid crystal films. \emph{Phys. Chem. Chem. Phys.}
  \textbf{2010}, \emph{12}, 11555--11564\relax
\mciteBstWouldAddEndPuncttrue
\mciteSetBstMidEndSepPunct{\mcitedefaultmidpunct}
{\mcitedefaultendpunct}{\mcitedefaultseppunct}\relax
\EndOfBibitem
\bibitem[Schulz et~al.(2011)Schulz, T\"auber, Schuster, Baumg\"artel, and von
  Borczyskowski]{schulz_influence_2011}
Schulz,~B.; T\"auber,~D.; Schuster,~J.; Baumg\"artel,~T.; von Borczyskowski,~C.
  Influence of mesoscopic structures on single molecule dynamics in thin
  smectic liquid crystal films. \emph{Soft Matter} \textbf{2011}, \emph{7},
  7431--7440\relax
\mciteBstWouldAddEndPuncttrue
\mciteSetBstMidEndSepPunct{\mcitedefaultmidpunct}
{\mcitedefaultendpunct}{\mcitedefaultseppunct}\relax
\EndOfBibitem
\bibitem[T\"auber et~al.(2009)T\"auber, Heidern\"atsch, Bauer, Radons,
  Schuster, and von Borczyskowski]{tauber_single_2009}
T\"auber,~D.; Heidern\"atsch,~M.; Bauer,~M.; Radons,~G.; Schuster,~J.; von
  Borczyskowski,~C. Single molecule tracking of the molecular mobility in
  thinning liquid films on thermally grown {SiO2}. \emph{Diffusion Fundamentals
  Journal} \textbf{2009}, \emph{11}, 107(10)\relax
\mciteBstWouldAddEndPuncttrue
\mciteSetBstMidEndSepPunct{\mcitedefaultmidpunct}
{\mcitedefaultendpunct}{\mcitedefaultseppunct}\relax
\EndOfBibitem
\bibitem[Israelachvili(2011)]{israelachvili_intermolecular_2011}
Israelachvili,~J.~N. \emph{Intermolecular and surface forces}, 3rd ed.;
  Academic Press: Amsterdam ; Heidelberg [etc], 2011\relax
\mciteBstWouldAddEndPuncttrue
\mciteSetBstMidEndSepPunct{\mcitedefaultmidpunct}
{\mcitedefaultendpunct}{\mcitedefaultseppunct}\relax
\EndOfBibitem
\bibitem[Seemann et~al.(2001)Seemann, Herminghaus, and
  Jacobs]{seemann_dewetting_2001}
Seemann,~R.; Herminghaus,~S.; Jacobs,~K. Dewetting Patterns and Molecular
  Forces: A Reconciliation. \emph{Phys. Rev. Lett.} \textbf{2001}, \emph{86},
  5534--5537\relax
\mciteBstWouldAddEndPuncttrue
\mciteSetBstMidEndSepPunct{\mcitedefaultmidpunct}
{\mcitedefaultendpunct}{\mcitedefaultseppunct}\relax
\EndOfBibitem
\bibitem[Jacobs et~al.(2008)Jacobs, Seemann, and
  Herminghaus]{jacobs_stability_2008}
Jacobs,~K.; Seemann,~R.; Herminghaus,~S. \emph{Polymer thin films}; Series in
  soft condensed matter; World Scientific: New Jersey, 2008; Vol.~1; pp
  243--266\relax
\mciteBstWouldAddEndPuncttrue
\mciteSetBstMidEndSepPunct{\mcitedefaultmidpunct}
{\mcitedefaultendpunct}{\mcitedefaultseppunct}\relax
\EndOfBibitem
\end{mcitethebibliography}
\end{document}


\subsection{Influence of confinement and heterogeneity on tracer diffusion coefficients}
In case of ultrathin films, neither single molecule tracking (SMT) nor fluorescence correlation spectroscopy (FCS) can resolve the vertical ($z$-)component of diffusion, but will detect only two-dimensional diffusion (the focal depth is about 1~$\mu m$ and thus at least two orders of magnitude larger than the film thickness $h$). We will explain the influence of film heterogeneity and confinement on the observed diffusion coefficients using a simple two-layer system with layer-dependent diffusion coefficients $D_1\neq D_2$ and exchange rates $f_{12}$ and $f_{21}$, where $f_{12}$ is the exchange rate from layer 2 into layer 1. Then, equilibrium probabilities $p_i$ to find the dye in layer $i$ are given as~\cite{bauer_investigations_2009, *bauer_how_2011, tauber_characterization_2011}
\begin{equation}
p_{i}=\frac{f_{ij}}{f_{ij}+f_{ji}}\, .
\label{equilibrium_prop}
\end{equation}
In the long time limit of the observation, the effective diffusion coefficient $D_{\rm eff}$ of the system can be calculated from the $p_i$ and the corresponding diffusion coefficients $D_i$~\cite{bauer_investigations_2009, *bauer_how_2011, tauber_characterization_2011}
\begin{equation}
D_{\rm eff}=\sum_i p_iD_i\, ,
\label{long_timeD}
\end{equation}
which is a weighted sum of the diffusion coefficients related to the involved heterogeneous regions. The observed diffusion coefficients from the two layer system depend on the relation of the exchange rates $f_{12}$ and $f_{21}$ to the observation time $\tau_{\rm obs}$~\cite{tauber_characterization_2011, bauer_investigations_2009, montiel_quantitative_2006,*heidernaetsch_development_2009}. For $\tau_{\rm obs}\gg1/f_{12}$ and $1/f_{21}$, the experiment will yield $D_{\rm eff}$. In case $\tau_{\rm obs}$ is much smaller than both inverse exchange rates, the layer dependent diffusion coefficients $D_1$ and $D_2$ will be observed with the respective amplitudes $p_1$ and $p_2$. In between these two cases, the temporal behavior of the system is complex and cannot be predicted easily~\cite{tauber_characterization_2011, bauer_investigations_2009, montiel_quantitative_2006,*heidernaetsch_development_2009}. 

The diffusion coefficient of the used Rhodamine dyes in bulk TEHOS is about 55~$\rm\mu m^2/s$~\cite{tauber_characterization_2011, grabowski_comparing_2007}. 
Assuming a homogeneous TEHOS film with thickness $h=100$~\AA\ and no structural change throughout the film, it is possible to evaluate the vertical mean first passage time (mfpt) for tracer molecules. For a reflecting boundary condition at the liquid-air boundary and an adsorbing boundary condition at the substrate, the vertical mfpt is shorter than $10^{-7}$~s for $D\geq5$~$\rm\mu m^2/s$~\cite{tauber_characterization_2011}. In other words, a tracer molecule desorbing from the substrate will take on average less than $10^{-7}$~s to be resorbed to the substrate under the assumption of a probability of 1 for resorption~\cite{tauber_characterization_2011}.

The temporal resolution of FCS is in principle suited to detect time as short as 0.1~$\rm\mu s=10^{-7}$~s. Although, the obtained autocorrelation curves from the FCS experiment did contain components of that order~\cite{tauber_characterization_2011}, the FCS curves were only analyzed down to 0.1~ms here. So the vertical mfpt could not be resolved directly. Nevertheless, the fast vertical passage of a 100~\AA\ thick TEHOS film has another influence on obtained diffusion coefficients. With FCS, translational diffusion coefficients are obtained from passages of the tracers through the focal area. The diameter of the focal area in the here used setup is 0.8 $\rm\mu m$. Using the {\it Einstein-Smoluchowski Equation} $D=\langle r^2\rangle/(4\tau)$ for two-dimensional diffusion~\cite{tauber_characterization_2011}, the mean square displacement during the mfpt can be calculated as $\langle r^2\rangle=4D\tau=2.2\cdot10^{-6}$~$\rm\mu m^2$. This is considerably smaller than the focal area. In a 100~\AA\ thick heterogeneous film, the diffusion coefficient obtained by FCS, for this reason, will be an average diffusion coefficient for the passage of the focal area. To resolve physical diffusion coefficients from the observed ones, the vertical structure of the film and the exchange rates between heterogeneous regions have to be known, which up to now is not the case.

With SMT, the smallest possible observation time is the inverse frame rate $\tau_{\rm frame}$, which in the here used experiment is 20~ms. This is several orders of magnitude larger than the above evaluated mfpt for a homogeneous film of thickness $h=100$~\AA. As a consequence also here the observed diffusion coefficients are strongly influenced by the vertical confinement. In particular, a bulk-like layer with diffusion coefficient $D\approx50$~$\rm\mu m^2/s$ cannot be resolved. Since $\tau_{\rm obs}\gg {\rm mfpt}$ even for a 100~\AA\ thick layer, it will not make much difference on the $D_{\rm eff}$, whether the tracer can diffuse with $D_{\rm bulk}$ in a 20~\AA\ thick layer or in a 100~\AA\ thick layer. Due to the fast mfpt, the transition rate out of this layer will mainly depend on the probability to absorb into the region with smaller mobility, and not on the thickness of the bulk-like layer.

$D_{\rm traj}$ obtained from trajectory analysis of SMT will suffer further bias from heterogeneity. Fast tracer molecules will be lost to some extend during tracking. From the evaluation of trajectories we know that tracer molecules will experience periods of slow and of fast diffusion within ultrathin TEHOS films. When a tracer is lost from tracking during a fast period, but can be tracked before and after this excursion into fast diffusion, the SMT experiment will show two separate trajectories (with small diffusion coefficient) instead of a single longer one. This one the one hand will increase the number of detected trajectories. On the other hand this will overestimate slow diffusion, because there now will be two trajectories with underestimated diffusion coefficients contributing to the distributions of $D_{\rm traj}$, while the lost period of fast diffusion contributes to neither one of them. As long as only some short events of fast diffusion are lost, the remaining distribution of $D_{\rm traj}$ will show a considerable amount of fast trajectories. If nearly all excursions of tracers into fast diffusion will be missed out, this will render only few remaining trajectories of considerable length which exceed the set lower limit on covered area to be taken into account for "mobile" $D_{\rm traj}$.

\subsection{Diffusion coefficients derived by FCS}
Due to the small focal area, the exciting laser power is considerable high in case of FCS experiments. Within our experiment it was 0.4~$\rm kW/cm^2$ (compared to 0.1 $\rm kW/cm^2$ for SMT). This results in enhanced photobleaching of the dye molecules. In particular, RhB used within SMT experiments bleached fast, yielding FCS curves within noise. R6G and OG were more stable, rendering reasonable autocorrelation curves as can be seen in \ref{fcs}. 

For fitting the FCS autocorrelation curves for translational diffusion, a two-dimensional geometry is assumed, since the film thickness of 100~\AA\ is much smaller as the focal depth of the illuminating laser beam/detection area.
A single component correlation function for diffusion in a two-dimensional geometry~\cite{schwille_fluorescence_2002, tauber_characterization_2011} did note fit to the experimental data. Fitting could be improved by using the following two-component correlation function~\cite{hac_diffusion_2005, tauber_characterization_2011}
\begin{equation}
G_{D,{\rm bi}}(\tau) = a_1\frac{1}{1+\tau/\tau_{D1}}+(1-a_1)\frac{1}{1+\tau/\tau_{D2}}\, .
\label{2Dcorr_bi}
\end{equation} 

From the characteristic time $\tau_D$ related to translational diffusion the diffusion coefficient $D$ can be derived by~\cite{schwille_fluorescence_2002, tauber_characterization_2011}
\begin{equation}
D = \frac{w_{xy}^2}{4\tau_D},
\label{D_wxy_tau}
\end{equation} 
where $w_{xy}$ is the lateral focus width (from center to 1/e decay of intensity) of the illumination/detection area on the sample, which was 0.4~$\rm\mu m$ in our case. Parameters from fits to the FCS curves using \ref{2Dcorr_bi} are shown in \ref{fcs_param}.
\begin{table}[ht]
  \caption{Fitting parameters of FCS autocorrelation curves according to \ref{2Dcorr_bi} and derived diffusion coefficients using \ref{D_wxy_tau} for 100~\AA\ thick TEHOS films.}
  \label{fcs_param}
  \begin{tabular}{cccccc}
    \hline
  tracer &$a_1$ &$\tau_{D1}$ [ms] &$D_1$ &$\tau_{D2}$ [s]& $D_2$\\
    \hline
    R6G&0.5&$9.4$&$0.06$ &0.28& $4$ \\
    OG &0.5&$5.0$&$0.14$ &0.70& $8$\\
    \hline
  \end{tabular}
\end{table}

\subsection{Diffusion coefficients obtained from probability distributions of diffusivities}
Diffusion coefficients $D_{\rm traj}$ calculated from mean square displacements along trajectories as shown in~\ref{diffusion} are strongly averaged. In particular, fast diffusion will not be resolved by this method. Probability distributions of diffusivities $d_{\rm diff}$ ( $d_{\rm diff}=r^2/(4\tau)$, i.e.\ scaled square displacements $r^2$ during a time lag $\tau$) are superior in respect to this and they also yield larger data sets~\cite{tauber_characterization_2011, schulz_optical_2010, *schulz_influence_2011}. 
\begin{figure}[hb]
  \includegraphics[width=2.1in]{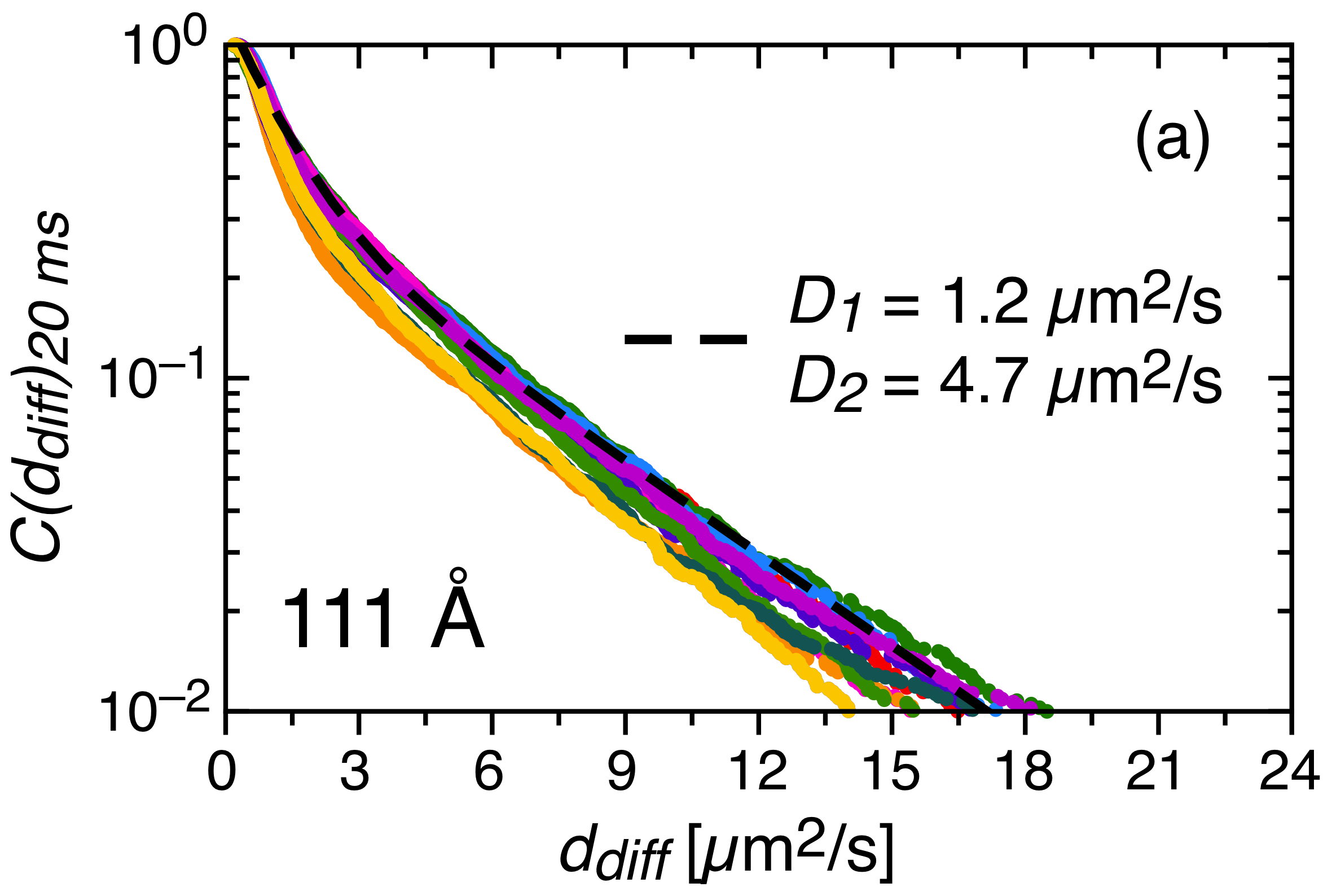}
  \includegraphics[width=2.1in]{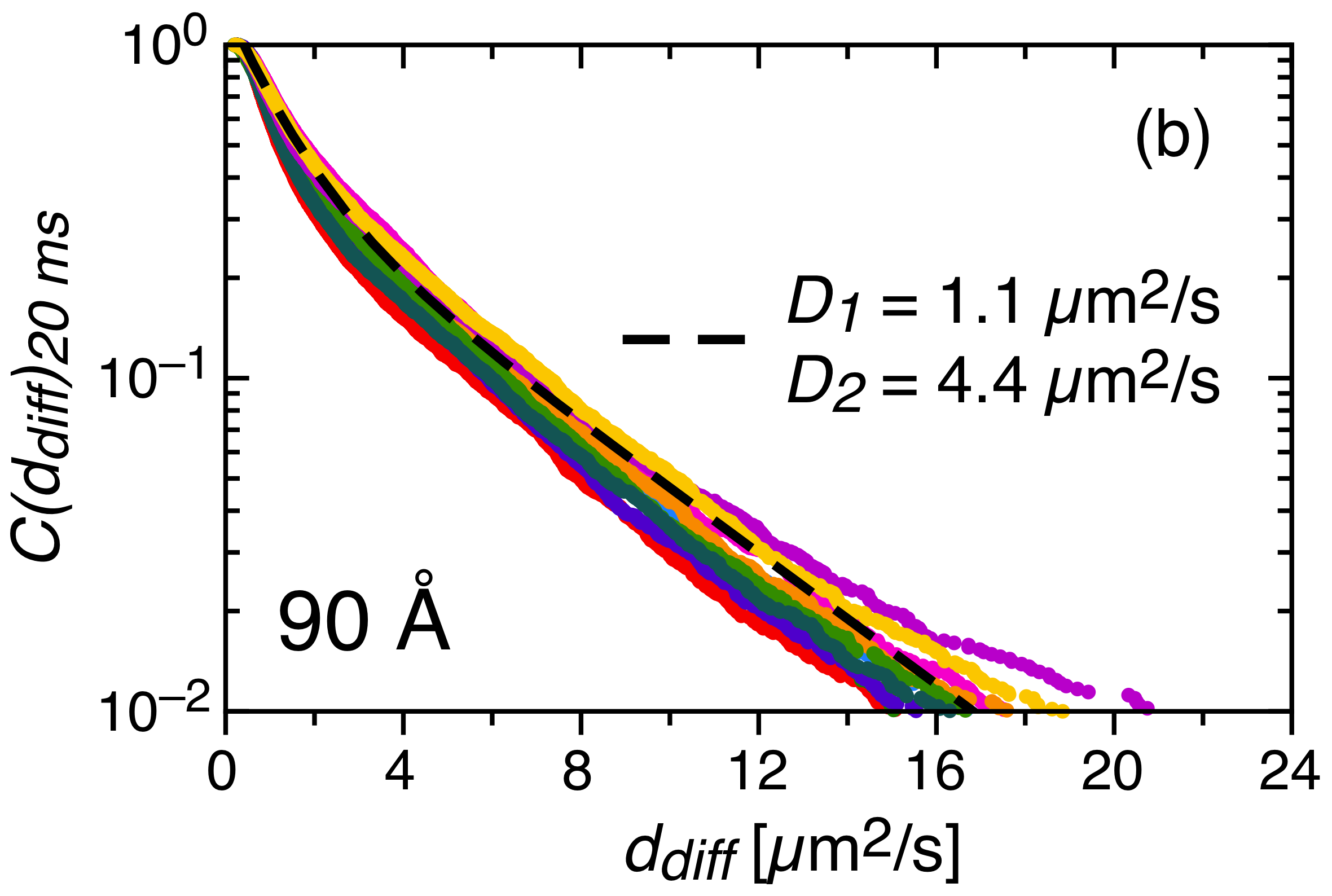}
  \includegraphics[width=2in]{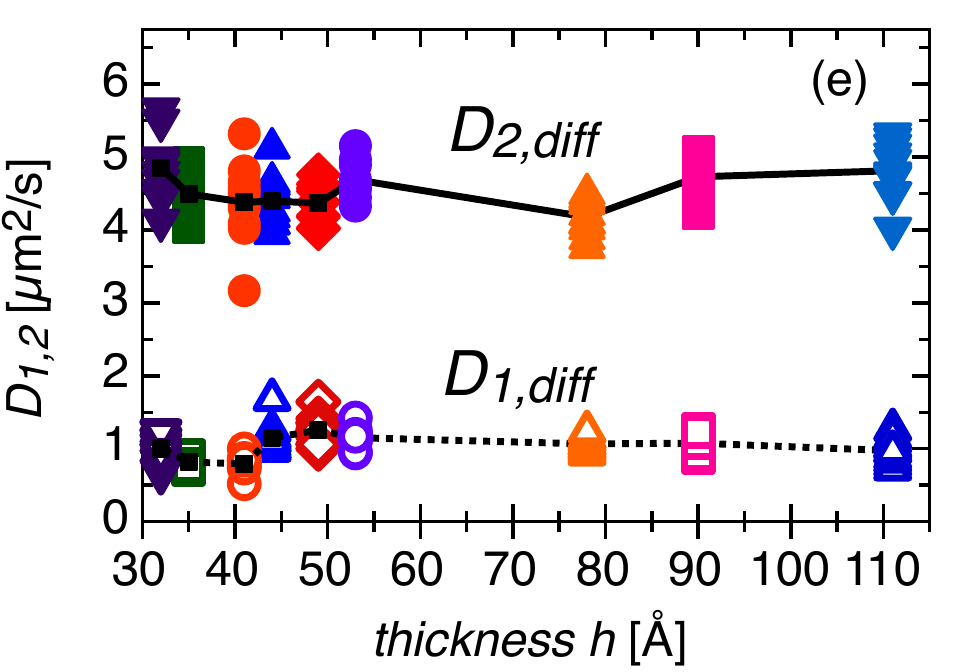}
  \includegraphics[width=2.1in]{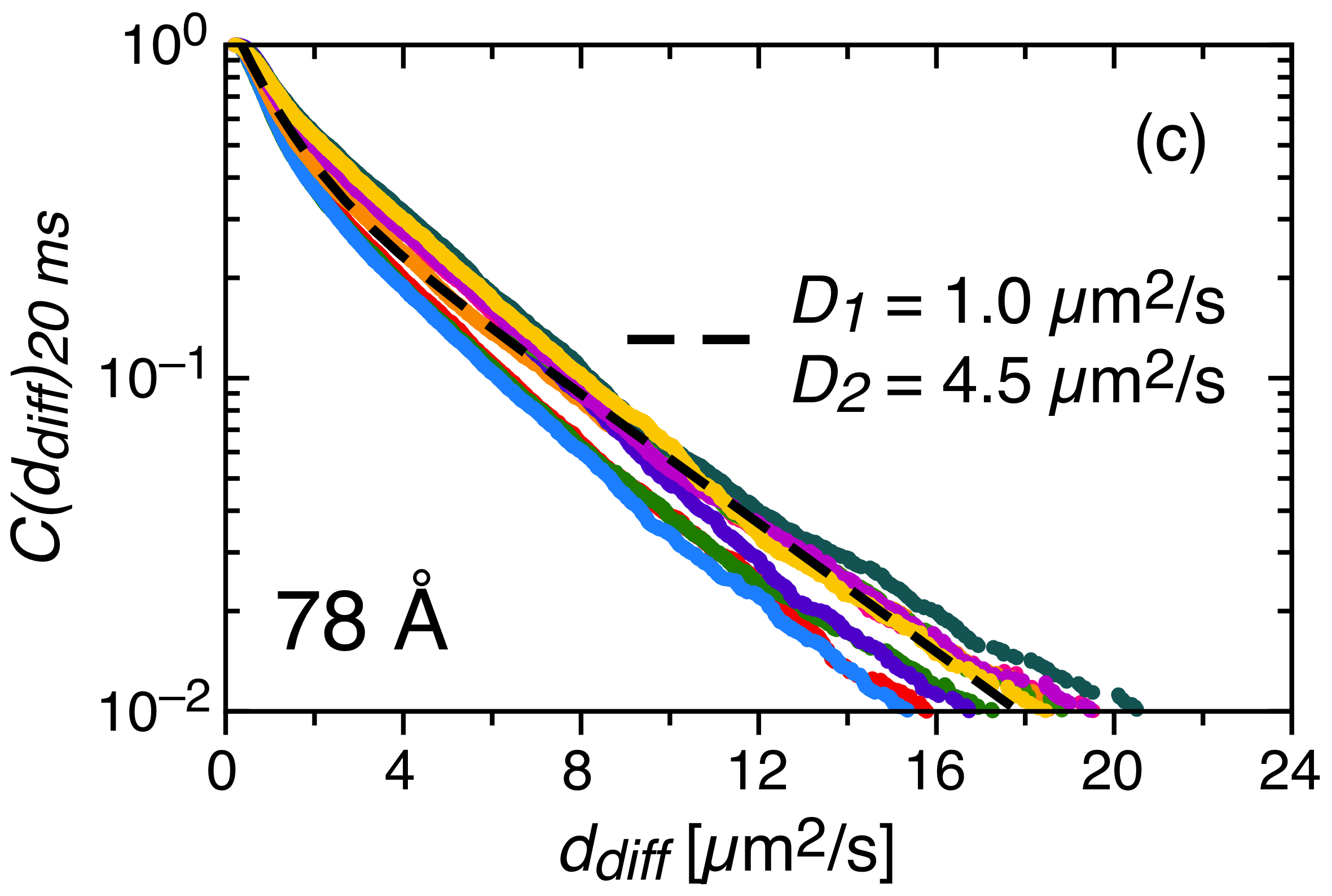}
  \includegraphics[width=2.1in]{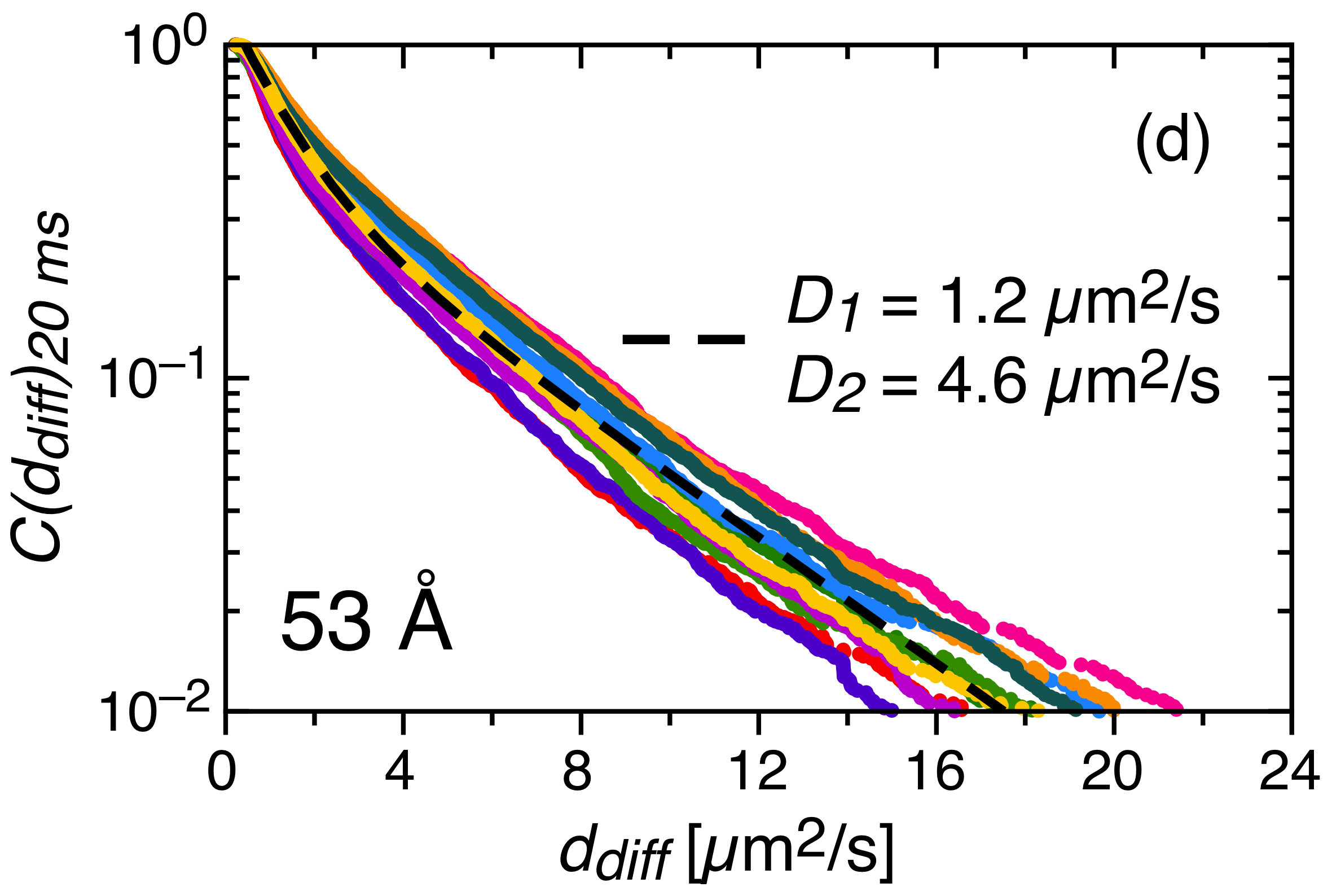}
  \includegraphics[width=2.in]{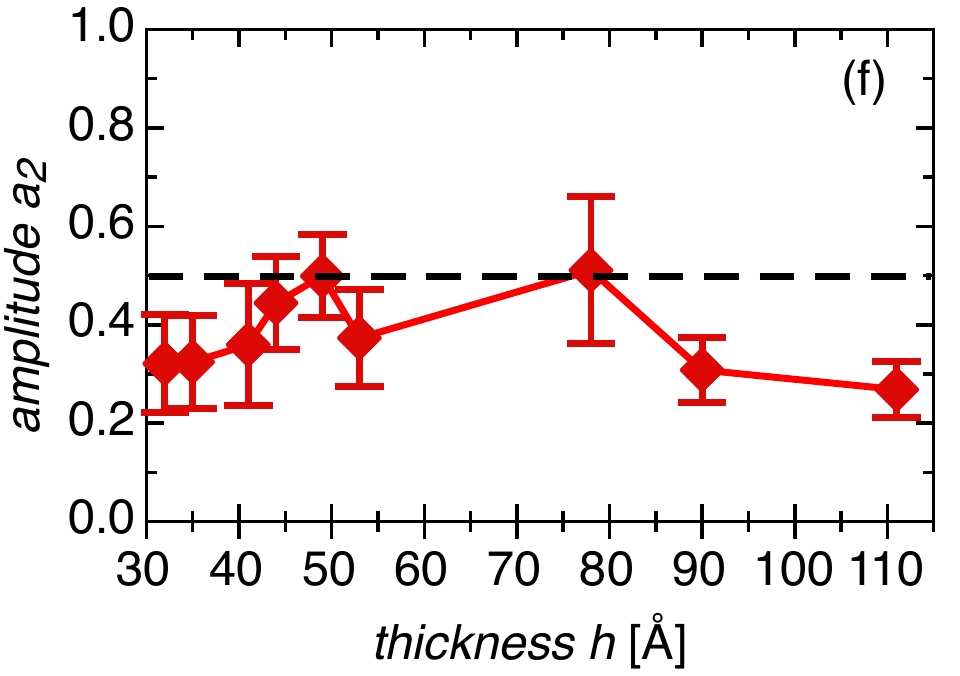}
  \caption{(a-d) Probability distributions $C(d_{\rm diff})_{\tau}$ of diffusivities $d_{\rm diff}$ from long term observation of RhB in a thinning TEHOS film, for film thickness $h$: (a) 111~\AA, (b) 90~\AA, (c) 78~\AA\ and (d) 53~\AA. The 10 different data sets for each film thickness are plotted with different colors. Dashed lines show examples of bi-exponential fits. (e) Diffusion coefficients $D_{\rm 1,diff}$, $D_{\rm 2,diff}$ from bi-exponential fits to probability distributions of diffusivities for each data set for decreasing $h$ from 111~\AA\ to 32~\AA. Mean values are connected by lines. (f) Corresponding amplitude $a_{2}$ of the fast component $D_{\rm 2, diff}$. Error bars denote standard deviations. (e,f) modified from~\cite{tauber_single_2009, tauber_characterization_2011}.}
  \label{diffusion2}
\end{figure}

In~\ref{diffusion2} (a-d) probability distributions of diffusivities are shown corresponding to the film thinning experiment shown in~\ref{diffusion}, for film thicknesses $h$ (a) 111~\AA, (b) 90~\AA, (c) 78~\AA\ and (d) 53~\AA. For the remaining five investigated film thicknesses between 49~\AA\ and 32~\AA, the corresponding probability distributions were previously published~\cite{tauber_single_2009, tauber_characterization_2011}. The $C(d_{\rm diff})_{\tau}$ contain only mobile $d_{\rm diff}$. Diffusivities related to completely immobile dye molecules had been excluded using a threshold related to the signal to noise ratio of the respective detected dye molecules~\cite{tauber_single_2009, tauber_characterization_2011}. This enables fitting the $C(d_{\rm diff})_{\tau}$ with a bi-exponential function according to~\cite{tauber_single_2009, tauber_characterization_2011}
\begin{equation}
C(d_{\rm diff})_{\tau}=a_1\exp\left(-\frac{d_{\rm diff}}{D_{\rm 1,diff}}\right)+a_2\exp\left(-\frac{d_{\rm diff}}{D_{\rm 2,diff}}\right)\, ,\quad a_1+a_2=1\, . 
\label{Prob2diff}
\end{equation}
Examples of such fits to $C(d_{\rm diff})_{\tau}$ are indicated as dashed lines together with the derived diffusion coefficients in~\ref{diffusion2} (a-d). \ref{diffusion2} (e) shows the obtained diffusion coefficients from fits to $C(d_{\rm diff})_{\tau}$ for $h$ decreasing from 111~\AA\ and 32~\AA\ using~\ref{Prob2diff}. Mean values of $D_{i,{\rm diff}}$ derived for each value of $h$ are connected by lines. They are almost constant during film thinning, yielding a fast component $D_{2,{\rm diff}}=4.5\pm 0.4$~$\rm\mu m^2/s$ and a slow component $D_{1,{\rm diff}}=1.0\pm 0.2$~$\rm\mu m^2/s$. The latter is influenced by the for this analysis set threshold for mobile tracers, which was chosen to ensure exclusion of all immobile tracers. Analysis of further SMT experiments on 60-80~\AA\ thick TEHOS films employing tri-exponential fits on probability distributions of diffusivities including immobile ones, yields $D_{1,\rm diff}=0.3\pm0.1$~$\rm\mu m^2/s$~\cite{tauber_characterization_2011}.
 In~\ref{diffusion2} (f) the mean amplitudes $a_2$ are shown for decreasing $h$ from 111~\AA\ and 32~\AA. Error bars denote standard deviations. The smaller $a_2$ for the two initial film thicknesses $h=111$~\AA\ and $h=90$~\AA\ can be explained by incomplete detection of fast diffusion, as is seen in comparison with results from FCS experiments on equal film thickness (see \ref{fcs_param}). The situation is different for $h\leq50$~\AA. There the smaller and almost constant number of detected trajectories (\ref{diffusion} a) points to a complete detection of tracer molecules. The decreasing values of $a_2$, therefore, are caused by a reduced mobility of tracer molecules, as is seen also by the change in the number of detected mobile trajectories in~\ref{diffusion} (b) and in the ratio of mobile to all trajectories in \ref{Evaporation} (b).

\subsection{Liquid flow}
\begin{figure}[h]
  \includegraphics[width=2.6in]{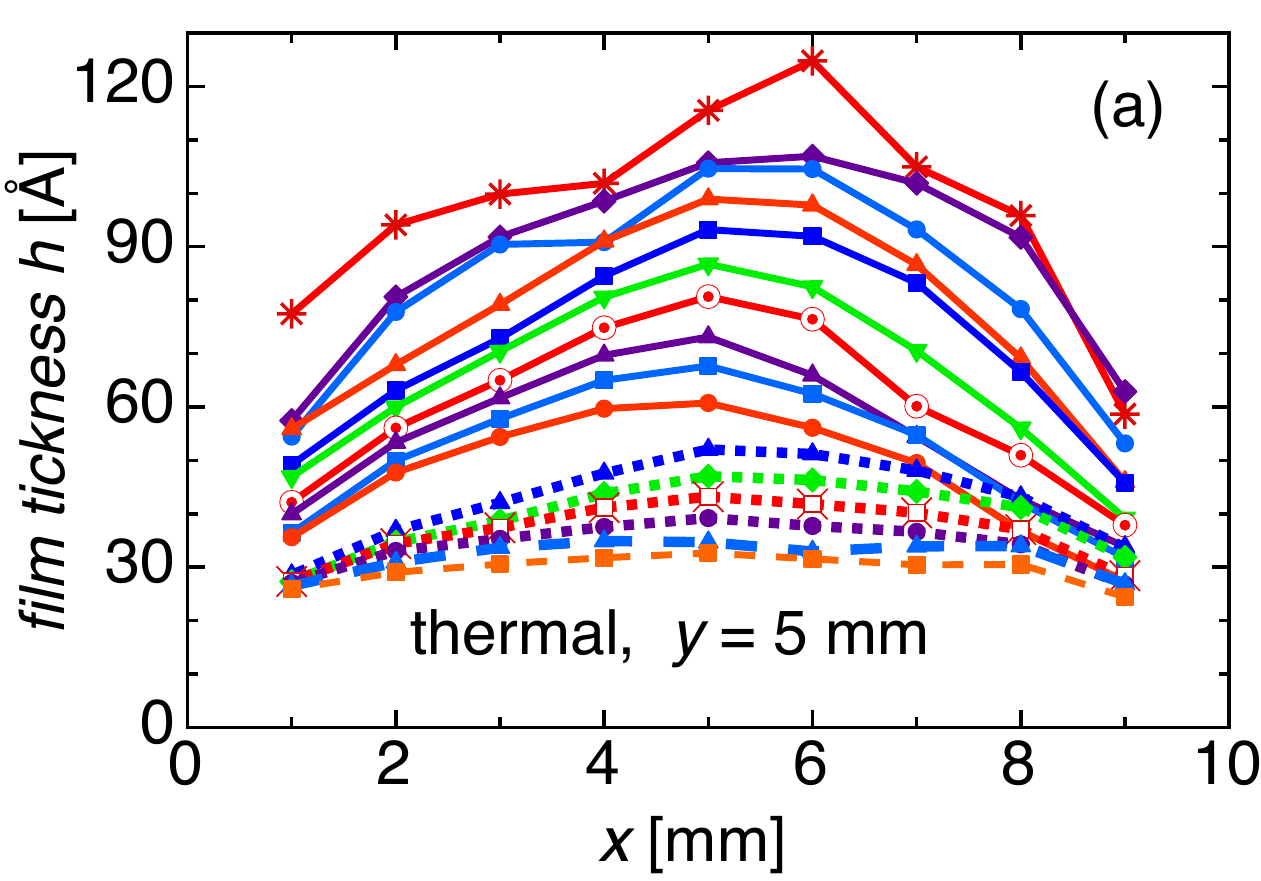}
  \hspace{0.2in}
  \includegraphics[width=2.6in]{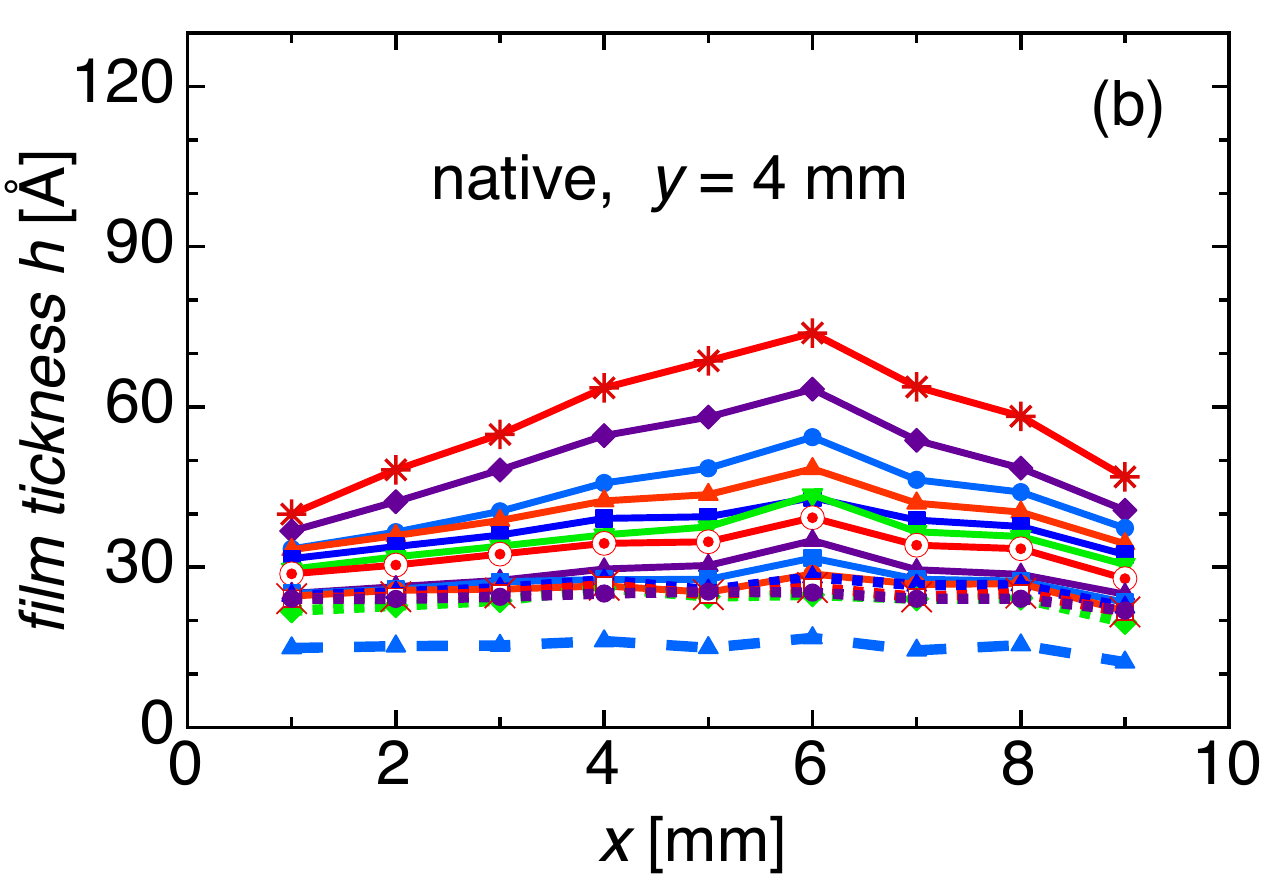}
  \caption{Height profiles at constant $y$-position of TEHOS films shown in \ref{Maps} during film thinning for films on (a) thermal ($y=5$ mm) and (b) native oxide ($y=4$ mm).}
  \label{profiles}
\end{figure}
\ref{tbl:thickness} gives the maximum thicknesses for the thinning TEHOS films on native and thermal oxide, for which  maps were shown in \ref{Maps}. For the film on thermal oxide, the position of the maximum film thickness changes during film thinning. In contrast, for the native oxide, the position of the maximum stays at $x=6$~mm, $y=4$~mm until  187~h after preparation. By this time, the film is flat as can be seen from the film profiles shown in \ref{profiles}. With a thickness below 35~\AA, it is in the regime of molecular layering, as can bee seen from the thinning rates in \ref{Evaporation} (a,b). For the flat film after 200~h, the maximum values are found at the edge of the sample, where touching the sample with pincers for mounting and demounting plays a role. For the flat film after 211~h from preparation the $h_{\rm max}$ are therefore shown in brackets.
\begin{table}[th]
  \caption{Maximum thickness (value and position) of TEHOS films shown in~\ref{Maps} during film thinning. Values in brackets are obtained at the edges within a flat film geometry.}
  \label{tbl:thickness}
  \begin{tabular}{|lccc|lccc|}
    \hline
   \multicolumn{4}{|c|}{TEHOS film on thermal oxide}&  \multicolumn{4}{|c|}{TEHOS film on native oxide}\\
    time [h] & $h_{\rm max}$ [\AA] & $x$ [mm] & $y$ [mm] & time [h] & $h_{\rm max}$ [\AA] & $x$ [mm] &$y$ [mm] \\
        \hline
     20  &133.4 & 3 & 3 & 18 & 73.8 & 6 & 4\\
     44 &149.6 & 6 & 3 & 43 & 63.3 & 6 & 4\\
     69 &108.2 & 6 & 4 & 66 & 54.3 & 6 & 4\\
     91 &99.5 & 6 & 4 & 89 & 48.4 & 6 & 4\\
   116 &93.2 & 5 & 5 & 114 & 43.0 & 6 & 4\\
   140 &86.7 & 5 & 5 & 138 & 43.5 & 6 & 4\\
   164 &80.7 & 5 & 5 & 162 & 39.3 & 6 & 4\\
   188 &73.1 & 5 & 5 & 187 & 35.0 & 6 & 4\\
   212 &67.7 & 5 & 5 & 211 & (33.9) & 9 & 2\\
   236 &60.7 & 5 & 5 & 235 & (34.7) & 9 & 2\\
   260 &54.5 & 4 & 6 & 259 & (34.8) & 9 & 2\\
   284 &49.5 & 4 & 6 & 283 & (31.1) & 1 & 1\\
   308 &45.4 & 6 & 7 & 306 & (32.1) & 9 & 2\\
   332 &41.9 & 6 & 7 & 331 & (28.1) & 1 & 1\\
   356 &38.5 & 5 & 7 & 354 & (22.3) & 9 & 2\\   
    \hline
  \end{tabular}
\end{table}

\subsection{Thinning rates including subsets}
\begin{figure}[th]
  \includegraphics[width=2.6in]{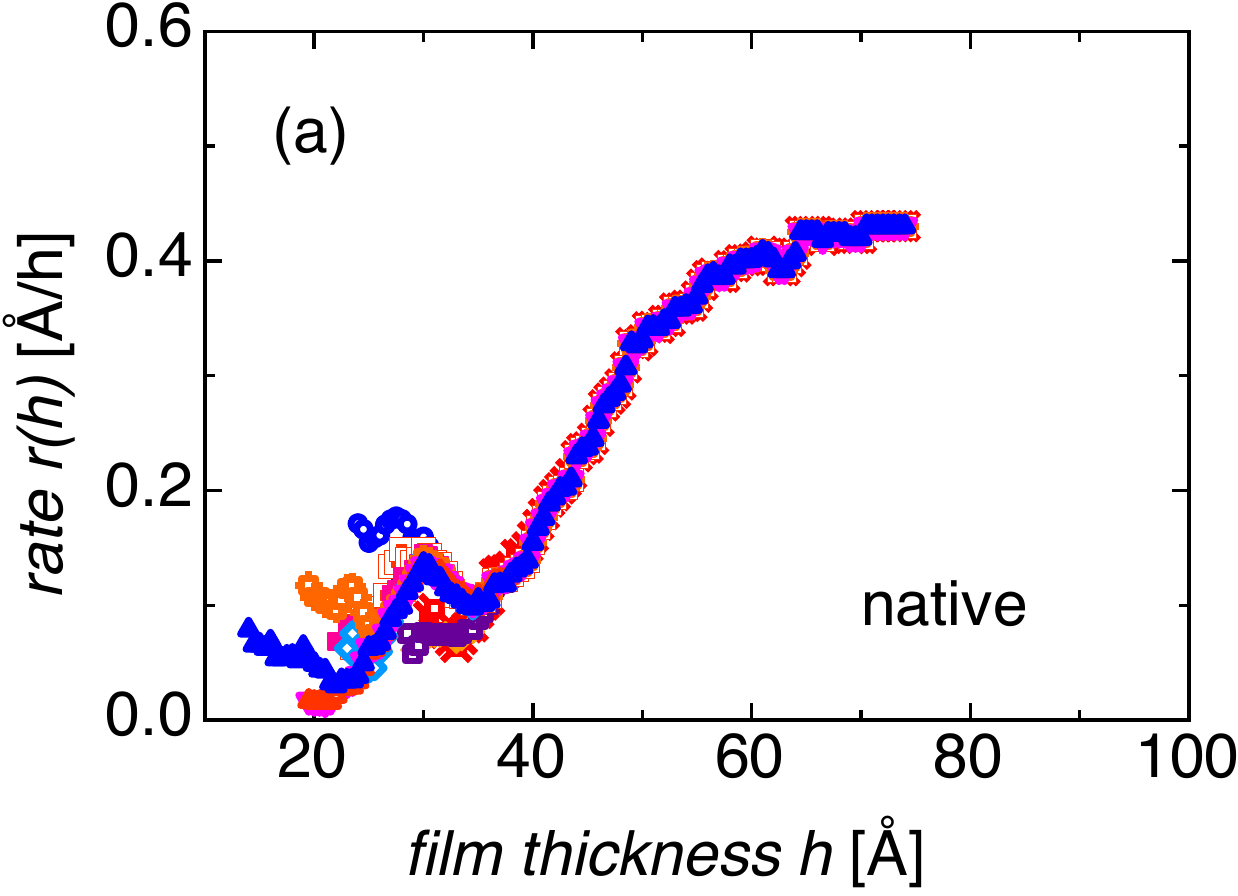}
  \includegraphics[width=2.6in]{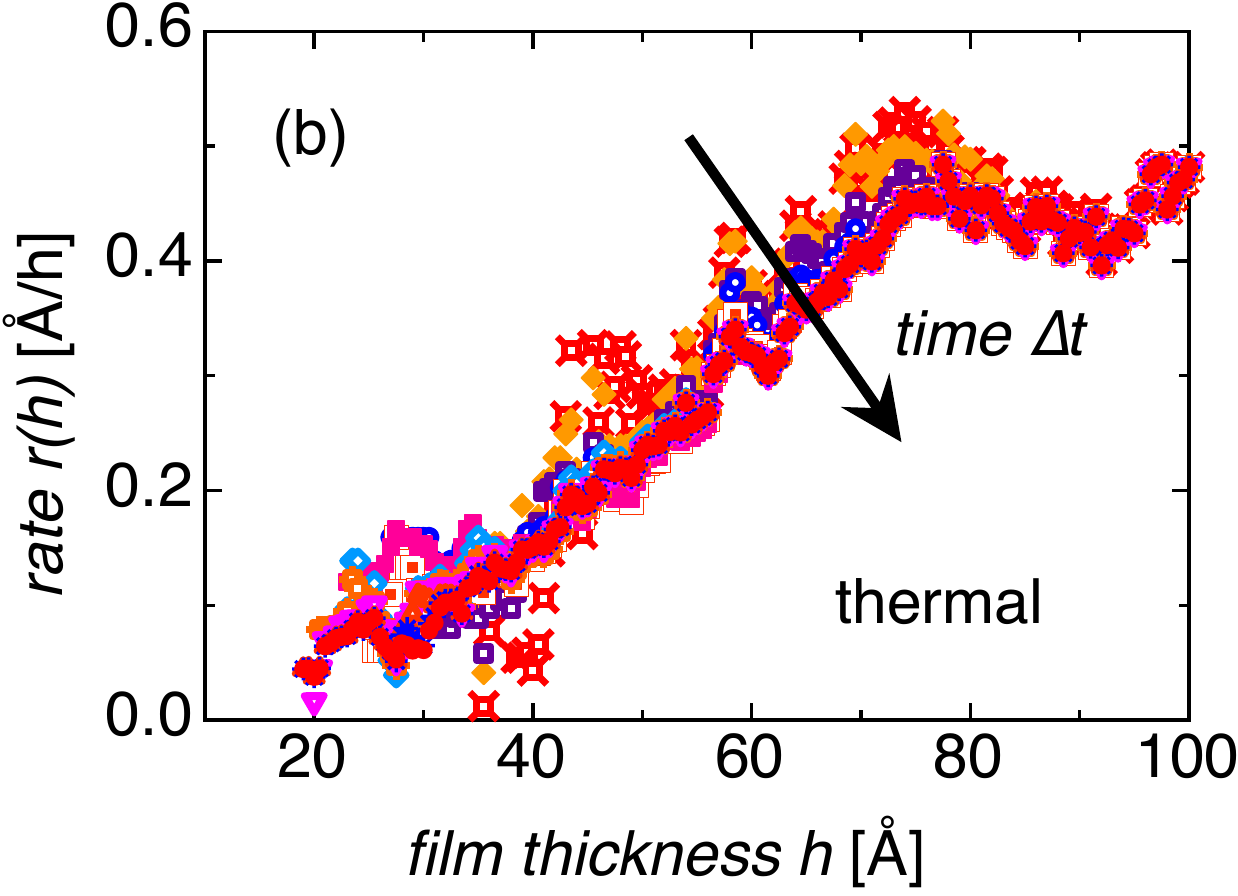}
  \caption{Film thinning rates for TEHOS films on (a) native oxide ($\blacktriangle$) and (b) 100 nm thermal oxide ($\bullet$) also shown in \ref{Evaporation} (a,c); including thinning rates calculated from several subsets (differing in colors and symbols) of the full data sets. The subsets contain increasing numbers of succeeding measurements after sample preparation, and thus cover increasing time spans $\Delta t$ of the total time of investigation.}
\label{Evaporation2}
\end{figure}

Average thinning rates calculated from different time spans $\Delta t$ of the total time of investigation for the films on (a) native and (b) thermal oxide shown in \ref{Evaporation} (a,c) native ($\blacktriangle$) and thermal ($\bullet$). For native oxide (\ref{Evaporation2} a), for $h<40$~\AA\ a scatter of the subsets about the thinning rate from the full data set is seen. This is due to large errors produced by averaging over not enough data by the particular binning method. Increased time of measurement produces more individual rates $r_i$ from thickness intervals $h_{i,2}-h_{i,1}$ yielding enough data for the relevant bins of thickness $h$ also for smaller $h$.

For the sample on thermal oxide (\ref{Evaporation2} b), the thickness dependent thinning rates decrease with increasing time span $\Delta t$ in the thickness range between 80 and 40~\AA, as is also seen for the other film on thermal oxide shown in \ref{Evaporation} (d).

\subsection{Derivation of interaction potentials from dielectric properties}
An approximation of van der Waals interactions may be obtained using the dielectric permittivities $\epsilon_i$ and refractive indices $n_i$ in the visible of the relevant materials (denoted by index $i$) together with the main electronic absorption frequency $\nu_e$ in the UV, as long as similar $\nu_e$ can be assumed for the used materials~\cite{israelachvili_intermolecular_2011}. The effective Hamaker coefficient $A_{12-23}$, describing the strength of interaction of material 1 with material 3 across medium 2 (for example \ce{SiO2} with air across TEHOS) can be approximated by~\cite{israelachvili_intermolecular_2011}
\begin{equation}
A_{12-23}\approx\frac{3}{4}k_BT\left(\frac{\epsilon_1-\epsilon_2}{\epsilon_1+\epsilon_2}\right)
\left(\frac{\epsilon_3-\epsilon_2}{\epsilon_3+\epsilon_2}\right)+\frac{3h\nu_e}{8\sqrt{2}}
\frac{(n_1^2-n_2^2)(n_3^2-n_2^2)}{\sqrt{n_1^2+n_2^2}\sqrt{n_3^2+n_2^2}
\left[\sqrt{n_1^2+n_2^2}+\sqrt{n_3^2+n_2^2}\right]}\, .
\label{Hamaker}
\end{equation}
The first term in the sum gives the contribution from permanent dipoles, while the second term is related to dispersion. Effective interface potentials $\Phi(h)$ for thin liquid films of thickness $h$ can be modeled as sum of short-range steric repulsion $C/h^8$ and the long-range non-retarded van der Waals potential $\Phi_{vdW}(h)$ according to~\cite{seemann_dewetting_2001,jacobs_stability_2008} 
\begin{equation}
\Phi(h)=\frac{C}{h^8}+\Phi_{vdW}(h)\, .
\label{interface_potential}
\end{equation}
In case of partially wetting polymer films, the strength $C$ of the steric repulsion can be derived from fits to experimentally obtained effective interface potentials~\cite{seemann_dewetting_2001, jacobs_stability_2008}. 

\begin{table}[b]
  \caption{Effective Hamaker coefficients [zJ $=10^{-21}$J] obtained using~\ref{Hamaker}, and components from permanent dipoles and from dispersion.}
  \label{tbl:effHam}
  \begin{tabular}{lccc}
    \hline
    material composition & permanent dipole part & dispersion part & effective Hamaker coefficient\\
    \hline
    Si/TEHOS/air & -0.7 & -41.7 & -42.5\\
    \ce{SiO2}/TEHOS/air & 0.2 & -2.4 & -2.3\\
    Si/hexane/air & -0.7& -37.1& -37.7\\
    \ce{SiO2}/hexane/air & 0.1& -9.0& -8.8\\
    Si/TEHOS/hexane & -0.1 & -6.1 & -6.2\\
    \ce{SiO2}/TEHOS/hexane & 0.02 & -0.3 & -0.3\\
    \hline
  \end{tabular}
\end{table}
Effective Hamaker coefficients for the used material compositions were calculated from optical properties using~\ref{Hamaker}. \ref{tbl:effHam} lists the obtained values, together with the corresponding contributions from permanent dipoles and from pure dispersion. Effective interface potentials $\Phi(h)$ for TEHOS films are calculated using~\ref{interface_potential}. Thereby, the strength of the short-range repulsion $C$ is estimated to be about $10^{-76}$~Jm$^{-6}$, since the interaction of the non-polar TEHOS with the polar \ce{SiO2} is expected to be weaker than that of polar polystyrene with polar \ce{SiO2} ($10^{-75}$~Jm$^{-6}$)~\cite{jacobs_stability_2008}, but stronger than the interaction of polystyrene with the non-polar OTS ($10^{-81}$~Jm$^{-6}$)\cite{jacobs_stability_2008}. For the silicon substrate with 100~nm thermal oxide, the simplified \ref{3Layer} 
can be used with $A_{SiO2}$ to calculate $\Phi_{vdW}(h)$, because for a thickness $d=100$~nm of the coating oxide layer, the contribution from $A_{Si}$ is negligible in \ref{4Layer}.
The average thickness of native oxide was determined by ellipsometry to be 2~nm. In case of native oxide,  \ref{4Layer} 
has to be used with a thickness $d=2$~nm of the coating layer. This yields a notable contribution from the underlying silicon.

\bibliography{VdW}